 \documentclass[onecolumn, resetfootnote]{aastex7}
\usepackage{newtxtext,newtxmath}
\usepackage[T1]{fontenc}
\usepackage{graphicx}	
\usepackage{amsmath}	
\usepackage{multirow}
\usepackage{threeparttable}
\usepackage{rotating}
\usepackage{color}
\usepackage{CJK}
\usepackage{soul}

\shorttitle{FRB~20240114A}
\shortauthors{Wang et al.}

\begin{document}
\begin{CJK*}{UTF8}{gbsn}

\title{Long-term simultaneous 2.25/8.60~GHz monitoring of the newly-discovered repeating FRB~20240114A}

\correspondingauthor{Zhen Yan (闫振), Zhi-Qiang Shen (沈志强)}

\author{Xiao-Wei Wang (王啸威)}
\affiliation{Shanghai Astronomical Observatory, Chinese Academy of Sciences, 80 Nandan Road, Shanghai 200030, People's Republic of China}
\affiliation{School of Astronomy and Space Sciences, University of Chinese Academy of Sciences, No. 19A Yuquan Road, Beijing 100049, People's Republic of China}
\email{xiaoweiwang@shao.ac.cn}

\author{Zhen Yan (闫振)}
\affiliation{Shanghai Astronomical Observatory, Chinese Academy of Sciences, 80 Nandan Road, Shanghai 200030, People's Republic of China}
\affiliation{School of Astronomy and Space Sciences, University of Chinese Academy of Sciences, No. 19A Yuquan Road, Beijing 100049, People's Republic of China}
\affiliation{Key Laboratory of Radio Astronomy and Technology, CAS, A20 Datun Road, Chaoyang District, Beijing 100101, People's Republic of China}
\email[show]{yanzhen@shao.ac.cn}

\author{Zhi-Qiang Shen (沈志强)}
\affiliation{Shanghai Astronomical Observatory, Chinese Academy of Sciences, 80 Nandan Road, Shanghai 200030, People's Republic of China}
\affiliation{School of Astronomy and Space Sciences, University of Chinese Academy of Sciences, No. 19A Yuquan Road, Beijing 100049, People's Republic of China}
\affiliation{Key Laboratory of Radio Astronomy and Technology, CAS, A20 Datun Road, Chaoyang District, Beijing 100101, People's Republic of China}
\affiliation{School of Physical Science and Technology, ShanghaiTech University, Shanghai 201210, People's Republic of China}
\email[show]{zshen@shao.ac.cn}

\author{Ke-Jia Lee (李柯伽)}
\affiliation{Department of Astronomy, School of Physics, Peking University, Beijing 100871, People's Republic of China}
\affiliation{Kavli Institute for Astronomy and Astrophysics, Peking University, Beijing 100871, People's Republic of China}
\email{}

\author{Ya-Jun Wu (吴亚军)}
\affiliation{Shanghai Astronomical Observatory, Chinese Academy of Sciences, 80 Nandan Road, Shanghai 200030, People's Republic of China}
\affiliation{Key Laboratory of Radio Astronomy and Technology, CAS, A20 Datun Road, Chaoyang District, Beijing 100101, People's Republic of China}
\email{}

\author{Rong-Bing Zhao (赵融冰)}
\affiliation{Shanghai Astronomical Observatory, Chinese Academy of Sciences, 80 Nandan Road, Shanghai 200030, People's Republic of China}
\affiliation{Key Laboratory of Radio Astronomy and Technology, CAS, A20 Datun Road, Chaoyang District, Beijing 100101, People's Republic of China}
\email{}

\author{Jie Liu (刘杰)}
\affiliation{Shanghai Astronomical Observatory, Chinese Academy of Sciences, 80 Nandan Road, Shanghai 200030, People's Republic of China}
\email{}

\author{Rui Wang (王睿)}
\affiliation{Shanghai Astronomical Observatory, Chinese Academy of Sciences, 80 Nandan Road, Shanghai 200030, People's Republic of China}
\affiliation{School of Astronomy and Space Sciences, University of Chinese Academy of Sciences, No. 19A Yuquan Road, Beijing 100049, People's Republic of China}
\email{}

\author{Kuo Liu (刘阔)}
\affiliation{Shanghai Astronomical Observatory, Chinese Academy of Sciences, 80 Nandan Road, Shanghai 200030, People's Republic of China}
\affiliation{Key Laboratory of Radio Astronomy and Technology, CAS, A20 Datun Road, Chaoyang District, Beijing 100101, People's Republic of China}
\email{}

\author{Yuan-Chuan Zou (邹远川)}
\affiliation{Department of Astronomy, School of Physics, Huazhong University of Science and Technology, Wuhan 430074, People's Republic of China}
\email{}

\author{Zhi-Peng Hang (黄志鹏)}
\affiliation{Department of Physics and Astronomy, Hubei University of Education, Wuhan 430205, People's Republic of China}
\email{}

\author{Chu-Yuan Zhang (张楚原)}
\affiliation{Shanghai Astronomical Observatory, Chinese Academy of Sciences, 80 Nandan Road, Shanghai 200030, People's Republic of China}
\affiliation{School of Astronomy and Space Sciences, University of Chinese Academy of Sciences, No. 19A Yuquan Road, Beijing 100049, People's Republic of China}
\email{}

\author{Fan Yang (杨帆)}
\affiliation{Shanghai Astronomical Observatory, Chinese Academy of Sciences, 80 Nandan Road, Shanghai 200030, People's Republic of China}
\affiliation{School of Astronomy and Space Sciences, University of Chinese Academy of Sciences, No. 19A Yuquan Road, Beijing 100049, People's Republic of China}
\email{}

\author{Zhen-Long Liao (廖振龙)}
\affiliation{Shanghai Astronomical Observatory, Chinese Academy of Sciences, 80 Nandan Road, Shanghai 200030, People's Republic of China}
\affiliation{School of Astronomy and Space Sciences, University of Chinese Academy of Sciences, No. 19A Yuquan Road, Beijing 100049, People's Republic of China}
\email{}

\author{Yang-Yang Lin (林扬洋)}
\affiliation{Shanghai Astronomical Observatory, Chinese Academy of Sciences, 80 Nandan Road, Shanghai 200030, People's Republic of China}
\affiliation{College of Science, Shanghai University, Shanghai 200444, People's Republic of China}
\email{}

\begin{abstract}

We report on the simultaneous monitoring of the repeating fast radio burst (FRB) 20240114A at 2.25 and 8.60~GHz, conducted 66 times between 2024 January 29 and 2025 February 15 with the Shanghai Tianma Radio Telescope (TMRT). In about 180 hours of observation, we detected 155 bursts at 2.25~GHz above a fluence threshold of 0.72~Jy~ms, but none at 8.60~GHz above a fluence threshold of 0.27~Jy~ms. FRB~20240114A exhibited frequency-dependent activity, as evidenced by the non-detections in 14.3 hours of observations at 2.25~GHz prior to 2024 February 24, despite its reported activity below 2~GHz. In contrast to its low-activity state reported below 1.4~GHz between 2024 June and December, FRB~20240114A exhibited high activity at 2.25~GHz in 2024 July with a mean burst rate of $1.72^{+0.18}_{-0.16}~\rm{hr}^{-1}$, followed by a low-activity state. We also detected a short-term reactivation at 2.25~GHz around 2025 January 20, about two weeks after renewed activity was reported below 1.4~GHz by other telescopes. The median burst width at 2.25~GHz is 3~ms, which is narrower than that at lower frequencies. The waiting time distribution peaks at 1019~s, and burst arrivals on hourly timescales consistent with a Poisson process. The isotropic-equivalent energy of bursts spans $10^{37} -10^{39}$~erg. The distribution of burst energy above the completeness threshold ($7.5\times10^{37}$~erg) follows a power-law relation with an index of $\gamma=-1.20\pm0.03\pm0.02$. Finally, we find that FRB~20240114A is at least two orders of magnitude less active at 8.60~GHz than at 2.25~GHz, and we constrain the broadband spectra of the detected bursts.
\end{abstract}


\section{Introduction}
\label{sec:intro}

Fast radio bursts (FRBs) are bright, microsecond- to millisecond-duration radio transients observed out to cosmological distances \citep[see][for FRB reviews]{Cordes2019, Petroff2019, Petroff2022}. Since the first FRB was discovered by \citet{Lorimer2007}, more than 800\footnote{\url{https://blinkverse.zero2x.org/overview}} FRB sources have been reported. However, the physical origin and emission mechanism of FRBs remain unclear. An important clue emerged when the Canadian Hydrogen Intensity Mapping Experiment Fast Radio Burst (CHIME/FRB) project and the Survey for Transient Astronomical Radio Emission 2 (STARE2) discovered an FRB-like burst from the Galactic magnetar SGR~1935+2154 \citep{CHIME2020, Bochenek2020}, suggesting a link between magnetars and FRBs. Additionally, \citet{Kirsten2022} localized FRB~20200120E to an old globular cluster in the nearby galaxy M81. This suggests that if magnetars power FRBs, they may have a variety of formation channels, such as accretion-induced collapse, core-collapse supernovae or binary mergers \citep{Margalit2020, Gourdji2020}. Meanwhile, other possible theoretical models remain under consideration, such as supergiant bursts from extragalactic neutron stars \citep{Cordes2016}, or interactions between the neutron star and asteroids/comets \citep{Geng2015, Dai2016}.

While most known FRBs appear to be one-off events, a subset \citep[$\sim$3\%,][]{CHIME2023} has been reported to repeat. Statistical studies of burst properties also reveal that repeating FRBs are generally narrower in bandwidths and shorter in burst durations compared with non-repeating FRBs \citep{Pleunis2021b, Curtin2024}. Some repeating FRBs exhibit hyperactive episodes, during which the mean burst rate can reach up to several hundred per hour \citep[e.g.,][]{Li2021, Zhang2022, Zhang2023, Jahns2023, Nimmo2023}. Monitoring repeaters, particularly hyperactive ones, enables detailed studies of burst energy distributions \citep[e.g.,][]{Li2021, Kirsten2024}, reveals a rich variety of polarization behaviors \citep[e.g.,][]{Luo2020, Jiang2024, Niu2024}, and provides insights into the dynamically evolving local environments of some FRBs \citep[e.g.,][]{Xu2022, Anna2023, Li2025}. Together, these studies may shed light on the mysterious origin and emission mechanisms of FRBs.

Multi-frequency observations are crucial for studying and constraining the frequency-dependent properties of FRBs. For example, bursts from repeating FRBs detected at high frequencies tend to have narrower widths compared with those observed at lower frequencies \citep{Michilli2018, Gajjar2018, Bethapudi2023}, and FRBs tend to exhibit higher activity at certain frequencies \citep[e.g.,][]{Houben2019, Pastor2021, Bethapudi2023}. Furthermore, multi-frequency monitoring also revealed that FRB~20180916B, which has a periodic activity cycle of $\sim$$16.3$~days \citep{CHIME2020b}, shows chromatic activity with earlier and narrower activity windows towards higher frequencies \citep{Pleunis2021b, Pastor2021, Bethapudi2023} and inspires various theoretical models \citep[e.g.,][]{Ioka2020, Lidong2021, Liqiao2021, Sridhar2021}.

On 2024 January 26, the CHIME/FRB collaboration reported the discovery of repeating FRB~20240114A with a dispersion measure (DM) of $\sim527.7~\rm{pc}~\rm{cm}^{-3}$ \citep{Shin2025}. In total, three bursts were detected by CHIME in the 400$-$800~MHz band within ten days (between 2024 January 14 and 24), with the brightest one having a fluence of $1014\pm108$~Jy~ms. This indicates that the source is highly active, given CHIME's daily exposure time of only about 5.5 minutes at its location. Subsequent MeerKAT observations \citep{Tian2024} and more precise Very Long Baseline Interferometry (VLBI) observations by the European VLBI Network (EVN) \citep[EVN-PRECISE team,][]{ATel16542} confidently localized FRB~20240114A to its host galaxy SDSS~J212739.84+041945.8, which has a redshift of $z=0.1300\pm0.0002$ obtained from the optical follow-up using the Gran Telescopio Canarias \citep[GTC,][]{ATel16613}. Additionally, follow-up multi-frequency observations have confirmed the hyperactivity of FRB~20240114A at low frequencies \citep[i.e., $<2$~GHz, see][]{ATel16430, ATel16432, ATel16434, ATel16446, ATel16452, ATel16494, ATel16542, ATel16547, ATel16565, Xie2024, Bhardwaj2025, Huang2025}. From late 2024 March onward, high-frequency detections of FRB~20240114 were reported \citep[from $\sim$$2~$GHz up to 6~GHz, see][]{ATel16597, ATel16599, ATel16620, Eppel25, zhang2025b}, indicating the source is also active at observing frequencies $>2$~GHz.

Although FRBs have been detected across a wide frequency range, from 110~MHz \citep{Pleunis2021b} to $\sim$8~GHz \citep{Gajjar2018}, the majority of detections have been made at relatively low frequencies ($<1.4$~GHz) and burst samples above 2~GHz remain sparse (see Table~\ref{table: above2ghz}). Before the discovery of FRB~20240114A, only two repeating FRBs, namely FRBs~20121102A and 20190520B, had more than 10 bursts detected above 2~GHz. Motivated by the current scarcity of high-frequency burst samples of FRBs, as well as FRB~20240114A's potential hyperactivity and its ability to produce bursts above 2~GHz, we conducted a long-term monitoring campaign of FRB~20240114A simultaneously at 2.25 and 8.60~GHz with the Shanghai Tianma Radio Telescope (TMRT). Here, we report the detection of 155 bursts from FRB~20240114A via this monitoring campaign between 2024 January 29 and 2025 February 15, with a total on-source time of 182.27~hr. All bursts were detected at 2.25~GHz. The details of the TMRT monitoring observations and burst search are described in Sec.~\ref{sec:2}. We present the observing results at 2.25~GHz in Sec.~\ref{sec:3}. The implications of our non-detection at 8.60~GHz and the potential frequency-dependent activity of FRB~20240114A are discussed in Sec.~\ref{sec:4}. In Sec.~\ref{sec:5}, we present the conclusions of this work.

\begin{deluxetable*}{llcccl} 
\tablecaption{A summary of repeating FRBs detected by observations above 2~GHz.}
\tablenum{1}
\label{table: above2ghz}
\tablehead{
    \multicolumn{1}{l}{Source} & \multicolumn{1}{l}{Reference} & \colhead{Frequency} & \colhead{Total duration} & \colhead{$N_{\rm{burst}}$} & \multicolumn{1}{l}{Telescope} \\
    \colhead{} & \colhead{} & \colhead{(GHz)} & \colhead{(hr)} & \colhead{} & \colhead{}
}
\startdata
\multirow{11}*{FRB~20121102A}     &  \citet{Scholz2016} &  1.6$-$2.4 &  15.3 & 5 & GBT \\
\cline{2-6}
                 &  \citet{Law2017} &  2.5$-$3.5 &  66 & 9 & VLA \\
\cline{2-6}
                 &  \citet{Spitler2018} &  4.6$-$5.1 &  22 & 3 & Effelsberg \\
\cline{2-6}
                &  \citet{Michilli2018} &  4.1$-$4.9 &  13 &  16 & AO \\
\cline{2-6}
                 &  \citet{Gajjar2018} &  4$-$8 &  6 & 21 & GBT \\
\cline{2-6}
                 &  \citet{Zhang2018} &  4$-$8 &  6 & 93\tablenotemark{a} & GBT \\
\cline{2-6}
                 &  \citet{Majid2020} &  2.19$-$2.31 &  5.7 & 6 & DSS-43 \\
\cline{2-6}
                 &  \citet{Pearlman2020} &  2.19$-$2.31 &  27.3 & 2 & DSS-63 \\
\cline{2-6}
                 & \multirow{3}*{\citet{Hilmarsson2021}}   &  4.1$-$4.9 &  \nodata\tablenotemark{b} &  13 & AO \\
                 &    &  2$-$4 &  $\sim$5 &  2 & VLA \\
                 &    &  4$-$8 &  115 &  1 & Effelsberg \\
\hline
FRB~20180916B   &  \citet{Bethapudi2023} &  4$-$8 &  79.66 &  8 & Effelsberg \\
\hline
\multirow{5}*{FRB~20190520B}   &  \multirow{2}*{\citet{Niu2022}} &  4.5$-$5.5 &  4.1 &  1 & VLA \\
                &                   &   2.5$-$3.5 & 3.4 & 5 & VLA \\ 
\cline{2-6}
               &  \citet{Feng22} &  4$-$8 &  \nodata\tablenotemark{b} &  3 & GBT \\
\cline{2-6}
                 &  \multirow{2}*{\citet{Anna2023}} &  4$-$8 &  $\sim$20 &  16 & GBT \\
                 &   &  0.704$-$4.032 &  $\sim$32 & 113 & Parkes \\
\hline
FRB~20200120E   &  \citet{Majid2021} &  2.197$-$2.309 &  4.1 & 1 & DSS-63 \\
\hline
FRB~20201124A   &  \citet{Ikebe2023} &  2.194$-$2.322 &  8 & 1 & UDSC \\
\hline
SGR~1935+2154   &  \citet{Liu2024} &  2.182$-$2.382 &  \nodata\tablenotemark{b} & 4 & NSRT \\
\hline
\multirow{3}*{FRB~20220912A}   &  \citet{ATel15734} &  2.214$-$2.386 & $\sim$6 & 1 & AO-12m \\
\cline{2-6}
                 &  \citet{ATel15791} &  2.2$-$2.3 &  2.15 & 2 & DSS-63 \\
\cline{2-6}
                &  \citet{Liu2024} &  2.182$-$2.382 &  \nodata\tablenotemark{b} & 2 & NSRT \\
\hline
\multirow{12}*{FRB~20240114A}   &  \citet{ATel16597} &  2.283$-$2.795 &  1 & 5 & NRT \\
\cline{2-6}
                 &  \citet{ATel16599} &  1.564$-$2.236 &  $\sim$200 & 3 & ATA \\
\cline{2-6}
                 &  \multirow{4}*{\citet{ATel16620}\tablenotemark{c}} &  1.95$-$2.6 & \multirow{4}*{4} & 18 & \multirow{4}*{EFF} \\
                 &  &  3.0$-$4.12 & & 4 &  \\
                 &  &  4.12$-$5.2 & & 1 & \\
                 &  &  5.2$-$6.0 & & 1 & \\
\cline{2-6}
                 &  \multirow{4}*{\citet{Eppel25}\tablenotemark{c}} &  1.9$-$2.6 & \multirow{4}*{$\sim$7} & 186 & \multirow{4}*{EFF} \\
                 &  &  3.0$-$4.1 &  & 133 &  \\
                 &  &  4.1$-$5.2 &  & 32 & \\
                 &  &  5.2$-$6.0 &  & 10 &  \\
\cline{2-6}
                 &  \citet{Huang2025} &  2.187$-$2.310 & $\sim$318 & 8 & KM40M \\
\cline{2-6}
                 &  This work &  2.2$-$2.3 & 182.27 & 155 & TMRT \\
\hline
FRB~20240619D   &  \citet{Tian2025} &  1.968$-$2.843 &  1 & 26 & MeerKAT \\
\enddata
\tablecomments{GBT: the Robert C. Byrd Green Bank Telescope; VLA: the Karl G. Jansky Very Large Array; Effelsberg: the Effelsberg Radio Telescope; AO: the Arecibo Observatory; DSS-43 and DSS-63: the NASA Deep Space Network 70~m telescope 43 and 63; Parkes: the Parkes/Murriyang radio telescope ; UDSC: the 64~m radio dish of Usuda Deep Space Center; NSRT: the Nanshan 26~m Radio Telescope; AO-12m: the Arecibo 12~m telescope; NRT: the Nan\c{c}ay Radio Telescope; ATA: the Allen Telescope Array; KM40M: the Kunming 40-Meter Radio Telescope; TMRT: the Tianma Radio Telescope; MeerKAT: the MeerKAT telescope.}
\vspace{-5pt}
\tablenotetext{a}{Including 21 bursts previously reported by \citet{Gajjar2018}.}
\vspace{-5pt}
\tablenotetext{b}{No available total on-source time in the reference work.}
\vspace{-5pt}
\tablenotetext{c}{Observations were conducted using the Ultra-Broad-Band receiver (UBB, 1.3$-$6~GHz) that divided into five sub-bands. The listed $N_{\rm{burst}}$ represents the number of detected bursts within certain sub-band.}
\end{deluxetable*}

\section{TMRT observations and burst search}
\label{sec:2}
The TMRT is a 65-m diameter, fully steerable radio telescope located in Shanghai, China. As early as 2018, pioneering simultaneous pulsar observations at 2.25 and 8.60 GHz with the TMRT had already achieved success \citep{Yan2018}.
The monitoring of FRB~20240114A was performed using the \textit{S}/\textit{X} dual-band cryogenic receiver on the TMRT, covering the frequency ranges of 2.2$-$2.3~GHz (\textit{S-}band) and 8.2$-$9.0~GHz (\textit{X-}band). During each observation, we applied the incoherent-dedispersion pulsar-search mode provided by the Digital Backend System \citep[DIBAS,][]{Yan2015, DIBAS} and recorded total intensity (Stokes I) data simultaneously at central frequencies of 2.25 and 8.60~GHz, with total bandwidths of 500 and 1000~MHz, respectively, in the 8-bit \textsc{PSRFITS} format \citep{PSRFITS}. Each band was divided into 512 frequency channels, yielding frequency/time resolutions of 0.9765625~MHz/65.536~$\mu$s and 1.953125~MHz/65.536~$\mu$s for \textit{S}-band and \textit{X}-band, respectively. Typically, the effective bandwidths are $\sim$100~MHz and $\sim$800~MHz for the \textit{S-}band and \textit{X-}band data, respectively, after removing frequency channels at edges of the band with low gain. 

Full-Stokes data were recorded for nine observations (see Table~\ref{table: obs_info}), but only Stokes I data were used in the following analysis due to the absence of noise diode scans for these observations. The system equivalent flux density (SEFD) of the TMRT is 46 and 48~Jy at 2.25 and 8.60~GHz \citep{Yan2018}, respectively, with typically a fractional uncertainty of 20\% \citep[estimated from multiple observations of the standard calibrator source 3C123 by][]{Zhao2019}. Between 2024 January 29 and 2025 February 15, we carried out 66 individual observations, with details summarized in Table~\ref{table: obs_info}. There was a large observation gap between 2024 February 24 and June 29 (O8$-$9), primarily due to the TMRT's full participation in China's Lunar and Deep Space Exploration. For observations on 2024 October 31 (O41) and November 11 (O43), only 2.25~GHz observation data is available due to data recording failures. As a result, we accumulated a total of 182.27 and 178.27 hours of observation at 2.25 and 8.60~GHz, respectively.

Before 2024 February 7 (O1$-$5, see Table~\ref{table: obs_info}), the telescope was pointed at the position of RA $=21^{\rm{h}}27^{\rm{m}}39.89^{\rm{s}}$, Dec $=+04^\circ21^{\prime}00.36^{\prime \prime}$ (J2000, with $\sim30^{\prime}$ uncertainties), initially reported by CHIME/FRB \citep{ATel16420}. Later, we then used the position of RA $=21^{\rm{h}}27^{\rm{m}}39.83^{\rm{s}}$, Dec $=+04^\circ19^{\prime}46.02^{\prime \prime}$ (J2000, with total uncertainty of $\sim1.5^{\prime \prime}$), provided by \citet{ATel16446} using MeerKAT, for observations between 2024 February 18 and 24 (O6$-$8). From 2024 June 29 onward, all observations were carried out using a much more precise position of RA $=21^{\rm{h}}27^{\rm{m}}39.835^{\rm{s}}$, Dec $=+04^\circ19^{\prime}45.634^{\prime \prime}$ (J2000, with 200~mas uncertainties) obtained by the EVN-PRECISE team via the VLBI observation \citep{ATel16542}. The CHIME/FRB and the MeerKAT positions are $\sim1.2^{\prime}$ and $\sim0.4^{\prime \prime}$ away from the EVN position, respectively. Considering that the primary beam of the TMRT are approximately $8.7^{\prime}$ and $2.4^{\prime}$ at 2.25 and 8.60~GHz\footnote{The primary beam width of the TMRT is calculated based on Equation~A1.12 in \citet{Handbook}, which involves the observing wavelength, the telescope diameter (65-m for the TMRT), and the telescope efficiency \citep[0.68 and 0.61 for the TMRT at 2.25 and 8.60~GHz, respectively,][]{Du2024}.}, the sensitivity of the initial five observations (O1$-$5) may be reduced due to the $\sim1.2^{\prime}$ position offset. Assuming a Gaussian beam response, this corresponds to a sensitivity loss of $\sim$5\% at 2.25~GHz and $\sim$50\% at 8.60~GHz, with the former being negligible.

\begin{figure*}
    \centering
    \includegraphics[width=\linewidth]{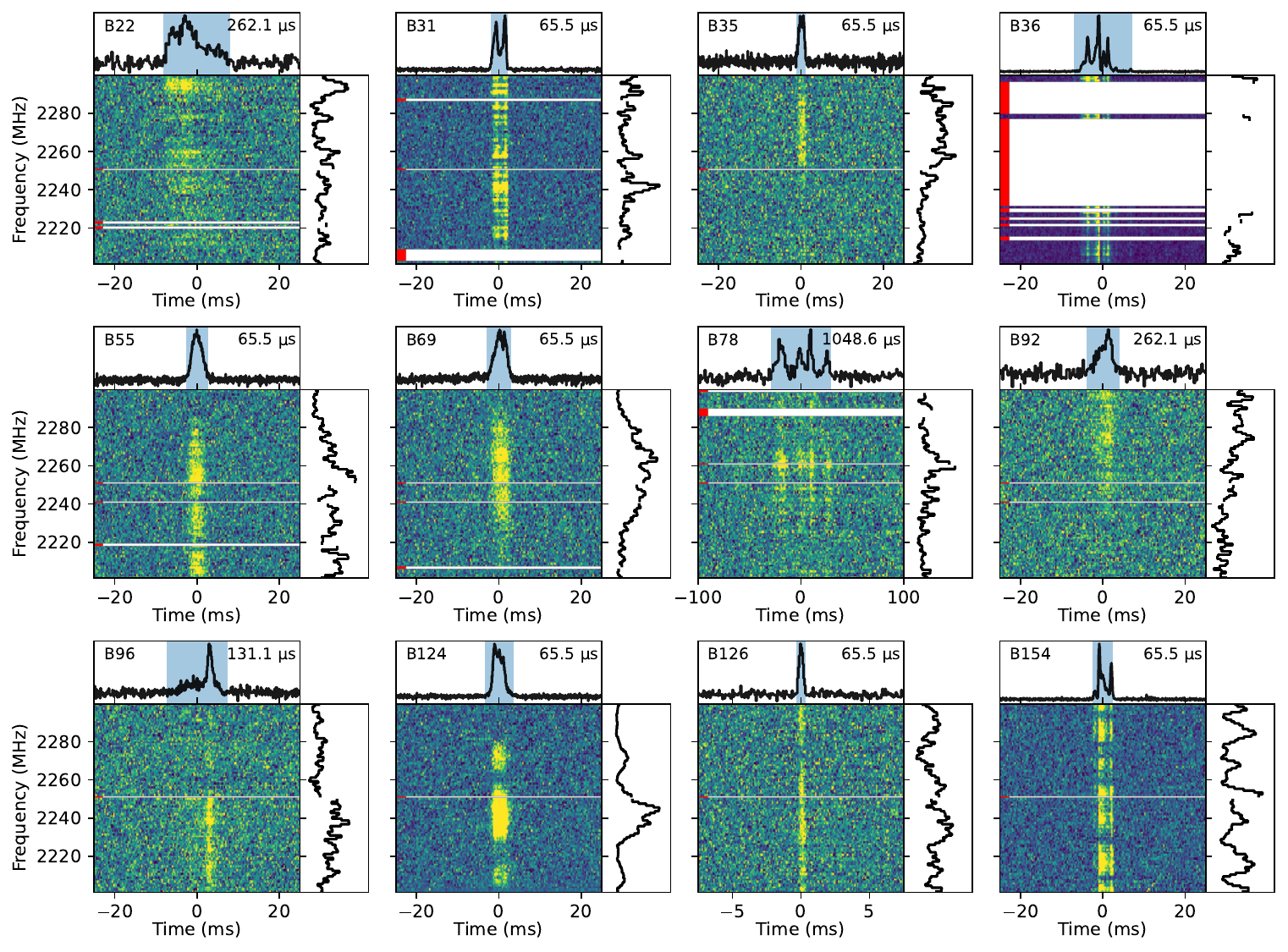}
    \caption{Dynamic spectra of a subset of bursts detected at 2.25~GHz. Each burst was incoherently dedispersed to a DM of 527.7~$\rm{pc~cm}^{-3}$. In each panel, the bottom sub-panel shows the dynamic spectrum with a frequency resolution of 0.977~MHz and a time resolution shown in the top-right corner of the top sub-panel. Each top sub-panel also contains the burst profile (averaging over the entire frequency band), with the burst ID in the top-left corner and a shaded area representing the on-burst region (see Sec.~\ref{sec:energy} for the definition of the on-burst region). The time-averaged spectrum of the on-burst region is displayed in each right sub-panel. The horizontal white lines in the dynamic spectrum, with red markers on the left side, represent masked frequency channels contaminated by RFI. For visual purposes, the intensity values of each dynamic spectrum are saturated at the 2nd and 98th percentiles.}
    \label{fig: subset bursts}
\end{figure*}

Before conducting the offline burst search, we first converted the \textsc{PSRFITS} data to 8-bit total intensity filterbank data using the \textsc{PRESTO}\footnote{\url{https://github.com/scottransom/presto}} routine \texttt{psrfits2fil.py}, and then extracted the frequency channels within the effective bandwidths using the \textsc{your}\footnote{\url{https://github.com/thepetabyteproject/your/}} \citep{your} routine \texttt{your\_writer.py}. The resulting data products have bandwidths of 98.6~MHz (2201.2$-$2299.8~MHz) for \textit{S-}band and 800.8~MHz (8199.6$-$9000.4~MHz) for \textit{X-}band, respectively. Radio frequency interference (RFI) mitigation and burst search were done using the \textsc{TransientX}\footnote{\url{https://github.com/ypmen/TransientX}} software \citep{transientx}. The burst search was done under the native time and frequency resolutions within the DM range of 450 to 650~$\rm{pc~cm}^{-3}$ in steps of 0.5~$\rm{pc~cm}^{-3}$. We searched for bursts with temporal widths ranging from 0.2 to 50~ms and applied a $7\sigma$ detection threshold, corresponding to fluence thresholds of 0.72 and 0.27~Jy~ms at 2.25 and 8.60~GHz, respectively, for a 1~ms width burst. A total of $\sim2.9 \times 10^4$ and $\sim2.0 \times 10^4$ candidates were generated at 2.25 and 8.60~GHz, respectively, across all observations, and each was then visually examined based on the \textsc{TransientX} candidate plot. We also inspected the zero-DM dynamic spectrum of each candidate using the \textsc{PRESTO} routine \texttt{waterfaller.py} to better distinguish between real burst signals and RFIs. To distinguish separated bursts from sub-bursts, we adopted the criterion that two bursts are considered separated if the root-mean-square (RMS) of the signal between them is comparable to that of the noise. As a result, we successfully detected 155 bursts at 2.25~GHz, while no bursts were detected in the simultaneously recorded 8.60~GHz data. In Fig.~\ref{fig: subset bursts}, we display a subset of bursts detected at 2.25~GHz. The dynamic spectra of all 155 bursts are presented in Fig.~\ref{fig:figureA1} of the Appendix.

\begin{deluxetable}{lccccc} 
\tablecaption{Basic observation information of FRB~20240114A with the TMRT}
\tablenum{2}
\label{table: obs_info}
\tablehead{
    \multicolumn{1}{l}{$\rm{ID}_{\rm{obs}}$} & \colhead{Start time\tablenotemark{a}} & \colhead{Duration} & \colhead{Full-pol} & \colhead{$N_{\rm{burst}}$} & \colhead{Rate\tablenotemark{b}} \\
    \colhead{} & \colhead{(UTC)} & \colhead{(s)} & \colhead{} & \colhead{} & \colhead{$(\rm{hr}^{-1})$}
}
\startdata
O1 &  2024-01-29T06:07:40.0 &  7200.5 &  N &  0 &  $<0.92$ \\
O2 &  2024-02-03T01:57:18.0 &  1147.8 &  N &  0 &  $<5.77$ \\
O3 &  2024-02-06T07:10:03.0 &  7194.1 &  N &  0 &  $<0.92$ \\
O4 &  2024-02-07T06:29:43.0 &  7200.5 &  N &  0 &  $<0.92$ \\
O5 &  2024-02-07T23:36:42.0 &  10800.8 &  N &  0 &  $<0.61$ \\
O6 &  2024-02-18T03:16:06.0 &  10800.8 &  N &  0 &  $<0.61$ \\
O7 &  2024-02-19T23:29:58.0 &  3600.3 &  N &  0 &  $<1.84$ \\
O8 &  2024-02-24T03:49:45.0 &  3600.3 &  N &  0 &  $<1.84$ \\
O9 &  2024-06-29T20:52:24.0 &  3600.3 &  N &  3 &  $3.00^{+2.94}_{-1.63}$ \\
O10 &  2024-06-30T19:59:51.0 &  3600.3 &  N &  1 &  $1.00^{+2.32}_{-0.83}$ \\
O11 &  2024-07-09T15:45:06.0 &  3600.3 &  N &  2 &  $2.00^{+2.66}_{-1.29}$ \\
O12 &  2024-07-10T14:10:14.0 &  25200.7 &  N &  12 &  $1.71^{+0.65}_{-0.49}$ \\
O13 &  2024-07-12T14:02:17.0 &  28799.9 &  N &  8 &  $1.00^{+0.49}_{-0.35}$ \\
O14 &  2024-07-14T15:32:06.0 &  23400.1 &  N &  1 &  $0.15^{+0.36}_{-0.13}$ \\
O15 &  2024-07-15T13:23:26.0 &  9000.1 &  N &  0 &  $<0.74$ \\
O16 &  2024-07-17T14:17:12.0 &  7200.5 &  N &  3 &  $1.50^{+1.47}_{-0.81}$ \\
O17 &  2024-07-18T13:41:39.0 &  28799.9 &  N &  18 &  $2.25^{+0.67}_{-0.52}$ \\
O18 &  2024-07-20T13:44:28.0 &  28799.9 &  N &  11 &  $1.38^{+0.55}_{-0.41}$ \\
O19 &  2024-07-21T13:17:22.0 &  28801.0 &  N &  33 &  $4.12^{+0.85}_{-0.71}$ \\
O20 &  2024-07-23T17:34:42.0 &  14400.0 &  Y &  12 &  $3.00^{+1.14}_{-0.85}$ \\
\nodata &  \nodata &  \nodata &  \nodata &  \nodata \\  
\enddata
\tablecomments{The complete version of this table is available in machine-readable format in the online journal.} The number of detected bursts, $N_{\rm{burst}}$, and inferred mean burst rates (or upper limits) are for 2.25~GHz observations.
\vspace{-5pt}
\tablenotetext{a}{Topocentric at the TMRT.}
\vspace{-5pt}
\tablenotetext{b}{The uncertainties or upper limits for burst rates correspond to $1\sigma$ Poisson errors \citep{poissonerr}.}
\end{deluxetable}

\begin{deluxetable*}{lccccccc} 
\tablecaption{Properties of 155 bursts from FRB~20240114A detected with the TMRT at 2.25~GHz}
\tablenum{3}
\label{table: burst properties}
\tablehead{
    \multicolumn{1}{l}{$\rm{ID}_{\rm{burst}}$} & \colhead{TOA\tablenotemark{a}} & \colhead{S/N$_{\rm{det}}$\tablenotemark{b}} & \colhead{S/N\tablenotemark{c}} & \colhead{Width\tablenotemark{d}} & \colhead{Fluence\tablenotemark{e, f}} & \colhead{Energy\tablenotemark{f}} & \colhead{Peak flux density\tablenotemark{f}} \\
    \colhead{} & \colhead{(MJD)} & \colhead{} & \colhead{} & \colhead{(ms)} & \colhead{$(\rm{Jy~ms})$} & \colhead{($10^{38}~\rm{erg}$)} & \colhead{(Jy)}
}
\startdata
B01 &  60490.877254813 &  9.0 &  7.8 &  $1.81\pm0.32$ &  $1.68\pm0.34$ &  $0.69\pm0.14$ &  $1.50\pm0.30$ \\
B02 &  60490.877781704 &  10.6 &  11.5 &  $4.24\pm0.46$ &  $3.27\pm0.65$ &  $1.34\pm0.27$ &  $1.47\pm0.29$ \\
B03 &  60490.877863290 &  12.9 &  15.5 &  $4.82\pm0.31$ &  $5.10\pm1.02$ &  $1.99\pm0.40$ &  $1.93\pm0.39$ \\
B04 &  60491.871781302 &  7.0 &  6.7 &  $8.50\pm1.39$ &  $2.66\pm0.53$ &  $1.09\pm0.22$ &  $1.33\pm0.27$ \\
B05 &  60500.668215640 &  7.7 &  7.1 &  $4.14\pm0.78$ &  $2.18\pm0.44$ &  $0.88\pm0.18$ &  $1.52\pm0.30$ \\
B06 &  60500.686660406 &  7.5 &  10.3 &  $9.09\pm1.60$ &  $4.43\pm0.89$ &  $1.78\pm0.36$ &  $1.71\pm0.34$ \\
B07 &  60501.614672263 &  7.4 &  5.4 &  $2.23\pm0.42$ &  $1.15\pm0.23$ &  $0.47\pm0.09$ &  $1.27\pm0.25$ \\
B08 &  60501.618415652 &  16.5 &  18.2 &  $3.18\pm0.24$ &  $4.58\pm0.92$ &  $1.88\pm0.38$ &  $1.90\pm0.38$ \\
B09 &  60501.628657695 &  11.2 &  26.1 &  $46.83\pm4.29$ &  $20.06\pm4.01$ &  $8.24\pm1.65$ &  $1.97\pm0.39$ \\
B10 &  60501.640032704 &  7.3 &  8.1 &  $5.94\pm0.91$ &  $2.73\pm0.55$ &  $1.12\pm0.22$ &  $1.39\pm0.28$ \\
B11 &  60501.674640852 &  9.3 &  7.9 &  $3.04\pm0.67$ &  $2.06\pm0.41$ &  $0.85\pm0.17$ &  $1.59\pm0.32$ \\
B12 &  60501.714895960 &  7.8 &  6.3 &  $2.14\pm0.43$ &  $1.35\pm0.27$ &  $0.56\pm0.11$ &  $1.53\pm0.31$ \\
B13 &  60501.736947322 &  7.8 &  6.7 &  $2.04\pm0.48$ &  $1.43\pm0.29$ &  $0.59\pm0.12$ &  $1.70\pm0.34$ \\
B14 &  60501.746368895 &  9.2 &  9.7 &  $2.28\pm0.25$ &  $2.08\pm0.42$ &  $0.85\pm0.17$ &  $1.32\pm0.26$ \\
B15 &  60501.750034617 &  10.8 &  8.9 &  $1.04\pm0.13$ &  $1.20\pm0.24$ &  $0.48\pm0.10$ &  $1.76\pm0.35$ \\
B16 &  60501.823491590 &  9.0 &  8.6 &  $2.52\pm0.48$ &  $2.12\pm0.42$ &  $0.83\pm0.17$ &  $1.46\pm0.29$ \\
B17 &  60501.825323426 &  29.2 &  34.7 &  $5.17\pm0.22$ &  $9.71\pm1.94$ &  $3.95\pm0.79$ &  $2.71\pm0.54$ \\
B18 &  60501.873196720 &  7.6 &  6.4 &  $3.04\pm0.56$ &  $1.68\pm0.34$ &  $0.69\pm0.14$ &  $1.54\pm0.31$ \\
B19 &  60503.591180076 &  8.2 &  6.6 &  $0.40\pm0.07$ &  $0.63\pm0.13$ &  $0.26\pm0.05$ &  $1.77\pm0.35$ \\
B20 &  60503.637351395 &  10.2 &  12.2 &  $4.74\pm0.67$ &  $3.39\pm0.68$ &  $1.35\pm0.27$ &  $1.47\pm0.29$ \\
\nodata &  \nodata &  \nodata &  \nodata &  \nodata &  \nodata &  \nodata &  \nodata \\
\enddata
\tablecomments{The complete version of this table is available in machine-readable format in the online journal.}
\vspace{-5pt}
\tablenotetext{a}{Corrected to the Solar System Barycenter at infinite frequency using a DM of 527.7~$\rm{pc~cm}^{-3}$ \citep{ATel16420}, a DM constant of $1/0.241~\rm{GHz^2~cm^3~pc^{-1}~ms}$, and the EVN position \citep{ATel16542}.}
\vspace{-5pt}
\tablenotetext{b}{The S/N reported by the searching pipeline.}
\vspace{-5pt}
\tablenotetext{c}{The S/N obtained after manual removal of RFI.}
\vspace{-5pt}
\tablenotetext{d}{Defined as the full width at half-maximum (FWHM) of a Gaussian function.}
\vspace{-5pt}
\tablenotetext{e}{Summing over the on-burst region (see Sec.~\ref{sec:energy}).}
\vspace{-5pt}
\tablenotetext{f}{Assuming a conservative uncertainty of 20\%, which is dominated by the uncertainty of SEFD.}
\end{deluxetable*}

\section{Burst analysis}
\label{sec:3}

We detected 155 bursts from FRB~20240114A at 2.25~GHz, providing a large, uniform burst sample for studying various burst properties of FRB~20240114A at observing frequencies $>2$~GHz. Here, we focus only on 2.25~GHz observations, while the implications of non-detection at 8.60~GHz will be discussed in Sec.~\ref{sec: xband}. In Table~\ref{table: burst properties}, we summarize the measured burst properties, including the barycentric time of arrival (TOA), temporal width, peak flux density, burst fluence, and isotropic-equivalent energy. In the following analysis, we adopt a DM of 527.7~pc~cm$^{-3}$ reported by \citet{Shin2025} to dedisperse the bursts. For bursts with S/N $<40$, the data were downsampled in time by factors of up to 16 before measuring burst properties, except for burst B126, the narrowest burst we detected, for which we retained the original time resolution of 65.536~$\mu$s. We fit each burst with single or multiple Gaussian functions, depending on the apparent burst morphology. The on-burst region was defined as the range between the full width at one tenth of maximum (FWTM) of the fitted Gaussian component in the case of single-component bursts. For bursts consisting of multiple components (sub-bursts), the on-burst region was defined as the interval between the left/right edge of the FWTM of the leftmost/rightmost component. The TOA of each burst was then defined as the midpoint of its on-burst region.

\subsection{Long-term burst rate}
\label{sec:burst rate}

\begin{figure*}
    \centering
    \includegraphics[width=\linewidth]{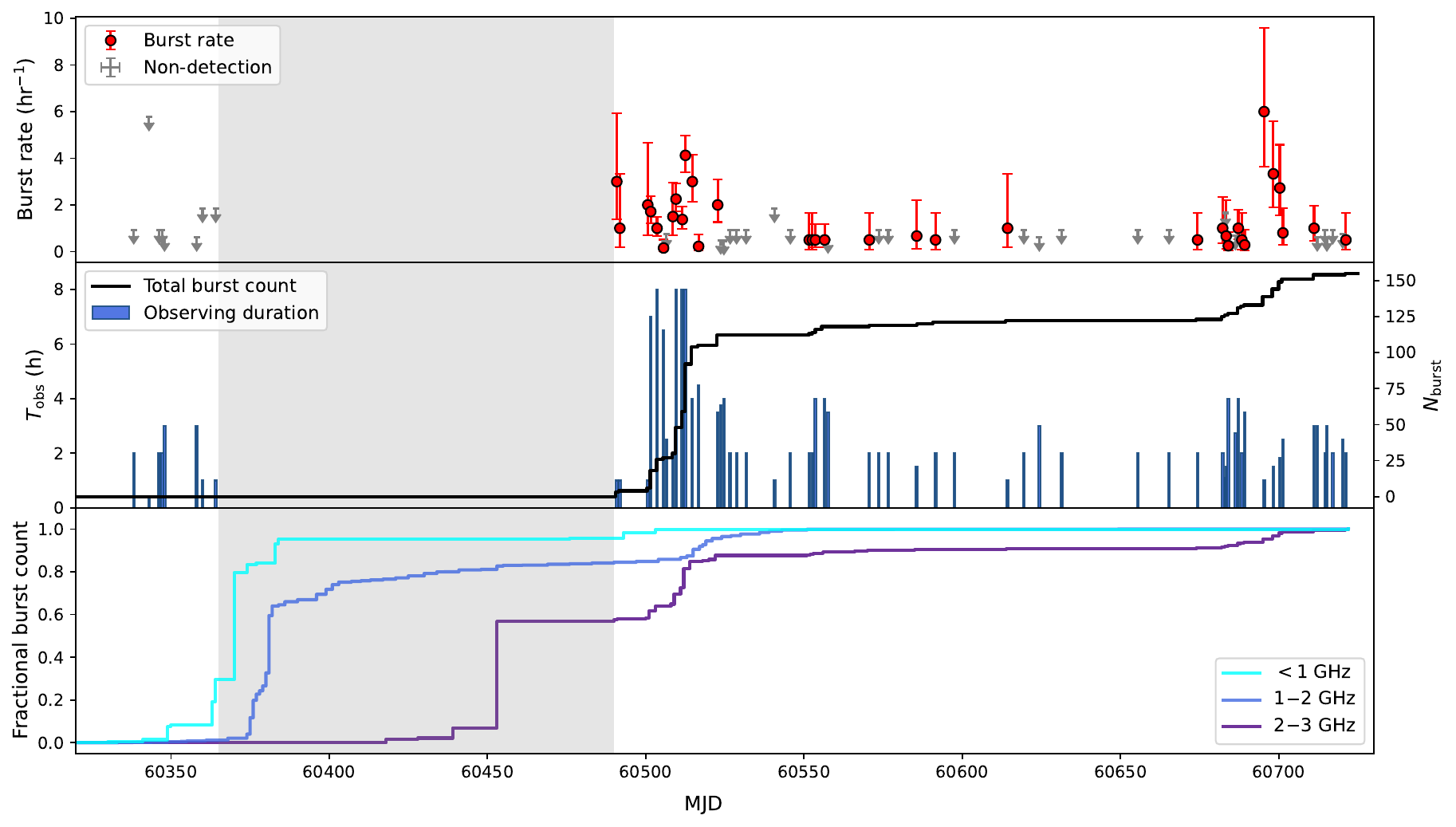}
    \caption{Top panel: average burst rate per TMRT 2.25~GHz observation (red circles), with 1$\sigma$ upper limits indicated for non-detections (gray arrows). Middle panel: cumulative number of bursts detected by TMRT (black solid line) and the duration of each observation (blue bars). Bottom panel: cumulative burst counts reported by various telescopes at different central frequencies: $<$1~GHz (cyan), 1$-$2~GHz (blue), and 2$-$3~GHz (violet). Except for our detections, other data are quoted from \citet{Shin2025, ATel16432, ATel16565, Atel16967, ATel16434, ATel16547, ATel16599, ATel16597, Huang2025, Tian2024, Panda2024, Kumar2024, Bhardwaj2025, Xie2024, Eppel25, ATel16620, zhang2025b}. The gray shaded region in all panels represent the TMRT observing gap between 2024 February 24 and June 29.}
    \label{fig:CDF}
\end{figure*}

We started the monitoring campaign of FRB~20240114A on 2024 January 29 with the TMRT, three days after the announcement by \citet{ATel16420} on the Astronomer's Telegram (ATel). By 2025 February 15, we had conducted 66 observations, yielding a total on-source time of 182.27~hrs at \textit{S-}band. The mean (or upper limit) burst rate per observation is summarized in Table~\ref{table: obs_info}, where we quote $1\sigma$ Poisson errors. In Fig.~\ref{fig:CDF}, we show the results of our 2.25~GHz monitoring campaign, including the cumulative number distribution (CDF) of detected bursts, the mean burst rate, and the duration per observation. The burst rate per observation exhibits epoch-to-epoch variations. The mean burst rate for FRB~20240114A at 2.25~GHz, inferred from the entire 182.27 hours of monitoring, is $0.85^{+0.07}_{-0.07}~\rm{hr}^{-1}$ above a fluence threshold of $0.72$~Jy~ms. The highest burst rate we observed was $6.00^{+3.60}_{-2.37}~\rm{hr}^{-1}$ on 2025 January 20, when six bursts were detected during a one-hour observation. 

In the early epochs of our monitoring campaign, i.e., from 2024 January 29 to February 24, we performed eight observations, but no bursts were detected. The $\sim$5\% sensitivity loss from the pointing offset in observations O1$-$5 (see Sec.~\ref{sec:2}) is negligible relative to the $\sim$20\% uncertainty in the SEFD of the TMRT. Based on the total on-source time of 14.3~hrs during this period, we place a $1\sigma$ upper limit of $<0.13$~hr$^{-1}$ on the mean burst rate at 2.25~GHz above a fluence threshold of $0.72$~Jy~ms. During our observing gap (see Sec.~\ref{sec:2}), \citet{ATel16597} detected five bursts on 2024 April 18 with the NRT in the 2283$-$2795~MHz band above a fluence threshold of 0.18~Jy~ms. Meanwhile, Effelsberg observations on 2024 May 9 and 23 resulted in the detection of 18 and 186 bursts within 4 and $\sim$7 hours, respectively, in the 1.9$-$2.6~GHz band \citep{ATel16620, Eppel25}, indicating an increase of burst activity at this frequency range. After the observation gap, we observed FRB~20240114A in a high-activity state between 2024 June 29 and July 31, during which 112 bursts were detected in 65 hours of on-source time, corresponding to a mean burst rate of $1.72^{+0.18}_{-0.16}~\rm{hr}^{-1}$. This indicates an increase of more than ten times in burst activity at 2.25~GHz compared with our observations before 2024 February 24. At low frequencies, however, \citet{Atel16967} reported that FRB~20240114A has been in a low-activity state since 2024 June. Based on observations of FRB~20240114A with five 25$-$32~m radio telescopes in Europe, they detected seven bursts in 500 hours of observations at 1.4~GHz and no bursts in 100 hours of observations at 0.33~GHz. For comparison, the observation campaign they performed before 2024 April resulted in the detection of 109 and 2 bursts at 1.4 and 0.33~GHz, respectively, for a total on-source time of 600~hrs \citep{ATel16565}.

We observed the end of the 2.25~GHz high-activity state around 2024 July 31. Specifically, we detected seven bursts during a continuous 3.5 hr observation on 2024 July 31, corresponding to a burst rate of $2.00^{+1.08}_{-0.74}~\rm{hr}^{-1}$. However, no bursts were detected in the subsequent two observations over the next two days, with a total on-source time of 7.75~hr, yielding an upper limit burst rate of $<0.23$~hr$^{-1}$. This indicates a decrease of an order of magnitude or more in the mean burst rate. From 2024 August to December, we observed FRB~20240114A in a low-activity state with a mean burst rate of $0.19^{+0.08}_{-0.06}~\rm{hr}^{-1}$. 

On 2025 January 1, the CHIME/FRB detected one burst from FRB~20240114A, which was the first detection of this source by CHIME/FRB after more than ten months of non-detections since 2024 March \citep{Shin2025}. Soon after, \citet{Atel16967} reported the detection of six bursts (above a fluence threshold of 10~Jy~ms) during a continuous 4.5~hr observation at 1.4~GHz on 2025 January 4. This represents an increase of nearly two orders of magnitude in 1.4~GHz burst activity compared with the low-activity state they observed between 2024 June and December, when only seven bursts were detected over 500 hours of observation at 1.4~GHz. At 2.25~GHz, we detected a single burst during a 2~hr observation on 2024 December 30. Motivated by the potentially renewed activity, we conducted high-cadence observations with the TMRT from 2025 January 7 to 14, which resulted in a total of 20.86 hours of on-source time and detected 10 bursts at 2.25~GHz. However, the inferred mean burst rate during this period was $0.48^{+0.20}_{-0.15}~\rm{hr}^{-1}$, which was comparable to that observed in the previous low-activity state, indicating no significant enhancement of burst activity at 2.25~GHz. About two weeks after the reported renewed activity of FRB~20240114A at 1.4~GHz, we observed a short-term increase in burst activity at 2.25~GHz. The burst rate jumped from $0.29^{+0.66}_{-0.24}~\rm{hr}^{-1}$ on 2025 January 14 to $6.00^{+3.60}_{-2.37}~\rm{hr}^{-1}$ on 2025 January 20, then declined to the low-activity level after several days.

\subsection{Burst fluence and energetics}
\label{sec:energy}
To estimate the burst fluence, we applied the radiometer equation \citep{Handbook}, from which the band-averaged burst profile (in S/N units) can be converted into Jy units as follows:
\begin{equation}
    S=\sigma_{\rm{s}}\cdot \rm{S/N}=\frac{\rm{SEFD\cdot S/N}}{\sqrt{n_{\rm{p}}\Delta\nu t_{\rm{obs}}}},
    \label{eq: radiometer}
\end{equation}
where $S$ is the flux, $\sigma_{\rm{s}}$ is the RMS of a manually selected off-burst region, $n_{\rm{p}}$ is the number of polarizations summed (2 for TMRT observations), $\Delta\nu$ is the effective bandwidth of the data (after masking frequency channels contaminated by RFIs), and $t_{\rm{obs}}$ denotes the integration time per bin. The burst fluence $F$ was then calculated by integrating the flux of the on-burst region with respect to time, i.e., $F=\int S\,\rm{d}t$. We adopted a conservative uncertainty of 20\% on the estimated burst fluences, which is dominated by the uncertainty in the SEFD of the TMRT (see Sec.~\ref{sec:2}). As a result, we obtained a wide spread of burst fluences spanning three orders of magnitude, i.e., from 0.53 to 277.41~Jy~ms.

Based on the obtained burst fluence, we calculated the isotropic-equivalent burst energy $E$ following \citep[see Equation 2 in][]{Hewitt2022}:
\begin{equation}
    \label{eq: energy}
    E = \frac{4\pi D_{\rm{L}}^2 F \Delta\nu}{1+z},
\end{equation}
where $F$ is the burst fluence, $\Delta\nu$ is the observing bandwidth (after RFI excision) within which the fluence was estimated, $z$ denotes the redshift of the source, and $D_{\rm{L}}$ is the corresponding luminosity distance. We adopted a redshift of $z=0.1300$ for the host galaxy of FRB~20240114A \citep{ATel16613}, resulting in a luminosity distance of 630.1~Mpc by assuming the cosmological parameters from \citet{Planck2016} implemented in the \textsc{astropy.cosmology}\footnote{\url{https://www.astropy.org/}} \citep{astropy}. The derived isotropic-equivalent burst energies range from $\sim$$10^{37}$ to $10^{39}~$erg. 

\begin{figure*}
    \centering
    \includegraphics[width=\linewidth]{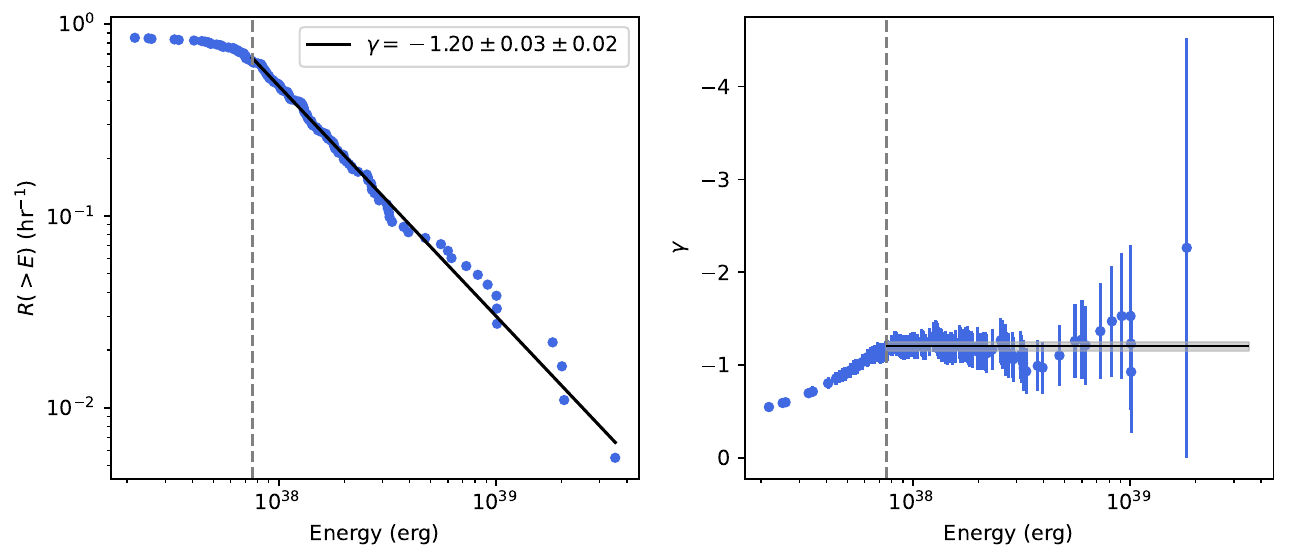}
    \caption{Left panel: the cumulative burst energy distribution. The solid black line denotes the best-fitting power-law function to bursts above the completeness threshold. Right panel: the power-law index $\gamma$ as a function of energy, determined via the maximum-likelihood estimation. The best-fitting power-law index $\gamma$ and its uncertainty are indicated by the horizontal black solid line and the gray shaded regions, respectively. The vertical gray dashed line in each panel represents the completeness threshold.}
    \label{fig:energy distribution}
\end{figure*}

The cumulative burst energy distribution is shown in Fig.~\ref{fig:energy distribution}, which exhibits a noticeable flattening towards the low-energy end. We attribute this to the incompleteness of our observations for low-energy bursts. To estimate the completeness threshold of our observations, we adopt a conservative S/N threshold of 10, yielding a completeness threshold of 1.8~Jy~ms, or $7.5\times10^{37}$~erg, for a burst width of 3~ms (the median value in our sample, see Sec.~\ref{sec: 3.3}). We use \textsc{curve\_fit} to fit a power-law relation, $R(>E)\propto E^{\gamma}$, to the cumulative burst energy distribution above this completeness threshold using a least-squares fitting method. The best-fitting power-law index is $\gamma=-1.20\pm0.03\pm0.02$, where the two uncertainties represent the fitting and systematic uncertainties, respectively. The fitting uncertainty is directly taken from \textsc{curve\_fit}. The systematic uncertainty, accounting for the 20\% fractional uncertainty in the burst energy, is determined using a bootstrapping method. Specifically, we formed a new dataset by resampling each data point based on its energy uncertainty and then performed the least-squares fitting. The bootstrapping step was repeated 1000 times, and we adopted the standard deviation of the fitting result as the bootstrapping uncertainty. For completeness, we also performed a maximum-likelihood estimation \citep[as described in][]{Crawford1970, James2019} of the power-law index, with the result shown in Fig.~\ref{fig:energy distribution}.

Compared to the cumulative energy distribution of FRB~20240114A studied at lower frequencies ($<2$~GHz), our best-fitting power-law index of $\gamma=-1.20\pm0.03\pm0.02$ is consistent with $\gamma=-1.24\pm0.11$ reported by \citet{Kumar2024} from 52 bursts detected in the 300$-$500~MHz band with the upgraded Giant Metrewave Radio Telescope (uGMRT), and is flatter than $\gamma=-1.8\pm0.2$ derived from a single-epoch MeerKAT observation in the 544$-$1088~MHz band \citep{Tian2024}. In contrast to the power-law model, \citet{Panda2024} fitted the energy distribution with a log-normal (LN) function, while \citet{zhang2025b} found a bimodal distribution consisting of two LN components, based on the uGMRT observations in the 300$-$500/550$-$750~MHz bands and the Five-hundred-meter Aperture Spherical radio Telescope (FAST) observations in the 1$-$1.5~GHz band, respectively.

\subsection{Burst width}
\label{sec: 3.3}
\begin{figure*}
    \centering
    \includegraphics[width=0.5\linewidth]{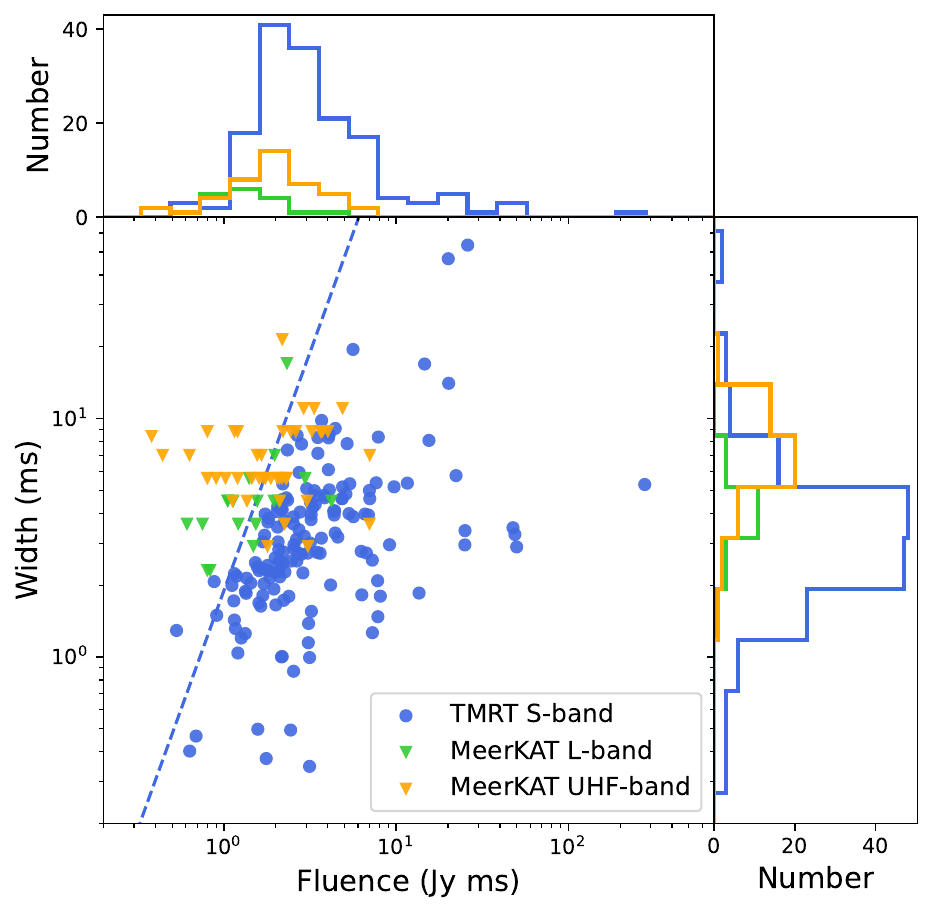}
    \caption{The width-fluence distribution for our sample of 155 bursts detected at 2.25~GHz (blue circles). The blue dashed line represents $7\sigma$ detection threshold of S/N of the TMRT for bursts with different temporal widths. We also show bursts detected by the MeerKAT \citep{Tian2024} at \textit{L}-band (856$-$1712~MHz) and the Ultra-High Frequency (UHF) band (544$-$1088~MHz), represented by green and orange triangles, respectively. The top and right panels show the histograms of fluence and width distributions, respectively.}
    \label{fig:flence_width_distribution}
\end{figure*}

We fit a Gaussian function to each burst profile and report the full width at half maximum (FWHM) of the fitted component as the measured temporal width. The temporal width and fluence distributions for the bursts in our sample are shown in Fig.~\ref{fig:flence_width_distribution}. For bursts from FRB~20240114A we detected at 2.25~GHz, the burst widths range from 0.35 to 53.41~ms, spanning two orders of magnitude. The widest burst in our sample is burst B78, which exhibits four distinct components (see Fig.~\ref{fig: subset bursts}) and has a temporal width of $53.41\pm4.64$~ms, which is also the widest burst ever reported for FRB~20240114A. The mean and median burst widths we obtained are 4.24 and 3.00~ms, respectively, which are comparable to those of other repeating FRBs \citep[e.g.][]{Pleunis2021a}. Previous studies have found that repeating FRBs tend to show narrower temporal widths at higher frequencies \citep[e.g.,][]{Gajjar2018, Michilli2018, Bethapudi2023}. For comparison, \citet{Tian2024} reported a mean burst width of 5.92~ms for bursts detected in the UHF-band (544-1088~MHz) and \textit{L}-band (856-1712~MHz) with MeerKAT. Thus, our result seems to confirm a similar trend for FRB~20240114A, suggesting that bursts tend to be narrower at higher frequencies. 

Through visual inspection of all detected bursts, we find no evident scattering feature. We select burst B126, which is the narrowest one in our sample and has a simple Gaussian-like shape, and constrain the scattering timescale ($\tau_{\rm{sc}}$) using the \textsc{scatfit}\footnote{\url{https://github.com/fjankowsk/scatfit}} \citep{scatfit} package. The bandwidth (2.2$-2.3$~GHz) is divided into four sub-bands, and $\tau_{\rm{sc}}$ are derived for three sub-bands with S/N $>10$, assuming a fixed scattering index of $-4$. The resulting $\tau_{\rm{sc}}$ is $\sim0.1$~ms, which is smaller than the expected intra-channel smearing of $\sim0.36$~ms for our observations ($\nu=2.25$~GHz and $\Delta\nu_{\rm{ch}}\approx0.98$~MHz), suggesting that the burst scattering is unresolved. We thus place an upper limit of $\tau_{\rm{sc}}<0.18$~ms at 2.25~GHz, corresponding to half the temporal width of burst B126, or equivalently $\tau_{\rm{sc}}<4.61$~ms at 1GHz assuming a $\nu^{-4}$ scaling. This upper limit is at least an order of magnitude less stringent than those obtained by \citet{Tian2024} and \citet{Shin2025} at observing frequencies below 2~GHz.

\subsection{Waiting time and burst clustering}
\label{sec:waiting time}
Previous studies have shown that the waiting time, defined as the time interval $\delta t$ between consecutive bursts, typically exhibits a bimodal distribution for several well-studied repeating FRBs, e.g., FRB~20121102A \citep{Li2021, Aggarwal2021, Hewitt2022, Jahns2023}, FRB~20200120E \citep{Nimmo2023}, FRB~20201124A \citep{Xu2022, Zhang2022}, and FRB~20220912A \citep{Zhang2023, Konijn2024}, particularly when the burst sample is sufficiently large. The left peak of the waiting time distribution, often thought to be associated with the characteristic timescale of the underlying emission process, is typically located at tens to hundreds of milliseconds. The right peak, which is related to the apparent burst rate, is typically found at timescales of several tens to hundreds of seconds, with its location primarily determined by the activity of the source and the sensitivity of the telescope.

\begin{figure*}
    \centering
    \includegraphics[width=0.5\linewidth]{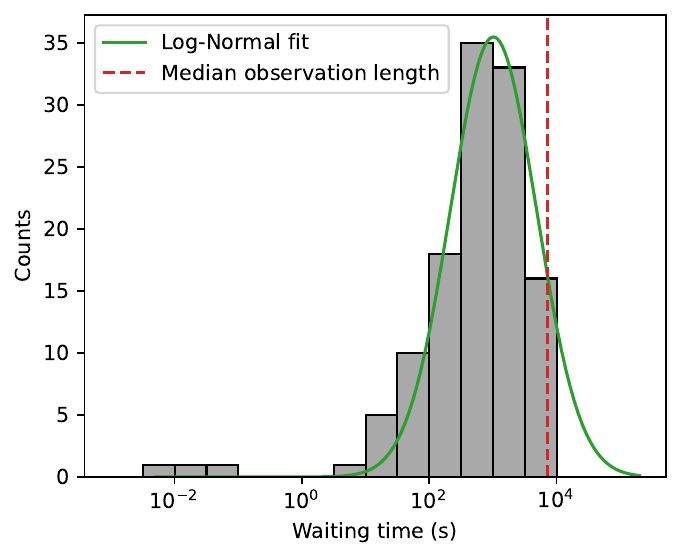}
    \caption{The distribution of waiting time of TMRT detected bursts from FRB~20240114A. Note that all waiting times are calculated using one TOA per burst, regardless of sub-bursts structure (see Sec.~\ref{sec:3} for the definition of TOA). The green line represents the best-fitting log-normal function for waiting times above 0.1~s, which peaks at $1019\pm66$~s. The red vertical dashed line indicates the median duration of TMRT observations ($\sim$7200~s).}
    \label{fig:waiting time}
\end{figure*}

We calculated the waiting times for bursts in our sample by subtracting the barycentric TOA differences between consecutive bursts in each observation and showed the joint distribution in Fig.~\ref{fig:waiting time}. The apparent cutoff in the distribution above $10^4$~s is primarily due to the limited during of each observation ($\sim$7200~s), which reduces sensitivity to waiting times longer than this timescale. We fit a log-normal function to waiting times greater than 0.1~s. The best-fitting result for the peak location is $1019\pm66$~s, with the uncertainty estimated using a bootstrapping method. This represents the location of the right peak. At observing frequencies below 2~GHz, \citet{Panda2024} and \citet{zhang2025b} reported the right peak to locate at 128.93~s and 7.11~s, respectively, based on the uGMRT observations in the 350$-$550/550$-$750~MHz bands and the FAST observations in the 1$-$1.5~GHz band. These values are approximately 8 and 143 times shorter, respectively, than our result for FRB~20240114A at 2.25~GHz. We also notice that three burst pairs (B28$-$29, B109$-$110, and B143$-$144) form a small group with waiting times shorter than 0.1~s, suggesting the existence of a left peak. The mean waiting time for this small group is $\sim$22~ms, which is comparable to values found for this \citep[34~ms,][]{zhang2025b} and other repeating FRBs, e.g., $\sim$24~ms for FRB~20121102A \citep{Hewitt2022} and $\sim$19~ms for FRB~20201124A \citep{Niu2022b}.

\begin{figure*}
    \centering
    \includegraphics[width=0.5\linewidth]{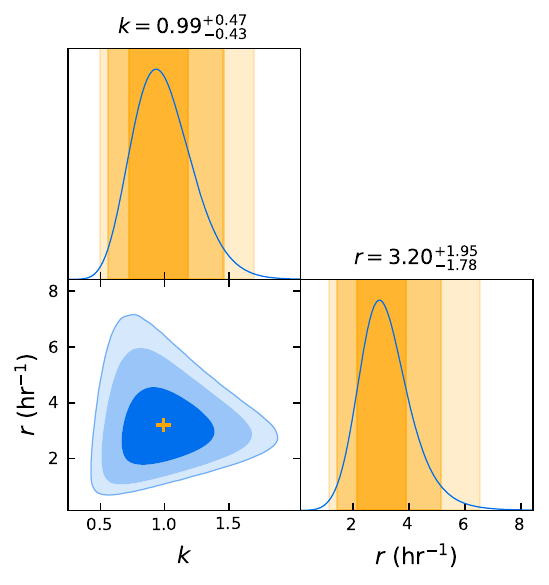}
    \caption{The two-dimensional posterior distribution of Weibull shape parameter $k$ and burst rate $r$ for the observation on 2024 July 23 (O20). The blue contours in the bottom-left sub-panel represent the 68\%, 95\%, and 99\% confidence regions, with the maximum likelihood point marked by the orange cross. The marginal distributions of $r$ and $k$ are shown in the bottom-right and top-left sub-panels, respectively. The shaded orange regions represent the 68\%, 95\% and 99\% confidence levels.}
    \label{fig:corner all}
\end{figure*}

To investigate whether bursts tend to cluster within observations, i.e., over hour-long timescales, we used the Weibull distribution to model the time intervals between subsequent bursts following \citet{Oppermann2018}:
\begin{equation}
    \small \mathcal{W}(\delta t \mid k, r) = k\, \delta t^{-1} \left[ \delta t\, r\, \Gamma\left(1 + 1/k\right) \right]^k \mathrm{e}^{-\left[\delta t\, r\, \Gamma\left(1 + 1/k\right)\right]^k}.
    \label{eq: weibull}
\end{equation}
Here, $r$ represents the Weibull burst rate, $\Gamma$ is the Gamma function, and $k$ is the shape parameter. When $k=1$, the Weibull distribution is equivalent to the Poisson distribution, and $k<1$ implies that bursts tend to cluster in time, with smaller values of $k$ indicating stronger clustering. We selected eight observations (O12, O13, O17, O18, O19, O20, O22, and O56) that detected more than five bursts, which provide better constraints on the Weibull parameters. We sampled the posterior distribution using the Markov Chain Monte Carlo (MCMC) method implemented in the \textsc{emcee}\footnote{\url{https://github.com/dfm/emcee}} \citep{emcee} package. The results are listed in Table~\ref{table: weibull individual}, and the resulting posterior distribution for the observation on 2024 July 23 (O20) is shown as an example in Fig.~\ref{fig:corner all}. Most $k$ values are consistent with $k=1$ within 95\% confidence levels, suggesting that the burst arrival times may follow a Poisson process in these observations, though the large uncertainties on the $k$ parameters limit the strength of this inference. Nevertheless, the result is consistent with previous findings for several active repeating FRBs \citep{Cruces2021, Jahns2023, Konijn2024}. In the case of the observation on 2024 July 31 (O22), however, the obtained $k=0.54^{+0.28}_{-0.25}$ suggests potential burst clustering during this 3.5~hr observation. We notice that the waiting time between bursts B110 and B109 detected in this observation is only $\sim$50~ms. Previous studies have shown that excluding bursts with waiting times shorter than 0.1~s from the dataset could significantly affect the estimated shape parameter $k$ \citep{Cruces2021, Jahns2023}. Moreover, given that the widest burst (B78) in our sample has a temporal width of $53.41\pm4.64$~ms, bursts B110 and B109 may represent two bright peaks from a single burst, with other fainter components remaining undetected. If this $\delta t < 0.1$~s event is excluded from the dataset of O22, the revised shape parameter shifts to $k=0.95^{+0.59}_{-0.52}$, which is consistent with $k=1$ and indicates no strong evidence of burst clustering.

\begin{deluxetable}{lccc}
\setlength{\tabcolsep}{12pt}
\tablecaption{Estimated Weibull distribution parameter of $k$ and $r$ of different observations.}
\tablenum{4}
\label{table: weibull individual}
\tablehead{
    \multicolumn{1}{l}{ID$_{\rm{obs}}$} & $N_{\rm{burst}}$ & \colhead{$k$} & \colhead{$r~(\rm{hr}^{-1})$}
}
\startdata
O12     & 12& $1.08^{+0.48}_{-0.45}$ &  $1.81^{+1.03}_{-0.93}$\\
O13     & 8& $0.99^{+0.54}_{-0.49}$ &  $1.12^{+0.87}_{-0.74}$\\
O17     & 18& $0.91^{+0.34}_{-0.33}$ &  $2.36^{+1.26}_{-1.16}$\\
O18     & 11& $1.19^{+0.54}_{-0.51}$ &  $1.44^{+0.79}_{-0.71}$\\
O19     & 33& $0.86^{+0.23}_{-0.21}$ &  $4.26^{+1.74}_{-1.64}$\\
O20     & 12& $0.99^{+0.47}_{-0.43}$ &  $3.20^{+1.95}_{-1.78}$\\
O22     & 7& $0.54^{+0.28}_{-0.25}$ &  $3.25^{+4.23}_{-2.97}$\\
O22\tablenotemark{a}     & 6& $0.95^{+0.59}_{-0.52}$ &  $2.12^{+2.08}_{-1.61}$\\
O56     & 6& $0.81^{+0.49}_{-0.43}$ &  $7.77^{+8.10}_{-6.37}$\\
\enddata
\tablecomments{Uncertainties are shown in 95\% confidence levels.}
\vspace{-5pt}
\tablenotetext{a}{Only considering waiting times grater than 0.1~s.}
\end{deluxetable}

\section{Discussions}
\label{sec:4} 

\subsection{Implications of non-detection at 8.60~GHz}
\label{sec: xband}

\begin{figure*}
    \centering
    \includegraphics[scale=0.75]{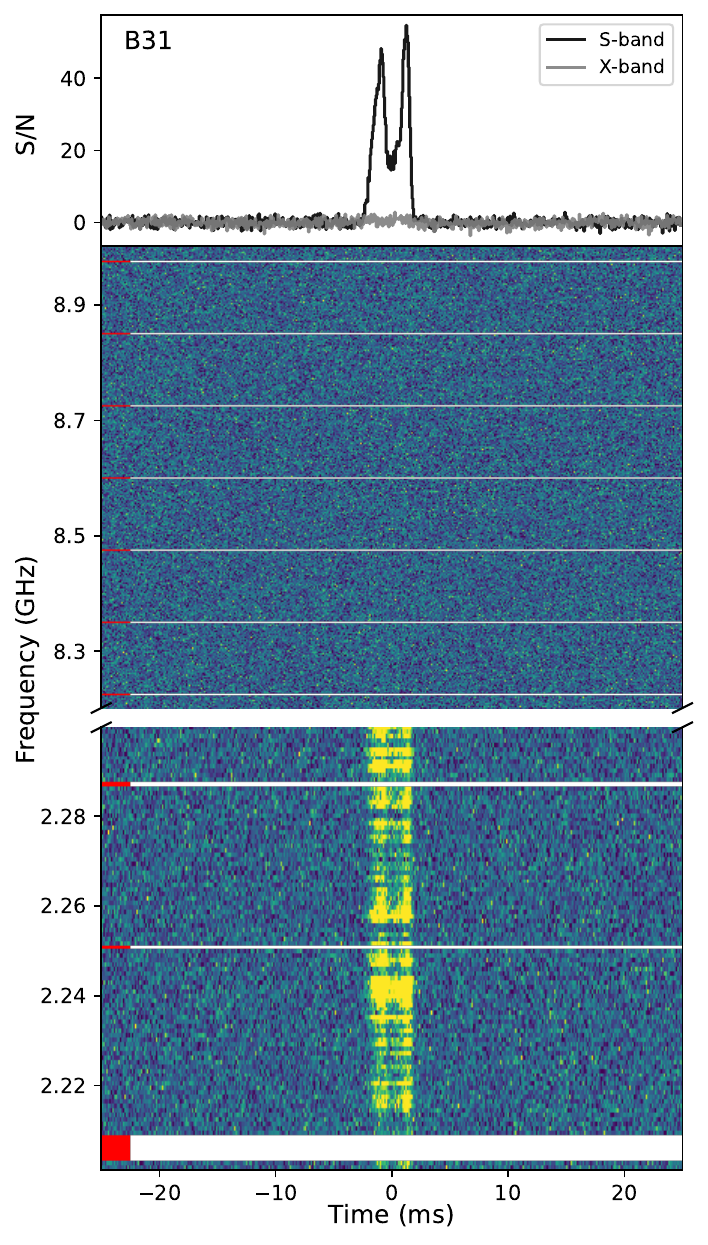}
    \caption{Top panel: the burst profile of burst B31 at 2.25~GHz (black line) and 8.60~GHz (gray line). Bottom panel: the dynamic spectrum of burst B31 at 2.25 and 8.60~GHz after correcting for the dispersion delay between the two observing frequencies. The horizontal white lines, with red markers on the left side, represent masked channels. For visual purposes, the intensity value of the dynamic spectrum is saturated at the 2nd and 98th percentiles.}
    \label{fig:simultaneous_SX}
\end{figure*}

Although FRBs have been detected up to $\sim8$~GHz \citep{Gajjar2018}, we did not detect any bursts from FRB~20240114A at 8.60~GHz above a fluence threshold of 0.27~Jy~ms in blind searches of the observing data. Additionally, we visually inspected the dynamic spectrum around the inferred burst TOA at 8.60~GHz (after correcting for the time delay caused by the DM) for bursts detected at 2.25~GHz, except for burst B122, which has no available 8.60~GHz data (see Sec.~\ref{sec:2}). However, no counterparts of these bursts were detected at 8.60~GHz. In Fig.~\ref{fig:simultaneous_SX}, we show the dynamic spectra for the bright burst B31 at 2.25~GHz and 8.60~GHz as an example. This burst was detected across the entire $\sim$100~MHz bandwidth at 2.25~GHz, but no evident emission was detected in the corresponding 8.60~GHz data.

Based on a total observing time of 178.27~hrs at 8.60~GHz, we place a $1\sigma$ upper limit of $<0.01~\rm{hr}^{-1}$ for the mean burst rate above a fluence threshold of 0.27~Jy~ms. When further focusing on the period when the source was in the high-activity state at 2.25~GHz (i.e., between 2024 June 29 and July 31, see Sec.\ref{sec:burst rate}), the total observing time is reduced to 65~hr, yielding a $1\sigma$ upper limit of $<0.028~\rm{hr}^{-1}$. Assuming the cumulative burst energy distribution at 8.60~GHz follows the same slope of $-1.20$ (see Sec.\ref{sec:energy}), we scale the $1\sigma$ upper limits accordingly. The resulting rates are $<3\times10^{-3}~\rm{hr}^{-1}$ combining all observations and $<9\times10^{-3}~\rm{hr}^{-1}$ during the high-activity period at 2.25~GHz. This suggests that FRB~20240114A is at least two orders of magnitude less active (for the same burst fluence) at 8.60~GHz compared to 2.25~GHz.

Simultaneous multi-frequency observations provide important constraints on the broadband spectra of FRBs \citep[see e.g.,][]{Law2017, Houben2019, Majid2020, Majid2021, Pelliciari2024}. The first such detection was made for FRB~20121102A by Arecibo at 1.4~GHz and the VLA at 3~GHz \citep{Law2017}. \citet{Law2017} assumed $S_\nu\propto\nu^{\alpha}$ (where $S_\nu$ denotes the burst flux density at an observing frequency and $\alpha$ is the spectral index) and obtained a positive spectral index of $\alpha=2.1$ for the single burst simultaneously detected at 1.4 and 3~GHz. Meanwhile, the non-detection of this and other bursts at 4.85~GHz with Effelsberg yielded an upper limit of $\alpha<-2.3$ between 3 and 4.85~GHz. Steep broadband spectra have been found for several repeating FRBs between \textit{S}-band and \textit{X}-band, including $\alpha<-2.6$ for FRB~20121102A \citep[between 2.25$-$8.36~GHz,][]{Majid2020}, $\alpha<-1.3$ for FRB~20200120E \citep[between 2.25$-$8.36~GHz,][]{Majid2021}, and $\alpha<-2.14$ for FRB~20201124A \citep[between 2.26$-$8.45~GHz,][]{Ikebe2023}. Assuming broadband emission for bursts we detected at 2.25~GHz, we derive the upper limit on the 2.25$-$8.60~GHz spectral index $\alpha$ for each detected burst via:
\begin{equation}
    \frac{F_{\rm{s}}}{F_{\rm{x,7\sigma}}}=(\frac{\nu_{\rm{s}}}{\nu_{\rm{x}}})^\alpha,
    \label{eq: spectral index limits}
\end{equation}
where $F_{\rm{s}}$ is the measured burst fluence at 2.25~GHz, $F_{\rm{x,7\sigma}}$ is the $7\sigma$ upper limit on the burst fluence at 8.60~GHz (assuming a temporal width identical to that at 2.25~GHz), $\nu_{\rm{s}}=2.25$~GHz and $\nu_{\rm{x}}=8.60$~GHz are the central frequencies of our observations, and $\alpha$ is the spectral index. The resulting upper limits on spectral indices range from $\alpha<-1.1$ to $-5.2$, with the most stringent constraint derived from the non-detection of the 8.60~GHz counterpart of burst B36, which is the brightest burst in our sample, with a temporal width of $5.29\pm0.15$~ms. For comparison, normal pulsars have a mean spectral index of $-1.60\pm0.03$ \citep{Jankowski2018}. Radio magnetars, proven to be capable of producing FRBs \citep{CHIME2020, Bochenek2020}, typically exhibit flatter spectra with $\alpha>-0.8$ \citep{Camilo2007, Torne2015, Torne2017}. Our results thus suggest that the bursts we detected at 2.25~GHz may be intrinsically narrow-band, similar to those observed for this source at low frequencies \citep[about 10\% fractional bandwidth in the 300$-$700~MHz band,][]{Kumar2024} and for other repeating FRBs \citep[e.g.][]{Spitler2016, Law2017, Majid2020, Gourdji2019, Zhang2023}.

\subsection{Varying frequency-dependent activity for FRB~20240114A}
\label{Sec:chromatic}

Quasi-simultaneous observations of FRB~20240114A with MeerKAT in the 544$-$1088~MHz and 856$-$1712~MHz bands on 2024 February 9 showed that it was three times more active at 816~MHz than at 1284~MHz, with burst rates of $44~\rm{hr}^{-1}$ and $18~\rm{hr}^{-1}$, respectively, above a completeness threshold of $\sim$1~Jy~ms \citep{Tian2024}. Furthermore, we detected no bursts at 2.25 GHz between 2024 January 29 and February 24, yielding a mean burst rate of $<0.13$~hr$^{-1}$ above a fluence threshold of 0.72~Jy~ms. These results suggest that the activity of FRB~20240114A decreases with increasing frequencies at that time. Such frequency-dependent activity has also been observed in other repeating FRBs, such as FRB~20180916B \citep{Pleunis2021b, Pastor2021, Bethapudi2023} and FRB~20190208A \citep{Hewitt24}. In 2024 May, however, observations with the Effelsberg telescope using the Ultra-Broad-Band receiver (UBB, covering 1.3$-$6~GHz) revealed that FRB~20240114A produced comparable, or even more, bursts in the 1.9$-$2.6~GHz band compared with the 1.3$-$1.9~GHz band \citep{ATel16620, Eppel25}. This indicates that the source may exhibit a flattening or even a turnover in its frequency-dependent activity towards low frequencies during this period. Additionally, as shown in the bottom panel of Fig.~\ref{fig:CDF}, burst detections at higher frequencies appear to exhibit a delayed onset, which may suggest a time-evolving preferred emission frequency for FRB~20240114A. 

\begin{figure*}
    \centering
    \includegraphics[width=0.5\linewidth]{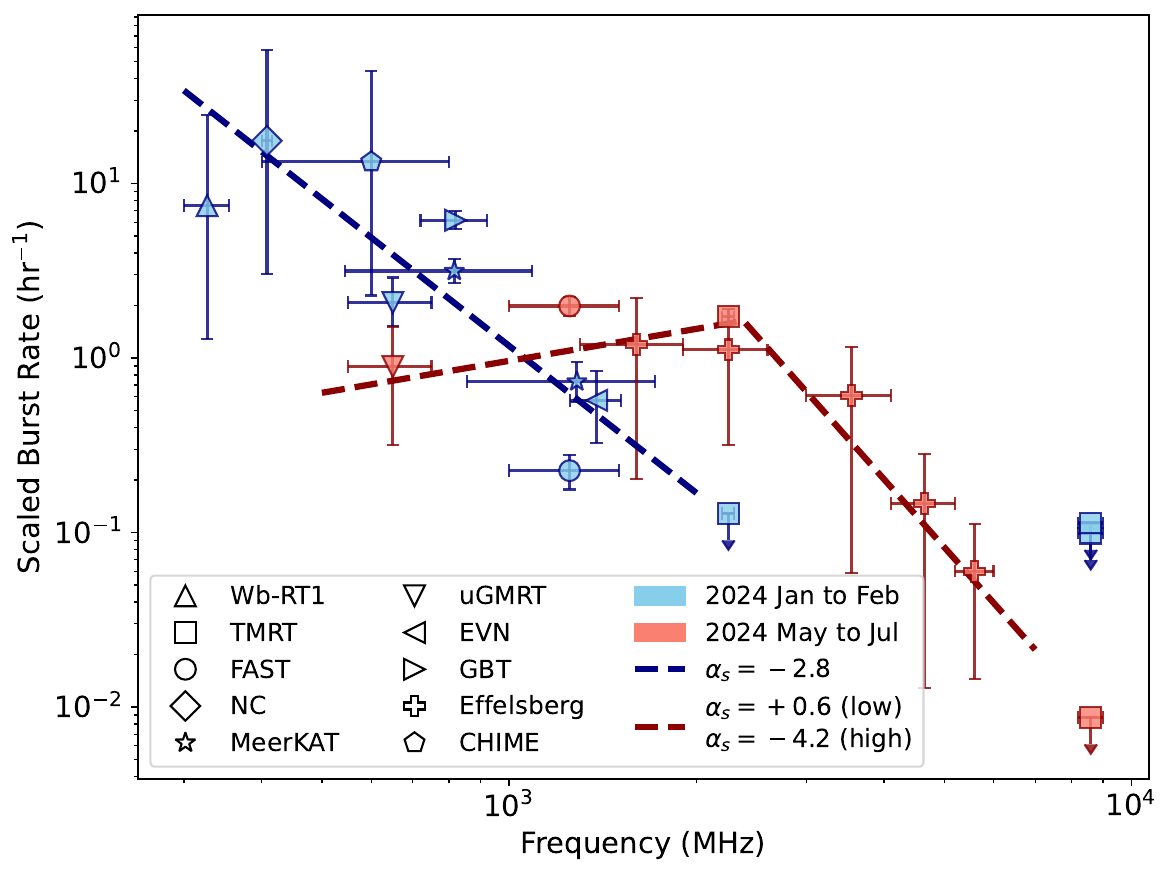}
    \caption{Scaled burst rates as a function of frequency, shown for two time periods: 2024 January-February (blue) and 2024 May-July (red). The horizontal bars indicate the frequency coverage of the corresponding observations. The fitted power-law models are plotted as dashed lines. The upper limit derived from the TMRT 8.60~GHz observations during 2024 January-February is split into two parts to account for the sub-optimal sensitivity in a subset of observations (O1-5, see Sec.~\ref{sec:2}) due to the pointing offsets.}
    \label{fig:multi_freq_burst_rate}
\end{figure*}

In order to make a simple quantification of the chromaticity in the burst rate, as well as its potentially time-evolving behavior, we select two time periods, i.e. (i) from 2024 January \citep[after the ATel announcement by][]{ATel16420} to February and (ii) from 2024 May to July. To take into account large sensitivity differences between various instruments, including the CHIME \citep{Shin2025}, the Westerbork RT1 telescope \citep[Wb-RT1,][]{ATel16432}, the FAST \citep{zhang2025b}, the Northern Cross ratio telescope \citep[NC,][]{ATel16434}, the Effelsberg \citep{ATel16620, Eppel25}, the MeerKAT \citep{Tian2024}, the uGMRT \citep{Kumar2024, Panda2024}, the GBT \citep{Xie2024}, the EVN \citep{Bhardwaj2025}, and the TMRT, the reported multi-frequency burst rates within these two periods are scaling to that expected from a fluence threshold of 0.72~Jy~ms, using the power-law index $\gamma=-1.2$ (assumed to be constant over time and frequency) and assuming a fluence threshold of S/N > 7 for 1~ms width burst (expect that we use the fluence threshold of 15.65~Jy~ms reported by \citet{Shin2025} for the CHIME/FRB observation and 0.64~Jy~ms reported by \citet{Kumar2024} for the uGMRT observation at 550$-$750~MHz on 2024 February 1 after correcting the pointing offset). For multiple observations that were conducted with a certain instrument in a time period at similar frequency ranges, their burst rates are then averaged to form one data point. As a result, the scaled burst rates at different frequencies are shown in Fig.~\ref{fig:multi_freq_burst_rate}. It seems that the scaled burst rate decreased with increasing frequencies in the first time period, which later became peaks at frequencies of $2-3$~GHz in the second time period. We model the frequency-dependence of the scaled burst rate using a power-law and a broken power-law model in the first and second time periods, respectively. The power-law indices are calculated via a bootstrapping method with 10000 trials, as we sample each scaled burst rate value within its uncertainty and perform the fitting, from which we determined a power-law index ($\alpha_{\rm{s}}$) of $-2.8\pm0.4$ for the first time period and $0.6\pm0.6$/$-4.2\pm2.1$ below and above a break frequency of $2.4\pm0.7$~GHz for the second time period. 

Observations of the persistent radio source (PRS), which is hypothesised to be a magnetised nebula surrounding the FRB source \citep[e.g.][]{Margalit2018, Bruni2025b}, associated with FRB~20240114A \citep{Bruni2025} may provide clues to the origin of the observed time-evolving burst activity. As reported by \citet{Zhang2025}, the PRS exhibited temporal-spectral evolution, as its \textit{L-}band (1.5~GHz) flux density increased from $64\pm14$~$\mu$Jy on 2024 February 9 to $131\pm10$~$\mu$Jy on July 23$-$29, while its spectrum evolved from a single power law to a broken power law with a spectral peak between $1-2$~GHz. The authors attributed the spectral peak in 2024 July to synchrotron self-absorption (SSA) and suggest that it might also explain the observed differences in multi-frequency activities of FRB~20240114A, as bursts in optically-thick regimes may be absorbed during propagation through the dense, magnetized plasma of the PRS. Therefore, the observed time-evolving frequency-dependent burst activity from 2024 Jan-Feb to May-July may result from a combination of intrinsic and external effects. While the overall burst rate could increase due to the enhanced activity of the central engine which powers the repeater, the apparent peak activity around 2$-$3~GHz may be caused by the SSA, indicating an evolving environment of the source. However, the current data are insufficient to confirm or rule out this interpretation. Alternative mechanisms, such as the extreme scattering events \citep[ESEs, e.g.][]{Cordes2017}, may also modulate the apparent burst activity at certain frequencies. Nevertheless, the time-evolving frequency-dependent burst activity of FRB~20240114A highlights the importance of wideband, long-term monitoring campaigns, such as those conducted for FRB~20180301A by \citet{Kumar2023} using the Parkes telescope and its Ultra-Wideband Low (UWL, covering 0.7$-$4~GHz) receiver, to further investigate such behaviors in this and other active repeaters in the future.

\section{Conclusions}
\label{sec:5}
We present the results of a long-term monitoring campaign of the newly discovered repeating FRB~20240114A \citep{Shin2025} simultaneously at 2.25 and 8.60~GHz with the TMRT. From 2024 January 29 to 2025 February 15, we performed 66 observations. We accumulated a total on-source time of 182.27~hrs at 2.25~GHz, during which 155 bursts were detected above a fluence threshold of 0.72~Jy~ms. No bursts were detected in 178.27 hours of observation at 8.60~GHz above a fluence threshold of 0.27~Jy~ms.

(i) During our monitoring, the source exhibited higher activity at 2.25~GHz in 2024 July, which we refer to as a high-activity state, with a mean burst rate of $1.72^{+0.18}_{-0.16}~\rm{hr}^{-1}$. Between 2024 August 1 and 2025 February 15, it remained mostly in a low-activity state with a mean burst rate approximately an order of magnitude lower, except for a short-term reactivation around 2025 January 20.

(ii) The mean and median burst widths we obtain at 2.25~GHz are 4.24 and 3~ms, respectively, narrower than those reported by \citet{Tian2024} at lower frequencies. The widest burst in our sample, B78 (see Fig.~\ref{fig: subset bursts}), has a temporal width of $53.41\pm4.64$~ms and is also the widest burst by far reported for this source.

(iii) The waiting time distribution peaks at $\sim$1019~s (see Fig.~\ref{fig:waiting time}), which is longer than those observed for this and other repeating FRBs at lower frequencies \citep[e.g.][]{Li2021, Xu2022, Nimmo2023, Zhang2023, Panda2024, zhang2025b}. We model the waiting time in individual observations (with more than five bursts) using a Weibull distribution, finding no evident burst clustering on hour-long timescales when excluding waiting times below 0.1~s (see Table~\ref{table: weibull individual}).

(iv) The isotropic-equivalent energy of the detected bursts spans $10^{37}$$-$$10^{39}$~erg. The cumulative burst energy distribution follows a power-law relation with a best-fitting index of $\gamma=-1.20\pm0.03\pm0.02$ above the completeness threshold of $7.5\times10^{37}$~erg (see Fig.~\ref{fig:energy distribution}).

(v) FRB~20240114A appears to be at least two orders of magnitude less active at 8.60~GHz than at 2.25~GHz. Based on the non-detection of any counterparts at 8.60~GHz, we derive upper limits on the broadband spectral index ranging from $\alpha<-1.1$ to $-5.2$ between 2.25 and 8.60~GHz.

(vi) The frequency dependence of the burst activity for FRB~20240114A may evolve with time (see Sec.~\ref{Sec:chromatic}), which requires further long-term, wideband observations to investigate such evolution in this and maybe in other active repeaters.

\section*{acknowledgments}
We would like to express our appreciation to the anonymous referee for the helpful comments and suggestions on the paper. This work is supported in part by the National Key R\&D Program of China (grant No. 2022YFA1603104), the National SKA Program of China (grant No. 2020SKA0120104), the National Natural Science Foundation of China (grant Nos. 12041301, U2031119), and the Natural Science Foundation of Shanghai (grant No. 20ZR1467600). This work is also supported by the State Key Laboratory of Radio Astronomy and Technology.

\section*{Data Availability}
The archive data for each of the TMRT bursts are available via China-VO PaperData repository at\dataset[DOI: 10.12149/101581]{https://doi.org/10.12149/101581}.

\bibliography{reference}{}

\begin{thebibliography}{}
\expandafter\ifx\csname natexlab\endcsname\relax\def\natexlab#1{#1}\fi
\providecommand{\url}[1]{\href{#1}{#1}}
\providecommand{\dodoi}[1]{doi:~\href{http://doi.org/#1}{\nolinkurl{#1}}}
\providecommand{\doeprint}[1]{\href{http://ascl.net/#1}{\nolinkurl{http://ascl.net/#1}}}
\providecommand{\doarXiv}[1]{\href{https://arxiv.org/abs/#1}{\nolinkurl{https://arxiv.org/abs/#1}}}

\bibitem[{{Aggarwal} {et~al.}(2021){Aggarwal}, {Agarwal}, {Lewis},
  {Anna-Thomas}, {Tremblay}, {Burke-Spolaor}, {McLaughlin}, \&
  {Lorimer}}]{Aggarwal2021}
{Aggarwal}, K., {Agarwal}, D., {Lewis}, E.~F., {et~al.} 2021, \apj, 922, 115,
  \dodoi{10.3847/1538-4357/ac2577}

\bibitem[{Aggarwal {et~al.}(2020)Aggarwal, Agarwal, Kania, Fiore, Thomas,
  Ransom, Demorest, Wharton, Burke-Spolaor, Lorimer, Mclaughlin, \&
  Garver-Daniels}]{your}
Aggarwal, K., Agarwal, D., Kania, J.~W., {et~al.} 2020, Journal of Open Source
  Software, 5, 2750, \dodoi{10.21105/joss.02750}

\bibitem[{{Anna-Thomas} {et~al.}(2023){Anna-Thomas}, {Connor}, {Dai}, {Feng},
  {Burke-Spolaor}, {Beniamini}, {Yang}, {Zhang}, {Aggarwal}, {Law}, {Li},
  {Niu}, {Chatterjee}, {Cruces}, {Duan}, {Filipovic}, {Hobbs}, {Lynch}, {Miao},
  {Niu}, {Ocker}, {Tsai}, {Wang}, {Xue}, {Yao}, {Yu}, {Zhang}, {Zhang}, {Zhu},
  \& {Zhu}}]{Anna2023}
{Anna-Thomas}, R., {Connor}, L., {Dai}, S., {et~al.} 2023, Science, 380, 599,
  \dodoi{10.1126/science.abo6526}

\bibitem[{{Astropy Collaboration} {et~al.}(2013){Astropy Collaboration},
  {Robitaille}, {Tollerud}, {Greenfield}, {Droettboom}, {Bray}, {Aldcroft},
  {Davis}, {Ginsburg}, {Price-Whelan}, {Kerzendorf}, {Conley}, {Crighton},
  {Barbary}, {Muna}, {Ferguson}, {Grollier}, {Parikh}, {Nair}, {Unther},
  {Deil}, {Woillez}, {Conseil}, {Kramer}, {Turner}, {Singer}, {Fox}, {Weaver},
  {Zabalza}, {Edwards}, {Azalee Bostroem}, {Burke}, {Casey}, {Crawford},
  {Dencheva}, {Ely}, {Jenness}, {Labrie}, {Lim}, {Pierfederici}, {Pontzen},
  {Ptak}, {Refsdal}, {Servillat}, \& {Streicher}}]{astropy}
{Astropy Collaboration}, {Robitaille}, T.~P., {Tollerud}, E.~J., {et~al.} 2013,
  \aap, 558, A33, \dodoi{10.1051/0004-6361/201322068}

\bibitem[{{Bethapudi} {et~al.}(2023){Bethapudi}, {Spitler}, {Main}, {Li}, \&
  {Wharton}}]{Bethapudi2023}
{Bethapudi}, S., {Spitler}, L.~G., {Main}, R.~A., {Li}, D.~Z., \& {Wharton},
  R.~S. 2023, \mnras, 524, 3303, \dodoi{10.1093/mnras/stad2009}

\bibitem[{{Bhardwaj} {et~al.}(2024){Bhardwaj}, {Kirichenko}, \& {Gil de
  Paz}}]{ATel16613}
{Bhardwaj}, M., {Kirichenko}, A., \& {Gil de Paz}, A. 2024, The Astronomer's
  Telegram, 16613, 1

\bibitem[{{Bhardwaj} {et~al.}(2025){Bhardwaj}, {Snelders}, {Hessels}, {Gil de
  Paz}, {Bhandari}, {Marcote}, {Kirichenko}, {Ould-Boukattine}, {Kirsten},
  {Bempong-Manful}, {Bezrukovs}, {Bray}, {Buttaccio}, {Corongiu}, {Feiler},
  {Gawronski}, {Giroletti}, {Hewitt}, {Lindqvist}, {Maccaferri}, {Moroianu},
  {Nimmo}, {Paragi}, {Puchalska}, {Wang}, {Williams-Baldwin}, \&
  {Yuan}}]{Bhardwaj2025}
{Bhardwaj}, M., {Snelders}, M.~P., {Hessels}, J.~W.~T., {et~al.} 2025, arXiv
  e-prints, arXiv:2506.11915, \dodoi{10.48550/arXiv.2506.11915}

\bibitem[{{Bochenek} {et~al.}(2020){Bochenek}, {Ravi}, {Belov}, {Hallinan},
  {Kocz}, {Kulkarni}, \& {McKenna}}]{Bochenek2020}
{Bochenek}, C.~D., {Ravi}, V., {Belov}, K.~V., {et~al.} 2020, \nat, 587, 59,
  \dodoi{10.1038/s41586-020-2872-x}

\bibitem[{{Bruni} {et~al.}(2024{\natexlab{a}}){Bruni}, {Piro}, {Yang}, {Quai},
  {Zhang}, {Palazzi}, {Nicastro}, {Feruglio}, {Tripodi}, {O'Connor}, {Gardini},
  {Savaglio}, {Rossi}, {Nicuesa Guelbenzu}, \& {Paladino}}]{Bruni2025b}
{Bruni}, G., {Piro}, L., {Yang}, Y.-P., {et~al.} 2024{\natexlab{a}}, \nat, 632,
  1014, \dodoi{10.1038/s41586-024-07782-6}

\bibitem[{{Bruni} {et~al.}(2024{\natexlab{b}}){Bruni}, {Piro}, {Yang},
  {Palazzi}, {Nicastro}, {Rossi}, {Savaglio}, {Maiorano}, \&
  {Zhang}}]{Bruni2025}
{Bruni}, G., {Piro}, L., {Yang}, Y.~P., {et~al.} 2024{\natexlab{b}}, arXiv
  e-prints, arXiv:2412.01478, \dodoi{10.48550/arXiv.2412.01478}

\bibitem[{{Camilo} {et~al.}(2007){Camilo}, {Ransom}, {Pe{\~n}alver},
  {Karastergiou}, {van Kerkwijk}, {Durant}, {Halpern}, {Reynolds}, {Thum},
  {Helfand}, {Zimmerman}, \& {Cognard}}]{Camilo2007}
{Camilo}, F., {Ransom}, S.~M., {Pe{\~n}alver}, J., {et~al.} 2007, \apj, 669,
  561, \dodoi{10.1086/521548}

\bibitem[{{CHIME/FRB Collaboration} {et~al.}(2020){CHIME/FRB Collaboration},
  {Andersen}, {Bandura}, {Bhardwaj}, {Bij}, {Boyce}, {Boyle}, {Brar},
  {Cassanelli}, {Chawla}, {Chen}, {Cliche}, {Cook}, {Cubranic}, {Curtin},
  {Denman}, {Dobbs}, {Dong}, {Fandino}, {Fonseca}, {Gaensler}, {Giri}, {Good},
  {Halpern}, {Hill}, {Hinshaw}, {H{\"o}fer}, {Josephy}, {Kania}, {Kaspi},
  {Landecker}, {Leung}, {Li}, {Lin}, {Masui}, {McKinven}, {Mena-Parra},
  {Merryfield}, {Meyers}, {Michilli}, {Milutinovic}, {Mirhosseini},
  {M{\"u}nchmeyer}, {Naidu}, {Newburgh}, {Ng}, {Patel}, {Pen},
  {Pinsonneault-Marotte}, {Pleunis}, {Quine}, {Rafiei-Ravandi}, {Rahman},
  {Ransom}, {Renard}, {Sanghavi}, {Scholz}, {Shaw}, {Shin}, {Siegel}, {Singh},
  {Smegal}, {Smith}, {Stairs}, {Tan}, {Tendulkar}, {Tretyakov}, {Vanderlinde},
  {Wang}, {Wulf}, \& {Zwaniga}}]{CHIME2020}
{CHIME/FRB Collaboration}, {Andersen}, B.~C., {Bandura}, K.~M., {et~al.} 2020,
  \nat, 587, 54, \dodoi{10.1038/s41586-020-2863-y}

\bibitem[{{Chime/Frb Collaboration} {et~al.}(2020){Chime/Frb Collaboration},
  {Amiri}, {Andersen}, {Bandura}, {Bhardwaj}, {Boyle}, {Brar}, {Chawla},
  {Chen}, {Cliche}, {Cubranic}, {Deng}, {Denman}, {Dobbs}, {Dong}, {Fandino},
  {Fonseca}, {Gaensler}, {Giri}, {Good}, {Halpern}, {Hessels}, {Hill},
  {H{\"o}fer}, {Josephy}, {Kania}, {Karuppusamy}, {Kaspi}, {Keimpema},
  {Kirsten}, {Landecker}, {Lang}, {Leung}, {Li}, {Lin}, {Marcote}, {Masui},
  {McKinven}, {Mena-Parra}, {Merryfield}, {Michilli}, {Milutinovic},
  {Mirhosseini}, {Naidu}, {Newburgh}, {Ng}, {Nimmo}, {Paragi}, {Patel}, {Pen},
  {Pinsonneault-Marotte}, {Pleunis}, {Rafiei-Ravandi}, {Rahman}, {Ransom},
  {Renard}, {Sanghavi}, {Scholz}, {Shaw}, {Shin}, {Siegel}, {Singh}, {Smegal},
  {Smith}, {Stairs}, {Tendulkar}, {Tretyakov}, {Vanderlinde}, {Wang}, {Wang},
  {Wulf}, {Yadav}, \& {Zwaniga}}]{CHIME2020b}
{Chime/Frb Collaboration}, {Amiri}, M., {Andersen}, B.~C., {et~al.} 2020, \nat,
  582, 351, \dodoi{10.1038/s41586-020-2398-2}

\bibitem[{{Chime/FRB Collaboration} {et~al.}(2023){Chime/FRB Collaboration},
  {Andersen}, {Bandura}, {Bhardwaj}, {Boyle}, {Brar}, {Cassanelli},
  {Chatterjee}, {Chawla}, {Cook}, {Curtin}, {Dobbs}, {Dong}, {Faber},
  {Fandino}, {Fonseca}, {Gaensler}, {Giri}, {Herrera-Martin}, {Hill}, {Ibik},
  {Josephy}, {Kaczmarek}, {Kader}, {Kaspi}, {Landecker}, {Lanman}, {Lazda},
  {Leung}, {Lin}, {Masui}, {McKinven}, {Mena-Parra}, {Meyers}, {Michilli},
  {Ng}, {Pandhi}, {Pearlman}, {Pen}, {Petroff}, {Pleunis}, {Rafiei-Ravandi},
  {Rahman}, {Ransom}, {Renard}, {Sand}, {Sanghavi}, {Scholz}, {Shah}, {Shin},
  {Siegel}, {Smith}, {Stairs}, {Su}, {Tendulkar}, {Vanderlinde}, {Wang},
  {Wulf}, \& {Zwaniga}}]{CHIME2023}
{Chime/FRB Collaboration}, {Andersen}, B.~C., {Bandura}, K., {et~al.} 2023,
  \apj, 947, 83, \dodoi{10.3847/1538-4357/acc6c1}

\bibitem[{{Cordes} \& {Chatterjee}(2019)}]{Cordes2019}
{Cordes}, J.~M., \& {Chatterjee}, S. 2019, \araa, 57, 417,
  \dodoi{10.1146/annurev-astro-091918-104501}

\bibitem[{{Cordes} \& {Wasserman}(2016)}]{Cordes2016}
{Cordes}, J.~M., \& {Wasserman}, I. 2016, \mnras, 457, 232,
  \dodoi{10.1093/mnras/stv2948}

\bibitem[{{Cordes} {et~al.}(2017){Cordes}, {Wasserman}, {Hessels}, {Lazio},
  {Chatterjee}, \& {Wharton}}]{Cordes2017}
{Cordes}, J.~M., {Wasserman}, I., {Hessels}, J.~W.~T., {et~al.} 2017, \apj,
  842, 35, \dodoi{10.3847/1538-4357/aa74da}

\bibitem[{{Crawford} {et~al.}(1970){Crawford}, {Jauncey}, \&
  {Murdoch}}]{Crawford1970}
{Crawford}, D.~F., {Jauncey}, D.~L., \& {Murdoch}, H.~S. 1970, \apj, 162, 405,
  \dodoi{10.1086/150672}

\bibitem[{{Cruces} {et~al.}(2021){Cruces}, {Spitler}, {Scholz}, {Lynch},
  {Seymour}, {Hessels}, {Gouiff{\'e}s}, {Hilmarsson}, {Kramer}, \&
  {Munjal}}]{Cruces2021}
{Cruces}, M., {Spitler}, L.~G., {Scholz}, P., {et~al.} 2021, \mnras, 500, 448,
  \dodoi{10.1093/mnras/staa3223}

\bibitem[{{Curtin} {et~al.}(2024){Curtin}, {Sand}, {Pleunis}, {Jain}, {Kaspi},
  {Michilli}, {Fonseca}, {Shin}, {Nimmo}, {Brar}, {Dong}, {Eadie}, {Gaensler},
  {Herrera-Martin}, {Ibik}, {Joseph}, {Kaczmarek}, {Leung}, {Main}, {Masui},
  {McKinven}, {Mena-Parra}, {Ng}, {Pandhi}, {Pearlman}, {Rafiei-Ravandi},
  {Sammons}, {Scholz}, {Smith}, \& {Stairs}}]{Curtin2024}
{Curtin}, A.~P., {Sand}, K.~R., {Pleunis}, Z., {et~al.} 2024, arXiv e-prints,
  arXiv:2411.02870, \dodoi{10.48550/arXiv.2411.02870}

\bibitem[{{Dai} {et~al.}(2016){Dai}, {Wang}, {Wu}, \& {Huang}}]{Dai2016}
{Dai}, Z.~G., {Wang}, J.~S., {Wu}, X.~F., \& {Huang}, Y.~F. 2016, \apj, 829,
  27, \dodoi{10.3847/0004-637X/829/1/27}

\bibitem[{{Du} {et~al.}(2024){Du}, {Zheng}, {Liu}, {Zhang}, {Zhang}, {Zhang},
  {Qin}, {Shen}, {Liu}, {Li}, \& {Wang}}]{Du2024}
{Du}, B., {Zheng}, Y., {Liu}, G., {et~al.} 2024, Astronomical Techniques and
  Instruments, 1, 247, \dodoi{10.61977/ati2024038}

\bibitem[{Eppel {et~al.}(2025)Eppel, Krumpe, Limaye, Intrarat,
  Wongphechauxsorn, Cruces, Herrmann, Jankowski, Jaroenjittichai, Spitler, \&
  Kadler}]{Eppel25}
Eppel, F., Krumpe, M., Limaye, P., {et~al.} 2025, Astronomy \&; Astrophysics,
  \dodoi{10.1051/0004-6361/202453563}

\bibitem[{{Feng} {et~al.}(2022){Feng}, {Li}, {Yang}, {Zhang}, {Zhu}, {Zhang},
  {Lu}, {Wang}, {Dai}, {Lynch}, {Yao}, {Jiang}, {Niu}, {Zhou}, {Xu}, {Miao},
  {Niu}, {Meng}, {Qian}, {Tsai}, {Wang}, {Xue}, {Yue}, {Yuan}, {Zhang}, \&
  {Zhang}}]{Feng22}
{Feng}, Y., {Li}, D., {Yang}, Y.-P., {et~al.} 2022, Science, 375, 1266,
  \dodoi{10.1126/science.abl7759}

\bibitem[{{Foreman-Mackey} {et~al.}(2013){Foreman-Mackey}, {Hogg}, {Lang}, \&
  {Goodman}}]{emcee}
{Foreman-Mackey}, D., {Hogg}, D.~W., {Lang}, D., \& {Goodman}, J. 2013, \pasp,
  125, 306, \dodoi{10.1086/670067}

\bibitem[{{Gajjar} {et~al.}(2018){Gajjar}, {Siemion}, {Price}, {Law},
  {Michilli}, {Hessels}, {Chatterjee}, {Archibald}, {Bower}, {Brinkman},
  {Burke-Spolaor}, {Cordes}, {Croft}, {Enriquez}, {Foster}, {Gizani},
  {Hellbourg}, {Isaacson}, {Kaspi}, {Lazio}, {Lebofsky}, {Lynch}, {MacMahon},
  {McLaughlin}, {Ransom}, {Scholz}, {Seymour}, {Spitler}, {Tendulkar},
  {Werthimer}, \& {Zhang}}]{Gajjar2018}
{Gajjar}, V., {Siemion}, A.~P.~V., {Price}, D.~C., {et~al.} 2018, \apj, 863, 2,
  \dodoi{10.3847/1538-4357/aad005}

\bibitem[{{Gehrels}(1986)}]{poissonerr}
{Gehrels}, N. 1986, \apj, 303, 336, \dodoi{10.1086/164079}

\bibitem[{{Geng} \& {Huang}(2015)}]{Geng2015}
{Geng}, J.~J., \& {Huang}, Y.~F. 2015, \apj, 809, 24,
  \dodoi{10.1088/0004-637X/809/1/24}

\bibitem[{{Gourdji} {et~al.}(2019){Gourdji}, {Michilli}, {Spitler}, {Hessels},
  {Seymour}, {Cordes}, \& {Chatterjee}}]{Gourdji2019}
{Gourdji}, K., {Michilli}, D., {Spitler}, L.~G., {et~al.} 2019, \apjl, 877,
  L19, \dodoi{10.3847/2041-8213/ab1f8a}

\bibitem[{{Gourdji} {et~al.}(2020){Gourdji}, {Rowlinson}, {Wijers}, \&
  {Goldstein}}]{Gourdji2020}
{Gourdji}, K., {Rowlinson}, A., {Wijers}, R.~A.~M.~J., \& {Goldstein}, A. 2020,
  \mnras, 497, 3131, \dodoi{10.1093/mnras/staa2128}

\bibitem[{{Hewitt} {et~al.}(2024{\natexlab{a}}){Hewitt}, {Huang}, {Hessels},
  {Cognard}, {Guillemot}, {Ould-Boukattine}, {Snelders}, \&
  {Kirsten}}]{ATel16597}
{Hewitt}, D.~M., {Huang}, J., {Hessels}, J.~W.~T., {et~al.} 2024{\natexlab{a}},
  The Astronomer's Telegram, 16597, 1

\bibitem[{{Hewitt} {et~al.}(2022){Hewitt}, {Snelders}, {Hessels}, {Nimmo},
  {Jahns}, {Spitler}, {Gourdji}, {Hilmarsson}, {Michilli}, {Ould-Boukattine},
  {Scholz}, \& {Seymour}}]{Hewitt2022}
{Hewitt}, D.~M., {Snelders}, M.~P., {Hessels}, J.~W.~T., {et~al.} 2022, \mnras,
  515, 3577, \dodoi{10.1093/mnras/stac1960}

\bibitem[{{Hewitt} {et~al.}(2024{\natexlab{b}}){Hewitt}, {Bhardwaj}, {Gordon},
  {Kirichenko}, {Nimmo}, {Bhandari}, {Cognard}, {Fong}, {Gil de Paz},
  {Gopinath}, {Hessels}, {Kirsten}, {Marcote}, {Bezrukovs}, {Blaauw}, {Bray},
  {Buttaccio}, {Cassanelli}, {Chawla}, {Corongiu}, {Deng}, {Didehbani}, {Dong},
  {Gawro{\'n}ski}, {Giroletti}, {Guillemot}, {Huang}, {Ivanov}, {Joseph},
  {Kaspi}, {Kharinov}, {Lazda}, {Lindqvist}, {Maccaferri}, {Mas-Ribas},
  {Masui}, {Mckinven}, {Melnikov}, {Michilli}, {Mikhailov}, {Nugent},
  {Ould-Boukattine}, {Paragi}, {Pearlman}, {Pen}, {Pleunis}, {Sand}, {Shah},
  {Shin}, {Snelders}, {Venturi}, {Wang}, {Williams-Baldwin}, {Yang}, \&
  {Yuan}}]{Hewitt24}
{Hewitt}, D.~M., {Bhardwaj}, M., {Gordon}, A.~C., {et~al.} 2024{\natexlab{b}},
  \apjl, 977, L4, \dodoi{10.3847/2041-8213/ad8ce1}

\bibitem[{{Hilmarsson} {et~al.}(2021){Hilmarsson}, {Michilli}, {Spitler},
  {Wharton}, {Demorest}, {Desvignes}, {Gourdji}, {Hackstein}, {Hessels},
  {Nimmo}, {Seymour}, {Kramer}, \& {Mckinven}}]{Hilmarsson2021}
{Hilmarsson}, G.~H., {Michilli}, D., {Spitler}, L.~G., {et~al.} 2021, \apjl,
  908, L10, \dodoi{10.3847/2041-8213/abdec0}

\bibitem[{{Hotan} {et~al.}(2004){Hotan}, {van Straten}, \&
  {Manchester}}]{PSRFITS}
{Hotan}, A.~W., {van Straten}, W., \& {Manchester}, R.~N. 2004, \pasa, 21, 302,
  \dodoi{10.1071/AS04022}

\bibitem[{{Houben} {et~al.}(2019){Houben}, {Spitler}, {ter Veen}, {Rachen},
  {Falcke}, \& {Kramer}}]{Houben2019}
{Houben}, L.~J.~M., {Spitler}, L.~G., {ter Veen}, S., {et~al.} 2019, \aap, 623,
  A42, \dodoi{10.1051/0004-6361/201833875}

\bibitem[{{Huang} {et~al.}(2025){Huang}, {Zhang}, {Xu}, {Hao}, {Lee}, {Zhang},
  {Wang}, {Cao}, {Zhou}, {Xu}, {Li}, {Xu}, {Wang}, {Jiang}, {Guo}, {Xue},
  {Shen}, {Wang}, {Men}, {Chen}, {Wu}, \& {Wang}}]{Huang2025}
{Huang}, Y.-X., {Zhang}, J.-S., {Xu}, H., {et~al.} 2025, arXiv e-prints,
  arXiv:2504.03569, \dodoi{10.48550/arXiv.2504.03569}

\bibitem[{{Ikebe} {et~al.}(2023){Ikebe}, {Takefuji}, {Terasawa}, {Eie},
  {Akahori}, {Murata}, {Hashimoto}, {Kisaka}, {Honma}, {Yoshiura}, {Suzuki},
  {Oyama}, {Sekido}, {Niinuma}, {Takeuchi}, {Yonekura}, \& {Enoto}}]{Ikebe2023}
{Ikebe}, S., {Takefuji}, K., {Terasawa}, T., {et~al.} 2023, \pasj, 75, 199,
  \dodoi{10.1093/pasj/psac101}

\bibitem[{{Ioka} \& {Zhang}(2020)}]{Ioka2020}
{Ioka}, K., \& {Zhang}, B. 2020, \apjl, 893, L26,
  \dodoi{10.3847/2041-8213/ab83fb}

\bibitem[{{Jahns} {et~al.}(2023){Jahns}, {Spitler}, {Nimmo}, {Hewitt},
  {Snelders}, {Seymour}, {Hessels}, {Gourdji}, {Michilli}, \&
  {Hilmarsson}}]{Jahns2023}
{Jahns}, J.~N., {Spitler}, L.~G., {Nimmo}, K., {et~al.} 2023, \mnras, 519, 666,
  \dodoi{10.1093/mnras/stac3446}

\bibitem[{{James} {et~al.}(2019){James}, {Ekers}, {Macquart}, {Bannister}, \&
  {Shannon}}]{James2019}
{James}, C.~W., {Ekers}, R.~D., {Macquart}, J.~P., {Bannister}, K.~W., \&
  {Shannon}, R.~M. 2019, \mnras, 483, 1342, \dodoi{10.1093/mnras/sty3031}

\bibitem[{{Jankowski} {et~al.}(2018){Jankowski}, {van Straten}, {Keane},
  {Bailes}, {Barr}, {Johnston}, \& {Kerr}}]{Jankowski2018}
{Jankowski}, F., {van Straten}, W., {Keane}, E.~F., {et~al.} 2018, \mnras, 473,
  4436, \dodoi{10.1093/mnras/stx2476}

\bibitem[{{Jankowski} {et~al.}(2023){Jankowski}, {Bezuidenhout}, {Caleb},
  {Driessen}, {Malenta}, {Morello}, {Rajwade}, {Sanidas}, {Stappers}, {Surnis},
  {Barr}, {Chen}, {Kramer}, {Wu}, {Buchner}, {Serylak}, \&
  {Prochaska}}]{scatfit}
{Jankowski}, F., {Bezuidenhout}, M.~C., {Caleb}, M., {et~al.} 2023, \mnras,
  524, 4275, \dodoi{10.1093/mnras/stad2041}

\bibitem[{{Jiang} {et~al.}(2024){Jiang}, {Xu}, {Niu}, {Lee}, {Zhu}, {Zhang},
  {Qu}, {Xu}, {Zhou}, {Cao}, {Wang}, {Wang}, {Cao}, {Zhang}, {Zhang}, {Gan},
  {Han}, {Hao}, {Huang}, {Jiang}, {Li}, {Li}, {Li}, {Li}, {Luo}, {Men}, {Qian},
  {Sun}, {Wang}, {Xu}, {Xu}, {Yang}, {Yao}, {Yue}, {Yu}, {Yuan}, \&
  {Zhu}}]{Jiang2024}
{Jiang}, J.~C., {Xu}, J.~W., {Niu}, J.~R., {et~al.} 2024, National Science
  Review, 12, nwae293, \dodoi{10.1093/nsr/nwae293}

\bibitem[{{Joshi} {et~al.}(2024){Joshi}, {Medina}, {Earwicker}, {Farah},
  {Gajjar}, {Sheikh}, {Pollak}, {Siemion}, {Cruz}, {Hickish}, {Premnath},
  {DeBoer}, {Donnachie}, {Singh}, {Davis}, {Snodgrass}, \& {Karn}}]{ATel16599}
{Joshi}, P., {Medina}, A., {Earwicker}, J.~T., {et~al.} 2024, The Astronomer's
  Telegram, 16599, 1

\bibitem[{{Kirsten} {et~al.}(2022){Kirsten}, {Marcote}, {Nimmo}, {Hessels},
  {Bhardwaj}, {Tendulkar}, {Keimpema}, {Yang}, {Snelders}, {Scholz},
  {Pearlman}, {Law}, {Peters}, {Giroletti}, {Paragi}, {Bassa}, {Hewitt},
  {Bach}, {Bezrukovs}, {Burgay}, {Buttaccio}, {Conway}, {Corongiu}, {Feiler},
  {Forss{\'e}n}, {Gawro{\'n}ski}, {Karuppusamy}, {Kharinov}, {Lindqvist},
  {Maccaferri}, {Melnikov}, {Ould-Boukattine}, {Possenti}, {Surcis}, {Wang},
  {Yuan}, {Aggarwal}, {Anna-Thomas}, {Bower}, {Blaauw}, {Burke-Spolaor},
  {Cassanelli}, {Clarke}, {Fonseca}, {Gaensler}, {Gopinath}, {Kaspi}, {Kassim},
  {Lazio}, {Leung}, {Li}, {Lin}, {Masui}, {Mckinven}, {Michilli}, {Mikhailov},
  {Ng}, {Orbidans}, {Pen}, {Petroff}, {Rahman}, {Ransom}, {Shin}, {Smith},
  {Stairs}, \& {Vlemmings}}]{Kirsten2022}
{Kirsten}, F., {Marcote}, B., {Nimmo}, K., {et~al.} 2022, \nat, 602, 585,
  \dodoi{10.1038/s41586-021-04354-w}

\bibitem[{{Kirsten} {et~al.}(2024){Kirsten}, {Ould-Boukattine}, {Herrmann},
  {Gawro{\'n}ski}, {Hessels}, {Lu}, {Snelders}, {Chawla}, {Yang}, {Blaauw},
  {Nimmo}, {Puchalska}, {Wolak}, \& {van Ruiten}}]{Kirsten2024}
{Kirsten}, F., {Ould-Boukattine}, O.~S., {Herrmann}, W., {et~al.} 2024, Nature
  Astronomy, 8, 337, \dodoi{10.1038/s41550-023-02153-z}

\bibitem[{{Konijn} {et~al.}(2024){Konijn}, {Hewitt}, {Hessels}, {Cognard},
  {Huang}, {Ould-Boukattine}, {Chawla}, {Nimmo}, {Snelders}, {Gopinath}, \&
  {Manaswini}}]{Konijn2024}
{Konijn}, D.~C., {Hewitt}, D.~M., {Hessels}, J. W.~T., {et~al.} 2024, \mnras,
  \dodoi{10.1093/mnras/stae2296}

\bibitem[{{Kumar} {et~al.}(2024{\natexlab{a}}){Kumar}, {Maan}, \&
  {Bhusare}}]{ATel16452}
{Kumar}, A., {Maan}, Y., \& {Bhusare}, Y. 2024{\natexlab{a}}, The Astronomer's
  Telegram, 16452, 1

\bibitem[{{Kumar} {et~al.}(2024{\natexlab{b}}){Kumar}, {Maan}, \&
  {Bhusare}}]{Kumar2024}
---. 2024{\natexlab{b}}, arXiv e-prints, arXiv:2406.12804,
  \dodoi{10.48550/arXiv.2406.12804}

\bibitem[{{Kumar} {et~al.}(2023){Kumar}, {Luo}, {Price}, {Shannon}, {Deller},
  {Bhandari}, {Feng}, {Flynn}, {Jiang}, {Uttarkar}, {Wang}, \&
  {Zhang}}]{Kumar2023}
{Kumar}, P., {Luo}, R., {Price}, D.~C., {et~al.} 2023, \mnras, 526, 3652,
  \dodoi{10.1093/mnras/stad2969}

\bibitem[{{Law} {et~al.}(2017){Law}, {Abruzzo}, {Bassa}, {Bower},
  {Burke-Spolaor}, {Butler}, {Cantwell}, {Carey}, {Chatterjee}, {Cordes},
  {Demorest}, {Dowell}, {Fender}, {Gourdji}, {Grainge}, {Hessels}, {Hickish},
  {Kaspi}, {Lazio}, {McLaughlin}, {Michilli}, {Mooley}, {Perrott}, {Ransom},
  {Razavi-Ghods}, {Rupen}, {Scaife}, {Scott}, {Scholz}, {Seymour}, {Spitler},
  {Stovall}, {Tendulkar}, {Titterington}, {Wharton}, \& {Williams}}]{Law2017}
{Law}, C.~J., {Abruzzo}, M.~W., {Bassa}, C.~G., {et~al.} 2017, \apj, 850, 76,
  \dodoi{10.3847/1538-4357/aa9700}

\bibitem[{{Li} \& {Zanazzi}(2021)}]{Lidong2021}
{Li}, D., \& {Zanazzi}, J.~J. 2021, \apjl, 909, L25,
  \dodoi{10.3847/2041-8213/abeaa4}

\bibitem[{{Li} {et~al.}(2021{\natexlab{a}}){Li}, {Wang}, {Zhu}, {Zhang},
  {Zhang}, {Duan}, {Zhang}, {Feng}, {Tang}, {Chatterjee}, {Cordes}, {Cruces},
  {Dai}, {Gajjar}, {Hobbs}, {Jin}, {Kramer}, {Lorimer}, {Miao}, {Niu}, {Niu},
  {Pan}, {Qian}, {Spitler}, {Werthimer}, {Zhang}, {Wang}, {Xie}, {Yue},
  {Zhang}, {Zhi}, \& {Zhu}}]{Li2021}
{Li}, D., {Wang}, P., {Zhu}, W.~W., {et~al.} 2021{\natexlab{a}}, \nat, 598,
  267, \dodoi{10.1038/s41586-021-03878-5}

\bibitem[{{Li} {et~al.}(2021{\natexlab{b}}){Li}, {Yang}, {Wang}, {Xu}, {Shao},
  {Liu}, \& {Dai}}]{Liqiao2021}
{Li}, Q.-C., {Yang}, Y.-P., {Wang}, F.~Y., {et~al.} 2021{\natexlab{b}}, \apjl,
  918, L5, \dodoi{10.3847/2041-8213/ac1922}

\bibitem[{{Li} {et~al.}(2025){Li}, {Zhang}, {Yang}, {Tsai}, {Yang}, {Law},
  {Anna-Thomas}, {Chen}, {Lee}, {Tang}, {Xiao}, {Xu}, {Yang}, {Chen}, {Feng},
  {Li}, {Mckinven}, {Niu}, {Shin}, {Wang}, {Zhang}, {Zhang}, {Zhou}, {Zhu},
  {Dai}, {Chang}, {Geng}, {Han}, {Hu}, {Li}, {Luo}, {Niu}, {Shi}, {Sun}, {Wu},
  {Zhu}, {Jiang}, \& {Zhang}}]{Li2025}
{Li}, Y., {Zhang}, S.~B., {Yang}, Y.~P., {et~al.} 2025, arXiv e-prints,
  arXiv:2503.04727, \dodoi{10.48550/arXiv.2503.04727}

\bibitem[{{Limaye} \& {Spitler}(2024)}]{ATel16620}
{Limaye}, P., \& {Spitler}, L. 2024, The Astronomer's Telegram, 16620, 1

\bibitem[{{Liu} {et~al.}(2024){Liu}, {Chen}, {Li}, {Yuan}, {Yuen}, {Liu},
  {Yan}, {Du}, \& {Zhai}}]{Liu2024}
{Liu}, Y.-L., {Chen}, M.-Z., {Li}, J., {et~al.} 2024, Research in Astronomy and
  Astrophysics, 24, 075008, \dodoi{10.1088/1674-4527/ad52c5}

\bibitem[{{Lorimer} {et~al.}(2007){Lorimer}, {Bailes}, {McLaughlin},
  {Narkevic}, \& {Crawford}}]{Lorimer2007}
{Lorimer}, D.~R., {Bailes}, M., {McLaughlin}, M.~A., {Narkevic}, D.~J., \&
  {Crawford}, F. 2007, Science, 318, 777, \dodoi{10.1126/science.1147532}

\bibitem[{{Lorimer} \& {Kramer}(2004)}]{Handbook}
{Lorimer}, D.~R., \& {Kramer}, M. 2004, {Handbook of Pulsar Astronomy}, Vol.~4

\bibitem[{{Luo} {et~al.}(2020){Luo}, {Wang}, {Men}, {Zhang}, {Jiang}, {Xu},
  {Wang}, {Lee}, {Han}, {Zhang}, {Caballero}, {Chen}, {Chen}, {Gan}, {Guo},
  {Hao}, {Huang}, {Jiang}, {Li}, {Li}, {Li}, {Luo}, {Pan}, {Pei}, {Qian},
  {Sun}, {Wang}, {Wang}, {Wen}, {Xu}, {Xu}, {Yan}, {Yan}, {Yu}, {Yuan},
  {Zhang}, \& {Zhu}}]{Luo2020}
{Luo}, R., {Wang}, B.~J., {Men}, Y.~P., {et~al.} 2020, \nat, 586, 693,
  \dodoi{10.1038/s41586-020-2827-2}

\bibitem[{{Majid} {et~al.}(2020){Majid}, {Pearlman}, {Nimmo}, {Hessels},
  {Prince}, {Naudet}, {Kocz}, \& {Horiuchi}}]{Majid2020}
{Majid}, W.~A., {Pearlman}, A.~B., {Nimmo}, K., {et~al.} 2020, \apjl, 897, L4,
  \dodoi{10.3847/2041-8213/ab9a4a}

\bibitem[{{Majid} {et~al.}(2021){Majid}, {Pearlman}, {Prince}, {Wharton},
  {Naudet}, {Bansal}, {Connor}, {Bhardwaj}, \& {Tendulkar}}]{Majid2021}
{Majid}, W.~A., {Pearlman}, A.~B., {Prince}, T.~A., {et~al.} 2021, \apjl, 919,
  L6, \dodoi{10.3847/2041-8213/ac1921}

\bibitem[{{Margalit} \& {Metzger}(2018)}]{Margalit2018}
{Margalit}, B., \& {Metzger}, B.~D. 2018, \apjl, 868, L4,
  \dodoi{10.3847/2041-8213/aaedad}

\bibitem[{{Margalit} {et~al.}(2020){Margalit}, {Metzger}, \&
  {Sironi}}]{Margalit2020}
{Margalit}, B., {Metzger}, B.~D., \& {Sironi}, L. 2020, \mnras, 494, 4627,
  \dodoi{10.1093/mnras/staa1036}

\bibitem[{{Men} \& {Barr}(2024)}]{transientx}
{Men}, Y., \& {Barr}, E. 2024, \aap, 683, A183,
  \dodoi{10.1051/0004-6361/202348247}

\bibitem[{{Michilli} {et~al.}(2018){Michilli}, {Seymour}, {Hessels}, {Spitler},
  {Gajjar}, {Archibald}, {Bower}, {Chatterjee}, {Cordes}, {Gourdji}, {Heald},
  {Kaspi}, {Law}, {Sobey}, {Adams}, {Bassa}, {Bogdanov}, {Brinkman},
  {Demorest}, {Fernandez}, {Hellbourg}, {Lazio}, {Lynch}, {Maddox}, {Marcote},
  {McLaughlin}, {Paragi}, {Ransom}, {Scholz}, {Siemion}, {Tendulkar}, {van
  Rooy}, {Wharton}, \& {Whitlow}}]{Michilli2018}
{Michilli}, D., {Seymour}, A., {Hessels}, J.~W.~T., {et~al.} 2018, \nat, 553,
  182, \dodoi{10.1038/nature25149}

\bibitem[{{Nimmo} {et~al.}(2023){Nimmo}, {Hessels}, {Snelders}, {Karuppusamy},
  {Hewitt}, {Kirsten}, {Marcote}, {Bach}, {Bansod}, {Barr}, {Behrend},
  {Bezrukovs}, {Buttaccio}, {Feiler}, {Gawro{\'n}ski}, {Lindqvist}, {Orbidans},
  {Puchalska}, {Wang}, {Winchen}, {Wolak}, {Wu}, \& {Yuan}}]{Nimmo2023}
{Nimmo}, K., {Hessels}, J.~W.~T., {Snelders}, M.~P., {et~al.} 2023, \mnras,
  520, 2281, \dodoi{10.1093/mnras/stad269}

\bibitem[{{Niu} {et~al.}(2022{\natexlab{a}}){Niu}, {Aggarwal}, {Li}, {Zhang},
  {Chatterjee}, {Tsai}, {Yu}, {Law}, {Burke-Spolaor}, {Cordes}, {Zhang},
  {Ocker}, {Yao}, {Wang}, {Feng}, {Niino}, {Bochenek}, {Cruces}, {Connor},
  {Jiang}, {Dai}, {Luo}, {Li}, {Miao}, {Niu}, {Anna-Thomas}, {Sydnor}, {Stern},
  {Wang}, {Yuan}, {Yue}, {Zhou}, {Yan}, {Zhu}, \& {Zhang}}]{Niu2022}
{Niu}, C.~H., {Aggarwal}, K., {Li}, D., {et~al.} 2022{\natexlab{a}}, \nat, 606,
  873, \dodoi{10.1038/s41586-022-04755-5}

\bibitem[{{Niu} {et~al.}(2022{\natexlab{b}}){Niu}, {Zhu}, {Zhang}, {Yuan},
  {Zhou}, {Zhang}, {Jiang}, {Han}, {Li}, {Lee}, {Wang}, {Feng}, {Li}, {Luo},
  {Wang}, {Dai}, {Miao}, {Niu}, {Xu}, {Zhang}, {Wang}, {Wang}, \&
  {Xu}}]{Niu2022b}
{Niu}, J.-R., {Zhu}, W.-W., {Zhang}, B., {et~al.} 2022{\natexlab{b}}, Research
  in Astronomy and Astrophysics, 22, 124004, \dodoi{10.1088/1674-4527/ac995d}

\bibitem[{{Niu} {et~al.}(2024){Niu}, {Wang}, {Jiang}, {Qu}, {Zhou}, {Zhu},
  {Lee}, {Han}, {Zhang}, {Li}, {Cao}, {Fang}, {Feng}, {Fu}, {Jiang}, {Jing},
  {Li}, {Li}, {Luo}, {Meng}, {Miao}, {Miao}, {Niu}, {Pan}, {Wang}, {Wang},
  {Wang}, {Wang}, {Wu}, {Wu}, {Xu}, {Xu}, {Xu}, {Xue}, {Yang}, {Yuan}, {Yue},
  {Zhao}, {Zhang}, {Zhang}, {Zhang}, {Zhang}, {Zhang}, \& {Zhu}}]{Niu2024}
{Niu}, J.~R., {Wang}, W.~Y., {Jiang}, J.~C., {et~al.} 2024, \apjl, 972, L20,
  \dodoi{10.3847/2041-8213/ad7023}

\bibitem[{{Oppermann} {et~al.}(2018){Oppermann}, {Yu}, \&
  {Pen}}]{Oppermann2018}
{Oppermann}, N., {Yu}, H.-R., \& {Pen}, U.-L. 2018, \mnras, 475, 5109,
  \dodoi{10.1093/mnras/sty004}

\bibitem[{{Ould-Boukattine} {et~al.}(2024{\natexlab{a}}){Ould-Boukattine},
  {Hessels}, {Kirsten}, {Hewitt}, {Snelders}, {Blaauw}, {Sluman}, {Mulder},
  {Herrmann}, {Gawronski}, {Puchalska}, \& {Gopinath}}]{ATel16432}
{Ould-Boukattine}, O.~S., {Hessels}, J.~W.~T., {Kirsten}, F., {et~al.}
  2024{\natexlab{a}}, The Astronomer's Telegram, 16432, 1

\bibitem[{{Ould-Boukattine} {et~al.}(2024{\natexlab{b}}){Ould-Boukattine},
  {Dijkema}, {Gawronski}, {Herrmann}, {Hessels}, {Kirsten}, {Snelders}, {Beer},
  {Bijlsma}, {Blaauw}, {Boons}, {Boven}, {Buchsteiner}, {Engelskirchen},
  {Fischer}, {Hewitt}, {Loge}, {Marcote}, {van der Meer}, {Mulder}, {Munk},
  {Nitsche}, {Ovinge}, {Puchalska}, {Sanders}, {Schmitz}, {Sluman}, {Telkamp},
  {Wolf}, \& {Yang}}]{ATel16565}
{Ould-Boukattine}, O.~S., {Dijkema}, T.~J., {Gawronski}, M., {et~al.}
  2024{\natexlab{b}}, The Astronomer's Telegram, 16565, 1

\bibitem[{{Ould-Boukattine} {et~al.}(2025){Ould-Boukattine}, {Blaauw},
  {Gawronski}, {Herrmann}, {Hessels}, {Hewitt}, {Huang}, {Kirsten}, {Moroianu},
  {Mulder}, {Pleunis}, {Puchalska}, {Snelders}, \& {Sluman}}]{Atel16967}
{Ould-Boukattine}, O.~S., {Blaauw}, R., {Gawronski}, M.~P., {et~al.} 2025, The
  Astronomer's Telegram, 16967, 1

\bibitem[{{Panda} {et~al.}(2024){Panda}, {Bhattacharyya}, {Dudeja}, {Kudale},
  \& {Roy}}]{ATel16494}
{Panda}, U., {Bhattacharyya}, S., {Dudeja}, C., {Kudale}, S., \& {Roy}, J.
  2024, The Astronomer's Telegram, 16494, 1

\bibitem[{{Panda} {et~al.}(2025){Panda}, {Roy}, {Bhattacharyya}, {Dudeja}, \&
  {Kudale}}]{Panda2024}
{Panda}, U., {Roy}, J., {Bhattacharyya}, S., {Dudeja}, C., \& {Kudale}, S.
  2025, \apj, 989, 15, \dodoi{10.3847/1538-4357/adeb74}

\bibitem[{{Pastor-Marazuela} {et~al.}(2021){Pastor-Marazuela}, {Connor}, {van
  Leeuwen}, {Maan}, {ter Veen}, {Bilous}, {Oostrum}, {Petroff}, {Straal},
  {Vohl}, {Attema}, {Boersma}, {Kooistra}, {van der Schuur}, {Sclocco},
  {Smits}, {Adams}, {Adebahr}, {de Blok}, {Coolen}, {Damstra}, {D{\'e}nes},
  {Hess}, {van der Hulst}, {Hut}, {Ivashina}, {Kutkin}, {Loose}, {Lucero},
  {Mika}, {Moss}, {Mulder}, {Norden}, {Oosterloo}, {Orr{\'u}}, {Ruiter}, \&
  {Wijnholds}}]{Pastor2021}
{Pastor-Marazuela}, I., {Connor}, L., {van Leeuwen}, J., {et~al.} 2021, \nat,
  596, 505, \dodoi{10.1038/s41586-021-03724-8}

\bibitem[{{Pearlman} {et~al.}(2020){Pearlman}, {Majid}, {Prince}, {Nimmo},
  {Hessels}, {Naudet}, \& {Kocz}}]{Pearlman2020}
{Pearlman}, A.~B., {Majid}, W.~A., {Prince}, T.~A., {et~al.} 2020, \apjl, 905,
  L27, \dodoi{10.3847/2041-8213/abca31}

\bibitem[{{Pelliciari} {et~al.}(2024{\natexlab{a}}){Pelliciari}, {Geminardi},
  {Bernardi}, {Pilia}, {Esposito}, \& {Naldi}}]{ATel16434}
{Pelliciari}, D., {Geminardi}, A., {Bernardi}, G., {et~al.} 2024{\natexlab{a}},
  The Astronomer's Telegram, 16434, 1

\bibitem[{{Pelliciari} {et~al.}(2024{\natexlab{b}}){Pelliciari}, {Geminardi},
  {Bernardi}, {Pilia}, {Esposito}, \& {Naldi}}]{ATel16547}
---. 2024{\natexlab{b}}, The Astronomer's Telegram, 16547, 1

\bibitem[{{Pelliciari} {et~al.}(2024{\natexlab{c}}){Pelliciari}, {Bernardi},
  {Pilia}, {Naldi}, {Maccaferri}, {Verrecchia}, {Casentini}, {Perri},
  {Kirsten}, {Bianchi}, {Bortolotti}, {Bruno}, {Dallacasa}, {Esposito},
  {Geminardi}, {Giarratana}, {Giroletti}, {Lulli}, {Maccaferri}, {Magro},
  {Mattana}, {Perini}, {Pupillo}, {Roma}, {Schiaffino}, {Setti}, {Tavani},
  {Trudu}, \& {Zanichelli}}]{Pelliciari2024}
{Pelliciari}, D., {Bernardi}, G., {Pilia}, M., {et~al.} 2024{\natexlab{c}},
  \aap, 690, A219, \dodoi{10.1051/0004-6361/202450271}

\bibitem[{Perera {et~al.}(2022)Perera, Perillat, Fernandez, Manoharan, Roshi,
  Salter, Smith, Vaddi, \& McGilvray}]{ATel15734}
Perera, B., Perillat, P., Fernandez, F., {et~al.} 2022, The Astronomer's
  Telegram, 15734, 1

\bibitem[{{Petroff} {et~al.}(2019){Petroff}, {Hessels}, \&
  {Lorimer}}]{Petroff2019}
{Petroff}, E., {Hessels}, J.~W.~T., \& {Lorimer}, D.~R. 2019, \aapr, 27, 4,
  \dodoi{10.1007/s00159-019-0116-6}

\bibitem[{{Petroff} {et~al.}(2022){Petroff}, {Hessels}, \&
  {Lorimer}}]{Petroff2022}
---. 2022, \aapr, 30, 2, \dodoi{10.1007/s00159-022-00139-w}

\bibitem[{{Planck Collaboration} {et~al.}(2016){Planck Collaboration}, {Ade},
  {Aghanim}, {Arnaud}, {Ashdown}, {Aumont}, {Baccigalupi}, {Banday},
  {Barreiro}, {Bartlett}, {Bartolo}, {Battaner}, {Battye}, {Benabed},
  {Beno{\^\i}t}, {Benoit-L{\'e}vy}, {Bernard}, {Bersanelli}, {Bielewicz},
  {Bock}, {Bonaldi}, {Bonavera}, {Bond}, {Borrill}, {Bouchet}, {Boulanger},
  {Bucher}, {Burigana}, {Butler}, {Calabrese}, {Cardoso}, {Catalano},
  {Challinor}, {Chamballu}, {Chary}, {Chiang}, {Chluba}, {Christensen},
  {Church}, {Clements}, {Colombi}, {Colombo}, {Combet}, {Coulais}, {Crill},
  {Curto}, {Cuttaia}, {Danese}, {Davies}, {Davis}, {de Bernardis}, {de Rosa},
  {de Zotti}, {Delabrouille}, {D{\'e}sert}, {Di Valentino}, {Dickinson},
  {Diego}, {Dolag}, {Dole}, {Donzelli}, {Dor{\'e}}, {Douspis}, {Ducout},
  {Dunkley}, {Dupac}, {Efstathiou}, {Elsner}, {En{\ss}lin}, {Eriksen},
  {Farhang}, {Fergusson}, {Finelli}, {Forni}, {Frailis}, {Fraisse},
  {Franceschi}, {Frejsel}, {Galeotta}, {Galli}, {Ganga}, {Gauthier}, {Gerbino},
  {Ghosh}, {Giard}, {Giraud-H{\'e}raud}, {Giusarma}, {Gjerl{\o}w},
  {Gonz{\'a}lez-Nuevo}, {G{\'o}rski}, {Gratton}, {Gregorio}, {Gruppuso},
  {Gudmundsson}, {Hamann}, {Hansen}, {Hanson}, {Harrison}, {Helou},
  {Henrot-Versill{\'e}}, {Hern{\'a}ndez-Monteagudo}, {Herranz}, {Hildebrandt},
  {Hivon}, {Hobson}, {Holmes}, {Hornstrup}, {Hovest}, {Huang}, {Huffenberger},
  {Hurier}, {Jaffe}, {Jaffe}, {Jones}, {Juvela}, {Keih{\"a}nen}, {Keskitalo},
  {Kisner}, {Kneissl}, {Knoche}, {Knox}, {Kunz}, {Kurki-Suonio}, {Lagache},
  {L{\"a}hteenm{\"a}ki}, {Lamarre}, {Lasenby}, {Lattanzi}, {Lawrence}, {Leahy},
  {Leonardi}, {Lesgourgues}, {Levrier}, {Lewis}, {Liguori}, {Lilje},
  {Linden-V{\o}rnle}, {L{\'o}pez-Caniego}, {Lubin}, {Mac{\'\i}as-P{\'e}rez},
  {Maggio}, {Maino}, {Mandolesi}, {Mangilli}, {Marchini}, {Maris}, {Martin},
  {Martinelli}, {Mart{\'\i}nez-Gonz{\'a}lez}, {Masi}, {Matarrese}, {McGehee},
  {Meinhold}, {Melchiorri}, {Melin}, {Mendes}, {Mennella}, {Migliaccio},
  {Millea}, {Mitra}, {Miville-Desch{\^e}nes}, {Moneti}, {Montier}, {Morgante},
  {Mortlock}, {Moss}, {Munshi}, {Murphy}, {Naselsky}, {Nati}, {Natoli},
  {Netterfield}, {N{\o}rgaard-Nielsen}, {Noviello}, {Novikov}, {Novikov},
  {Oxborrow}, {Paci}, {Pagano}, {Pajot}, {Paladini}, {Paoletti}, {Partridge},
  {Pasian}, {Patanchon}, {Pearson}, {Perdereau}, {Perotto}, {Perrotta},
  {Pettorino}, {Piacentini}, {Piat}, {Pierpaoli}, {Pietrobon}, {Plaszczynski},
  {Pointecouteau}, {Polenta}, {Popa}, {Pratt}, {Pr{\'e}zeau}, {Prunet},
  {Puget}, {Rachen}, {Reach}, {Rebolo}, {Reinecke}, {Remazeilles}, {Renault},
  {Renzi}, {Ristorcelli}, {Rocha}, {Rosset}, {Rossetti}, {Roudier},
  {Rouill{\'e} d'Orfeuil}, {Rowan-Robinson}, {Rubi{\~n}o-Mart{\'\i}n},
  {Rusholme}, {Said}, {Salvatelli}, {Salvati}, {Sandri}, {Santos},
  {Savelainen}, {Savini}, {Scott}, {Seiffert}, {Serra}, {Shellard}, {Spencer},
  {Spinelli}, {Stolyarov}, {Stompor}, {Sudiwala}, {Sunyaev}, {Sutton},
  {Suur-Uski}, {Sygnet}, {Tauber}, {Terenzi}, {Toffolatti}, {Tomasi},
  {Tristram}, {Trombetti}, {Tucci}, {Tuovinen}, {T{\"u}rler}, {Umana},
  {Valenziano}, {Valiviita}, {Van Tent}, {Vielva}, {Villa}, {Wade}, {Wandelt},
  {Wehus}, {White}, {White}, {Wilkinson}, {Yvon}, {Zacchei}, \&
  {Zonca}}]{Planck2016}
{Planck Collaboration}, {Ade}, P.~A.~R., {Aghanim}, N., {et~al.} 2016, \aap,
  594, A13, \dodoi{10.1051/0004-6361/201525830}

\bibitem[{{Pleunis} {et~al.}(2021{\natexlab{a}}){Pleunis}, {Michilli}, {Bassa},
  {Hessels}, {Naidu}, {Andersen}, {Chawla}, {Fonseca}, {Gopinath}, {Kaspi},
  {Kondratiev}, {Li}, {Bhardwaj}, {Boyle}, {Brar}, {Cassanelli}, {Gupta},
  {Josephy}, {Karuppusamy}, {Keimpema}, {Kirsten}, {Leung}, {Marcote}, {Masui},
  {Mckinven}, {Meyers}, {Ng}, {Nimmo}, {Paragi}, {Rahman}, {Scholz}, {Shin},
  {Smith}, {Stairs}, \& {Tendulkar}}]{Pleunis2021b}
{Pleunis}, Z., {Michilli}, D., {Bassa}, C.~G., {et~al.} 2021{\natexlab{a}},
  \apjl, 911, L3, \dodoi{10.3847/2041-8213/abec72}

\bibitem[{{Pleunis} {et~al.}(2021{\natexlab{b}}){Pleunis}, {Good}, {Kaspi},
  {Mckinven}, {Ransom}, {Scholz}, {Bandura}, {Bhardwaj}, {Boyle}, {Brar},
  {Cassanelli}, {Chawla}, {(Adam) Dong}, {Fonseca}, {Gaensler}, {Josephy},
  {Kaczmarek}, {Leung}, {Lin}, {Masui}, {Mena-Parra}, {Michilli}, {Ng},
  {Patel}, {Rafiei-Ravandi}, {Rahman}, {Sanghavi}, {Shin}, {Smith}, {Stairs},
  \& {Tendulkar}}]{Pleunis2021a}
{Pleunis}, Z., {Good}, D.~C., {Kaspi}, V.~M., {et~al.} 2021{\natexlab{b}},
  \apj, 923, 1, \dodoi{10.3847/1538-4357/ac33ac}

\bibitem[{Rajwade {et~al.}(2022)Rajwade, Wharton, Majid, Mickaliger, Stappers,
  Breton, Lyne, Keith, Naudet, Pearlman, {et~al.}}]{ATel15791}
Rajwade, K., Wharton, R., Majid, W., {et~al.} 2022, The Astronomer's Telegram,
  15791, 1

\bibitem[{{Scholz} {et~al.}(2016){Scholz}, {Spitler}, {Hessels}, {Chatterjee},
  {Cordes}, {Kaspi}, {Wharton}, {Bassa}, {Bogdanov}, {Camilo}, {Crawford},
  {Deneva}, {van Leeuwen}, {Lynch}, {Madsen}, {McLaughlin}, {Mickaliger},
  {Parent}, {Patel}, {Ransom}, {Seymour}, {Stairs}, {Stappers}, \&
  {Tendulkar}}]{Scholz2016}
{Scholz}, P., {Spitler}, L.~G., {Hessels}, J.~W.~T., {et~al.} 2016, \apj, 833,
  177, \dodoi{10.3847/1538-4357/833/2/177}

\bibitem[{{Shin} \& {CHIME/FRB Collaboration}(2024)}]{ATel16420}
{Shin}, K., \& {CHIME/FRB Collaboration}. 2024, The Astronomer's Telegram,
  16420, 1

\bibitem[{{Shin} {et~al.}(2025){Shin}, {Curtin}, {Fine}, {Pandhi}, {Andrew},
  {Bhardwaj}, {Chatterjee}, {Cook}, {Fonseca}, {Gaensler}, {Hessels}, {Jain},
  {Kaspi}, {Kharel}, {Lanman}, {Lazda}, {Leung}, {Main}, {Masui}, {Michilli},
  {Ng}, {Nimmo}, {Pearlman}, {Pen}, {Pleunis}, {Rafiei-Ravandi}, {Sammons},
  {Sand}, {Scholz}, {Smith}, \& {Stairs}}]{Shin2025}
{Shin}, K., {Curtin}, A., {Fine}, M., {et~al.} 2025, arXiv e-prints,
  arXiv:2505.13297, \dodoi{10.48550/arXiv.2505.13297}

\bibitem[{{Snelders} {et~al.}(2024){Snelders}, {Bhandari}, {Kirsten},
  {Hessels}, {Marcote}, {Hewitt}, {Gawronski}, {Puchalska}, {Ould-Boukattine},
  {Gopinath}, {Nimmo}, {Karuppusamy}, {Herrmann}, {Yang}, {Blaauw},
  {Buttaccio}, {Maccaferri}, {Bach}, {Feiler}, {Bray}, {Williams}, {Wrigley},
  {Keimpema}, {Paragi}, {Burgay}, {Corongiu}, {Giroletti}, {Kramer}, {Pilia},
  {Spitler}, {Surcis}, {Trudu}, {Yuan}, {Wang}, \& {Bezrukovs}}]{ATel16542}
{Snelders}, M.~P., {Bhandari}, S., {Kirsten}, F., {et~al.} 2024, The
  Astronomer's Telegram, 16542, 1

\bibitem[{{Spitler} {et~al.}(2016){Spitler}, {Scholz}, {Hessels}, {Bogdanov},
  {Brazier}, {Camilo}, {Chatterjee}, {Cordes}, {Crawford}, {Deneva}, {Ferdman},
  {Freire}, {Kaspi}, {Lazarus}, {Lynch}, {Madsen}, {McLaughlin}, {Patel},
  {Ransom}, {Seymour}, {Stairs}, {Stappers}, {van Leeuwen}, \&
  {Zhu}}]{Spitler2016}
{Spitler}, L.~G., {Scholz}, P., {Hessels}, J.~W.~T., {et~al.} 2016, \nat, 531,
  202, \dodoi{10.1038/nature17168}

\bibitem[{{Spitler} {et~al.}(2018){Spitler}, {Herrmann}, {Bower}, {Chatterjee},
  {Cordes}, {Hessels}, {Kramer}, {Michilli}, {Scholz}, {Seymour}, \&
  {Siemion}}]{Spitler2018}
{Spitler}, L.~G., {Herrmann}, W., {Bower}, G.~C., {et~al.} 2018, \apj, 863,
  150, \dodoi{10.3847/1538-4357/aad332}

\bibitem[{{Sridhar} {et~al.}(2021){Sridhar}, {Metzger}, {Beniamini},
  {Margalit}, {Renzo}, {Sironi}, \& {Kovlakas}}]{Sridhar2021}
{Sridhar}, N., {Metzger}, B.~D., {Beniamini}, P., {et~al.} 2021, \apj, 917, 13,
  \dodoi{10.3847/1538-4357/ac0140}

\bibitem[{{Tian} {et~al.}(2024{\natexlab{a}}){Tian}, {Pastor-Marazuela},
  {Stappers}, {Rajwade}, {Caleb}, {Bezuidenhout}, {Barr}, \&
  {Kramer}}]{ATel16446}
{Tian}, J., {Pastor-Marazuela}, I., {Stappers}, B., {et~al.}
  2024{\natexlab{a}}, The Astronomer's Telegram, 16446, 1

\bibitem[{{Tian} {et~al.}(2024{\natexlab{b}}){Tian}, {Rajwade},
  {Pastor-Marazuela}, {Stappers}, {Bezuidenhout}, {Caleb}, {Jankowski}, {Barr},
  \& {Kramer}}]{Tian2024}
{Tian}, J., {Rajwade}, K.~M., {Pastor-Marazuela}, I., {et~al.}
  2024{\natexlab{b}}, \mnras, 533, 3174, \dodoi{10.1093/mnras/stae2013}

\bibitem[{{Tian} {et~al.}(2025){Tian}, {Pastor-Marazuela}, {Rajwade},
  {Stappers}, {Shaji}, {Hanmer}, {Caleb}, {Bezuidenhout}, {Jankowski},
  {Breton}, {Barr}, {Kramer}, {Groot}, {Bloemen}, {Vreeswijk}, {Pieterse},
  {Woudt}, {Fender}, {Wijnands}, \& {Buckley}}]{Tian2025}
{Tian}, J., {Pastor-Marazuela}, I., {Rajwade}, K.~M., {et~al.} 2025, \mnras,
  540, 1685, \dodoi{10.1093/mnras/staf793}

\bibitem[{{Torne} {et~al.}(2015){Torne}, {Eatough}, {Karuppusamy}, {Kramer},
  {Paubert}, {Klein}, {Desvignes}, {Champion}, {Wiesemeyer}, {Kramer},
  {Spitler}, {Thum}, {Gusten}, {Schuster}, \& {Cognard}}]{Torne2015}
{Torne}, P., {Eatough}, R.~P., {Karuppusamy}, R., {et~al.} 2015, \mnras, 451,
  L50, \dodoi{10.1093/mnrasl/slv063}

\bibitem[{{Torne} {et~al.}(2017){Torne}, {Desvignes}, {Eatough}, {Karuppusamy},
  {Paubert}, {Kramer}, {Cognard}, {Champion}, \& {Spitler}}]{Torne2017}
{Torne}, P., {Desvignes}, G., {Eatough}, R.~P., {et~al.} 2017, \mnras, 465,
  242, \dodoi{10.1093/mnras/stw2757}

\bibitem[{{Uttarkar} {et~al.}(2024){Uttarkar}, {Kumar}, {Lower}, \&
  {Shannon}}]{ATel16430}
{Uttarkar}, P.~A., {Kumar}, P., {Lower}, M.~E., \& {Shannon}, R.~M. 2024, The
  Astronomer's Telegram, 16430, 1

\bibitem[{{Xie} {et~al.}(2024){Xie}, {Feng}, {Li}, {Zhang}, {Zhou}, {Qu},
  {Cui}, {Fang}, {Xu}, {Miao}, {Yuan}, {Tsai}, {Wang}, {Niu}, {Chen}, {Xue}, \&
  {Zhang}}]{Xie2024}
{Xie}, J.-T., {Feng}, Y., {Li}, D., {et~al.} 2024, arXiv e-prints,
  arXiv:2410.10172, \dodoi{10.48550/arXiv.2410.10172}

\bibitem[{{Xu} {et~al.}(2022){Xu}, {Niu}, {Chen}, {Lee}, {Zhu}, {Dong},
  {Zhang}, {Jiang}, {Wang}, {Xu}, {Zhang}, {Fu}, {Filippenko}, {Peng}, {Zhou},
  {Zhang}, {Wang}, {Feng}, {Li}, {Brink}, {Li}, {Lu}, {Yang}, {Caballero},
  {Cai}, {Chen}, {Dai}, {Djorgovski}, {Esamdin}, {Gan}, {Guhathakurta}, {Han},
  {Hao}, {Huang}, {Jiang}, {Li}, {Li}, {Li}, {Li}, {Li}, {Liu}, {Luo}, {Men},
  {Niu}, {Peng}, {Qian}, {Song}, {Stern}, {Stockton}, {Sun}, {Wang}, {Wang},
  {Wang}, {Wang}, {Wu}, {Xiao}, {Xiong}, {Xu}, {Xu}, {Yang}, {Yang}, {Yao},
  {Yi}, {Yue}, {Yu}, {Yu}, {Yuan}, {Zhang}, {Zhang}, {Zhang}, {Zhao}, {Zheng},
  {Zhu}, \& {Zou}}]{Xu2022}
{Xu}, H., {Niu}, J.~R., {Chen}, P., {et~al.} 2022, \nat, 609, 685,
  \dodoi{10.1038/s41586-022-05071-8}

\bibitem[{{Yan} {et~al.}(2015){Yan}, {Shen}, {Wu}, {Manchester}, {Weltevrede},
  {Wu}, {Zhao}, {Yuan}, {Lee}, {Fan}, {Hong}, {Jiang}, {Li}, {Liang}, {Ling},
  {Liu}, {Qian}, {Zhang}, {Zhong}, \& {Ye}}]{Yan2015}
{Yan}, Z., {Shen}, Z.-Q., {Wu}, X.-J., {et~al.} 2015, \apj, 814, 5,
  \dodoi{10.1088/0004-637X/814/1/5}

\bibitem[{{Yan} {et~al.}(2018){Yan}, {Shen}, {Manchester}, {Ng}, {Weltevrede},
  {Wang}, {Wu}, {Yuan}, {Wu}, {Zhao}, {Liu}, {Zhao}, \& {Liu}}]{Yan2018}
{Yan}, Z., {Shen}, Z.-Q., {Manchester}, R.~N., {et~al.} 2018, \apj, 856, 55,
  \dodoi{10.3847/1538-4357/aaae64}

\bibitem[{{Yan} {et~al.}(2024){Yan}, {Shen}, {Wu}, {Zhao}, {Liu}, {Huang},
  {Wang}, {Wang}, {Liu}, {Li}, {Wang}, {Zhong}, {Jiang}, \& {Xia}}]{DIBAS}
{Yan}, Z., {Shen}, Z., {Wu}, Y., {et~al.} 2024, Universe, 10, 195,
  \dodoi{10.3390/universe10050195}

\bibitem[{{Zhang} {et~al.}(2025{\natexlab{a}}){Zhang}, {Wang}, {Wang}, {Gao},
  {Wu}, {Li}, {Zhu}, {Zhang}, {Lee}, {Han}, {Tsai}, {Wang}, {Huang}, {Zou},
  {Zhou}, {Lu}, {Xie}, {Fang}, {Cao}, {Miao}, {Zhu}, {Chen}, {Cheng}, {Ke},
  {Zhang}, {Zhang}, {Cao}, {Tian}, {Wu}, {Zhang}, {Niu}, {Zhou}, {Xu}, {Wang},
  {Chen}, {Chen}, {Cui}, {Feng}, {G{\"u}gercino{\u{g}}lu}, {Huang}, {Li}, {Li},
  {Li}, {Li}, {Lin}, {Liu}, {Luo}, {Luo}, {Niu}, {Qu}, {Qu}, {Menberu Tedila},
  {Wang}, {Wang}, {Wang}, {Wang}, {Weng}, {Wu}, {Xu}, {Yang}, {Yang}, {Yew},
  {Yu}, {Zhang}, \& {Zhao}}]{zhang2025b}
{Zhang}, J.-S., {Wang}, T.-C., {Wang}, P., {et~al.} 2025{\natexlab{a}}, arXiv
  e-prints, arXiv:2507.14707.
\newblock \doarXiv{2507.14707}

\bibitem[{{Zhang} {et~al.}(2025{\natexlab{b}}){Zhang}, {Yu}, {Yan}, {Xing}, \&
  {Zhang}}]{Zhang2025}
{Zhang}, X., {Yu}, W., {Yan}, Z., {Xing}, Y., \& {Zhang}, B.
  2025{\natexlab{b}}, arXiv e-prints, arXiv:2501.14247,
  \dodoi{10.48550/arXiv.2501.14247}

\bibitem[{{Zhang} {et~al.}(2018){Zhang}, {Gajjar}, {Foster}, {Siemion},
  {Cordes}, {Law}, \& {Wang}}]{Zhang2018}
{Zhang}, Y.~G., {Gajjar}, V., {Foster}, G., {et~al.} 2018, \apj, 866, 149,
  \dodoi{10.3847/1538-4357/aadf31}

\bibitem[{{Zhang} {et~al.}(2022){Zhang}, {Wang}, {Feng}, {Zhang}, {Li}, {Tsai},
  {Niu}, {Luo}, {Yao}, {Zhu}, {Han}, {Lee}, {Zhou}, {Niu}, {Jiang}, {Wang},
  {Zhang}, {Xu}, {Wang}, \& {Xu}}]{Zhang2022}
{Zhang}, Y.-K., {Wang}, P., {Feng}, Y., {et~al.} 2022, Research in Astronomy
  and Astrophysics, 22, 124002, \dodoi{10.1088/1674-4527/ac98f7}

\bibitem[{{Zhang} {et~al.}(2023){Zhang}, {Li}, {Zhang}, {Cao}, {Feng}, {Wang},
  {Qu}, {Niu}, {Zhu}, {Han}, {Jiang}, {Lee}, {Li}, {Luo}, {Niu}, {Tsai},
  {Wang}, {Wang}, {Wu}, {Xu}, {Yang}, {Zhang}, {Zhou}, \& {Zhu}}]{Zhang2023}
{Zhang}, Y.-K., {Li}, D., {Zhang}, B., {et~al.} 2023, \apj, 955, 142,
  \dodoi{10.3847/1538-4357/aced0b}

\bibitem[{{Zhao} {et~al.}(2019){Zhao}, {Yan}, {Wu}, {Shen}, {Manchester},
  {Liu}, {Qiao}, {Xu}, \& {Lee}}]{Zhao2019}
{Zhao}, R.-S., {Yan}, Z., {Wu}, X.-J., {et~al.} 2019, \apj, 874, 64,
  \dodoi{10.3847/1538-4357/ab05de}

\end{thebibliography}
\bibliographystyle{aasjournal}



\restartappendixnumbering
\clearpage
\appendix
\label{sec:A}

Here, we present the dynamic spectra of each burst (in total 155) detected at 2.25~GHz from FRB~20240114A in Fig.~\ref{fig:figureA1}. 

\renewcommand{\thefigure}{A\arabic{figure}}
\begin{figure}[htb!]
	\centering
	\includegraphics[width=\linewidth]{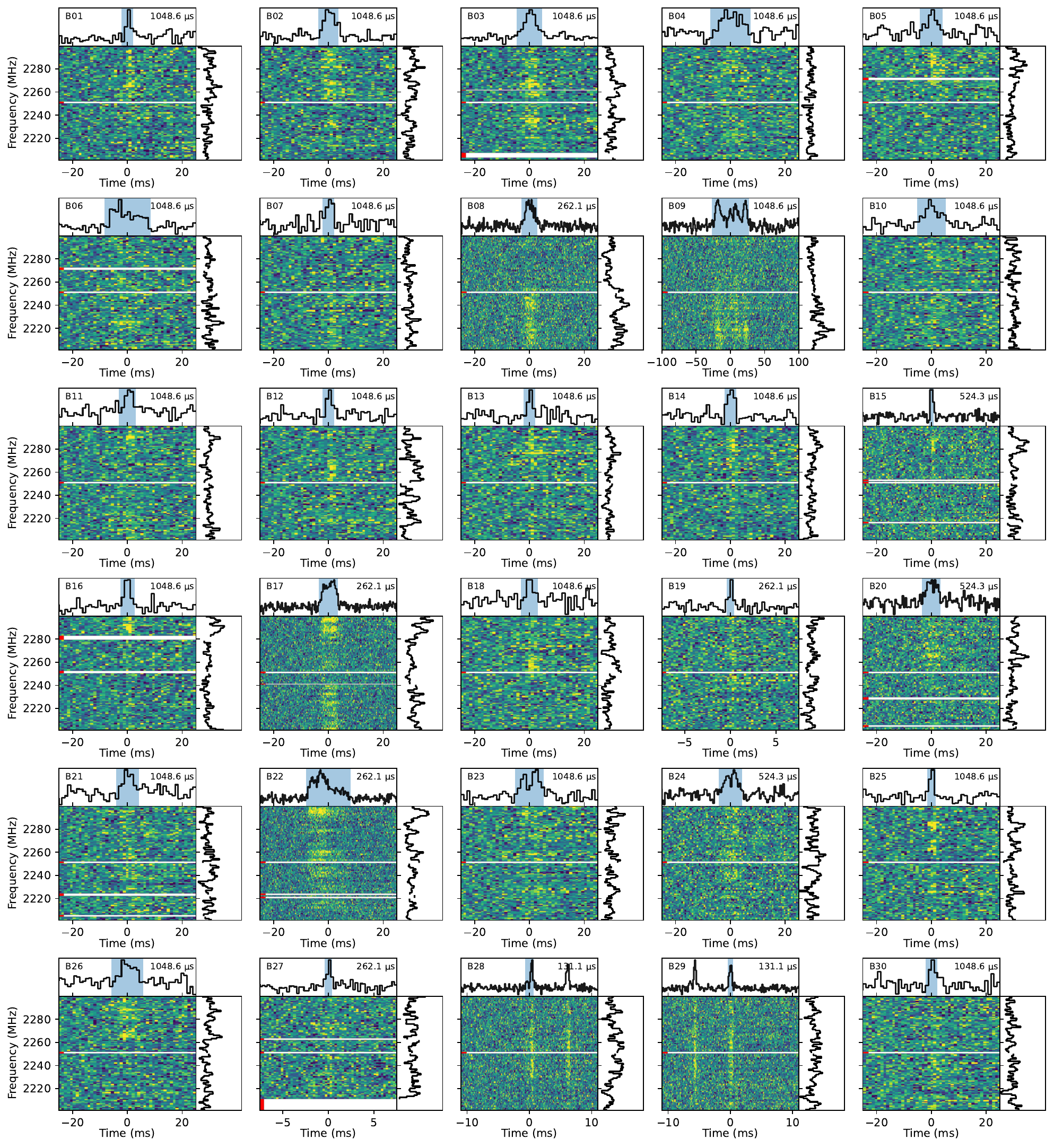}
	\caption{Dynamic spectra of 155 bursts detected at 2.25~GHz from FRB~20240114A.}
	\label{fig:figureA1}
\end{figure}
\addtocounter{figure}{-1}
\begin{figure*}
	\centering
	\includegraphics[width=\linewidth]{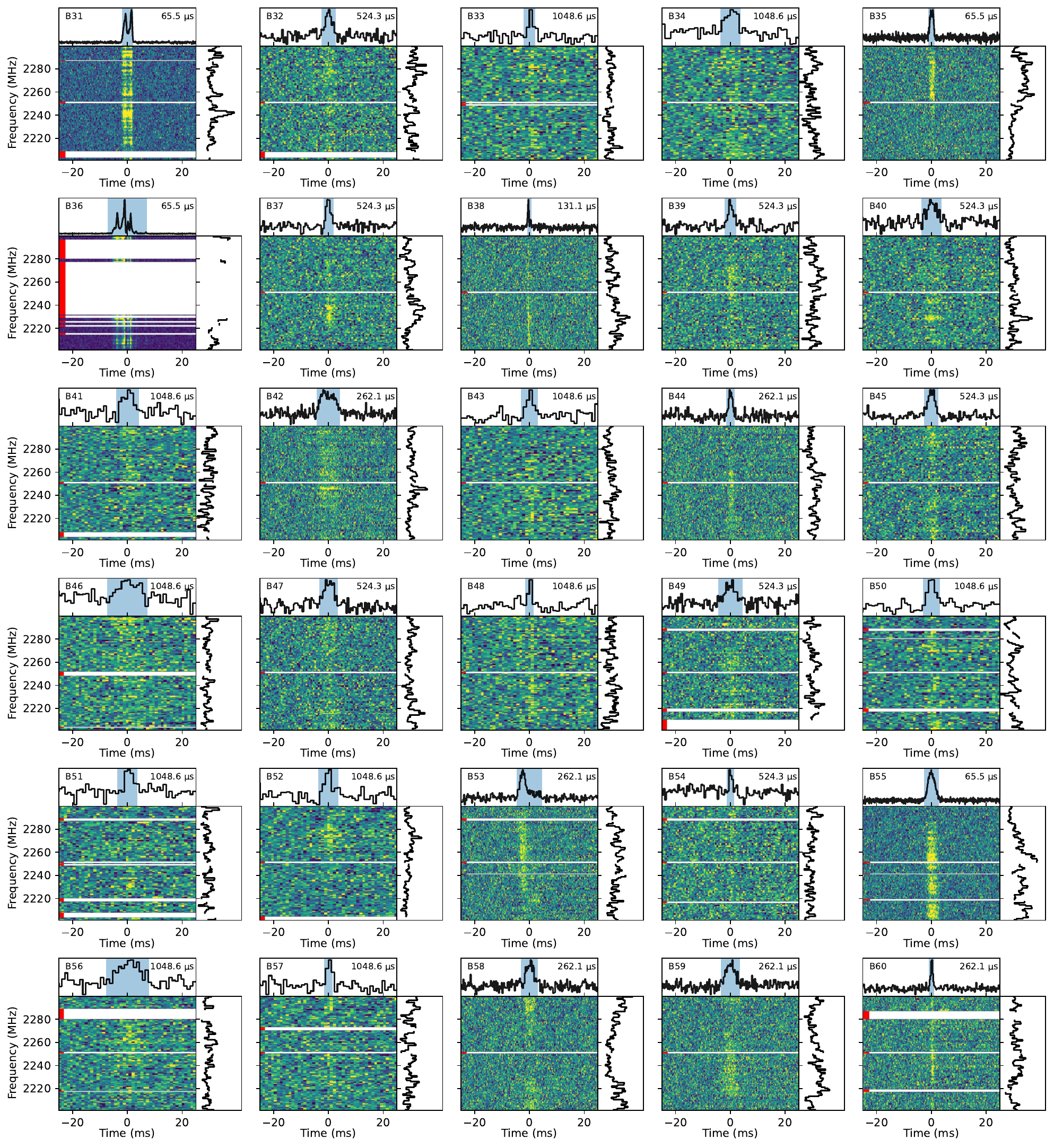}
	\caption{continued.}
\end{figure*}
\addtocounter{figure}{-1}
\begin{figure*}
	\centering
	\includegraphics[width=\linewidth]{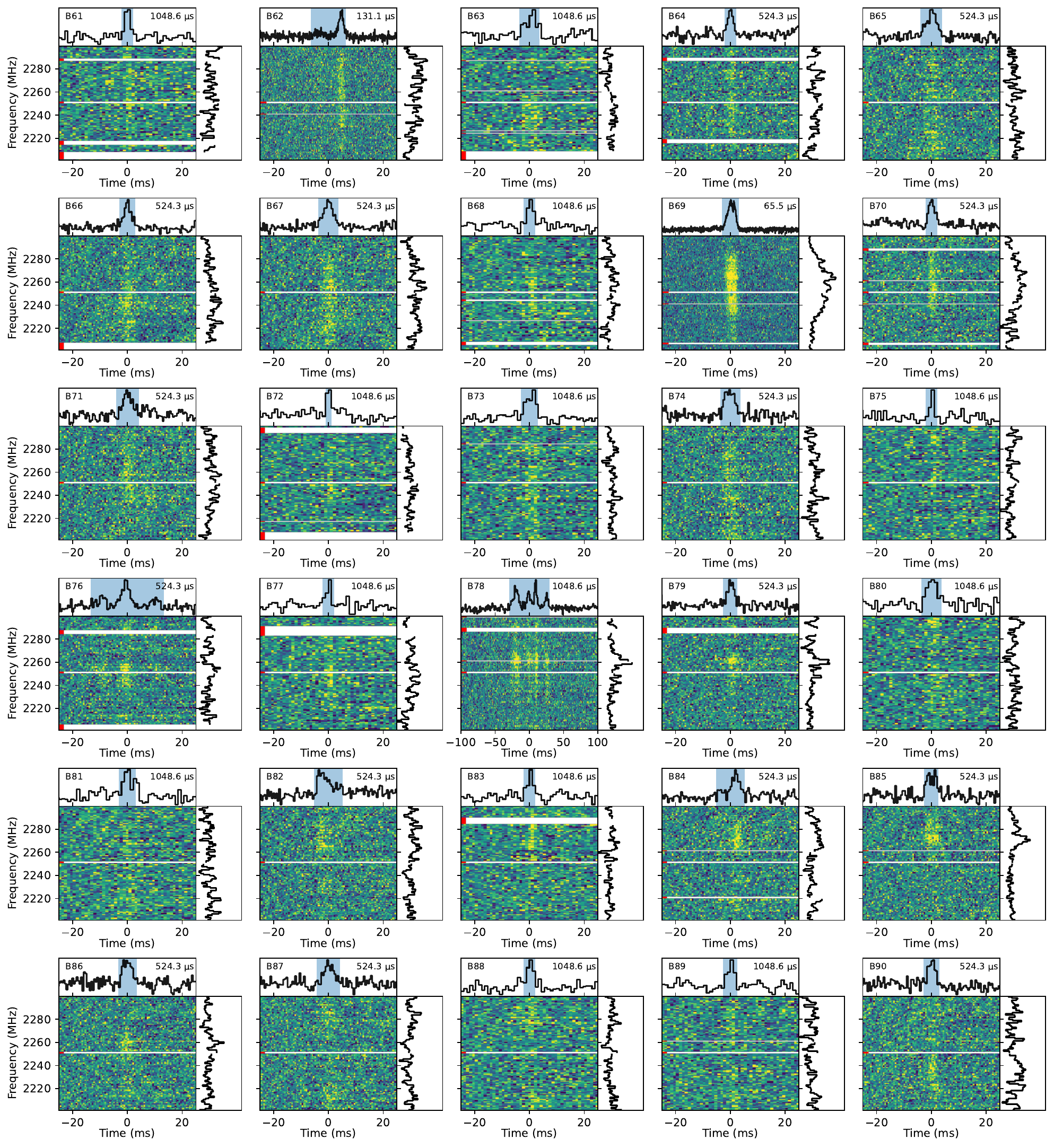}
	\caption{continued.}
\end{figure*}
\addtocounter{figure}{-1}
\begin{figure*}
	\centering
	\includegraphics[width=\linewidth]{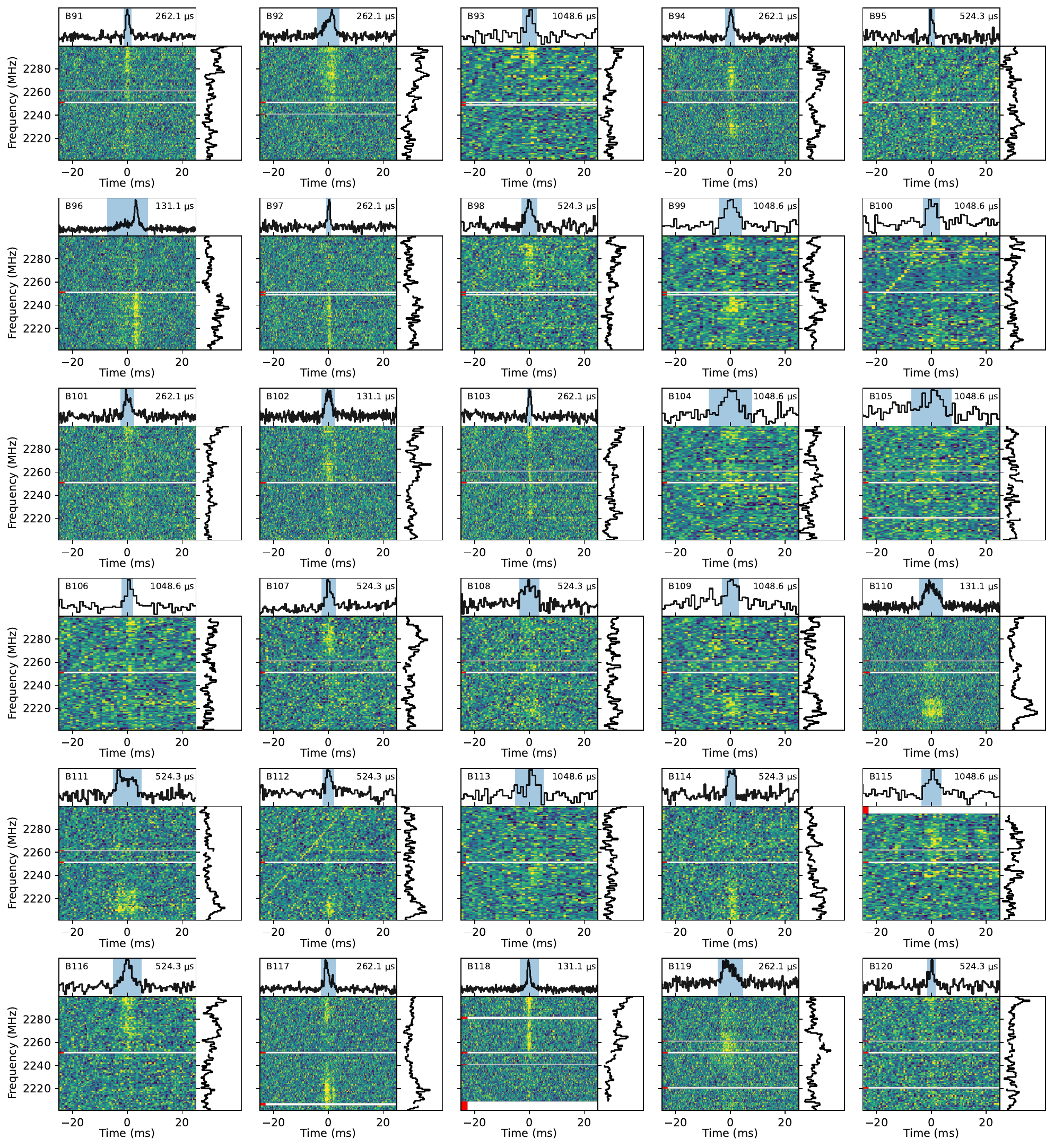}
	\caption{continued.}
\end{figure*}
\addtocounter{figure}{-1}
\begin{figure*}
	\centering
	\includegraphics[width=\linewidth]{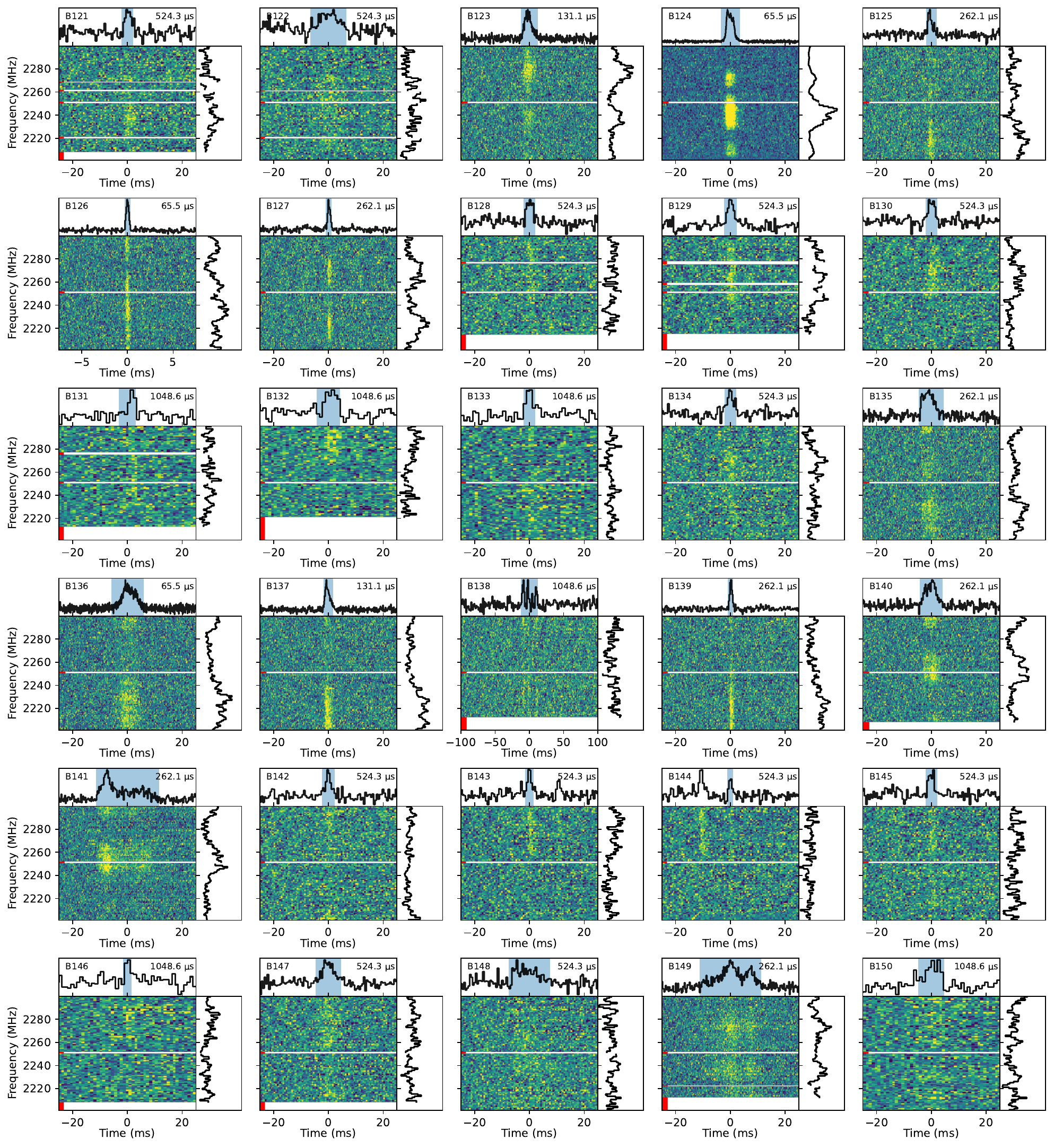}
	\caption{continued.}
\end{figure*}
\addtocounter{figure}{-1}
\clearpage
\begin{figure*}
	\centering
	\includegraphics[width=\linewidth]{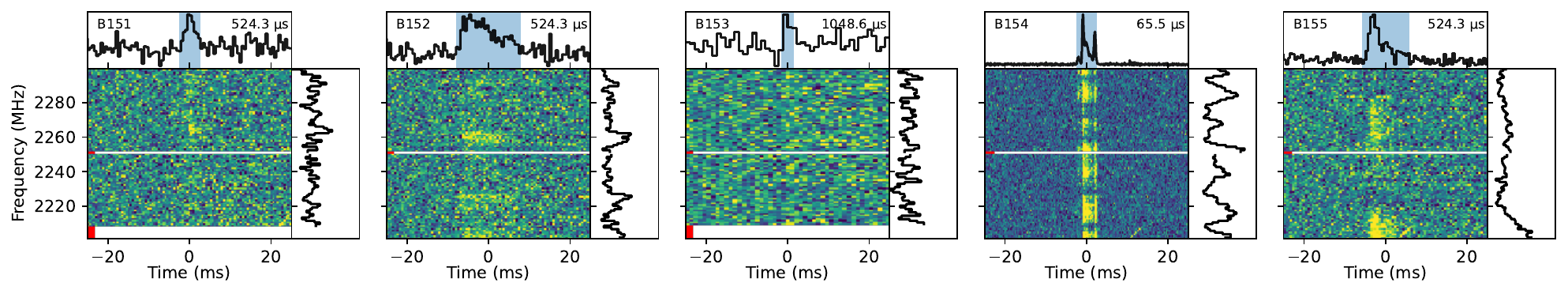}
	\caption{continued.}
\end{figure*}


\renewcommand{\thetable}{A\arabic{table}}
\startlongtable
\begin{deluxetable*}{lccccc} 
\tablecaption{Basic observation information of FRB~20240114A with the TMRT.}
\label{table: B1}
\tablehead{
    \multicolumn{1}{l}{$\rm{ID}_{\rm{obs}}$} & \colhead{Start time\tablenotemark{a}} & \colhead{Duration} & \colhead{Full-pol} & \colhead{$N_{\rm{burst}}$} & \colhead{Rate\tablenotemark{b}} \\
    \colhead{} & \colhead{(UTC)} & \colhead{(s)} & \colhead{} & \colhead{} & \colhead{$(\rm{hr}^{-1})$}
}
\startdata
O1 &  2024-01-29T06:07:40.0 &  7200.5 &  N &  0 &  $<0.92$ \\
O2 &  2024-02-03T01:57:18.0 &  1147.8 &  N &  0 &  $<5.77$ \\
O3 &  2024-02-06T07:10:03.0 &  7194.1 &  N &  0 &  $<0.92$ \\
O4 &  2024-02-07T06:29:43.0 &  7200.5 &  N &  0 &  $<0.92$ \\
O5 &  2024-02-07T23:36:42.0 &  10800.8 &  N &  0 &  $<0.61$ \\
O6 &  2024-02-18T03:16:06.0 &  10800.8 &  N &  0 &  $<0.61$ \\
O7 &  2024-02-19T23:29:58.0 &  3600.3 &  N &  0 &  $<1.84$ \\
O8 &  2024-02-24T03:49:45.0 &  3600.3 &  N &  0 &  $<1.84$ \\
O9 &  2024-06-29T20:52:24.0 &  3600.3 &  N &  3 &  $3.00^{+2.94}_{-1.63}$ \\
O10 &  2024-06-30T19:59:51.0 &  3600.3 &  N &  1 &  $1.00^{+2.32}_{-0.83}$ \\
O11 &  2024-07-09T15:45:06.0 &  3600.3 &  N &  2 &  $2.00^{+2.66}_{-1.29}$ \\
O12 &  2024-07-10T14:10:14.0 &  25200.7 &  N &  12 &  $1.71^{+0.65}_{-0.49}$ \\
O13 &  2024-07-12T14:02:17.0 &  28799.9 &  N &  8 &  $1.00^{+0.49}_{-0.35}$ \\
O14 &  2024-07-14T15:32:06.0 &  23400.1 &  N &  1 &  $0.15^{+0.36}_{-0.13}$ \\
O15 &  2024-07-15T13:23:26.0 &  9000.1 &  N &  0 &  $<0.74$ \\
O16 &  2024-07-17T14:17:12.0 &  7200.5 &  N &  3 &  $1.50^{+1.47}_{-0.81}$ \\
O17 &  2024-07-18T13:41:39.0 &  28799.9 &  N &  18 &  $2.25^{+0.67}_{-0.52}$ \\
O18 &  2024-07-20T13:44:28.0 &  28799.9 &  N &  11 &  $1.38^{+0.55}_{-0.41}$ \\
O19 &  2024-07-21T13:17:22.0 &  28801.0 &  N &  33 &  $4.12^{+0.85}_{-0.71}$ \\
O20 &  2024-07-23T17:34:42.0 &  14400.0 &  Y &  12 &  $3.00^{+1.14}_{-0.85}$ \\
O21 &  2024-07-25T17:07:26.0 &  16199.5 &  Y &  1 &  $0.22^{+0.52}_{-0.18}$ \\
O22 &  2024-07-31T18:01:50.0 &  12600.4 &  N &  7 &  $2.00^{+1.08}_{-0.74}$ \\
O23 &  2024-08-01T17:16:44.0 &  13499.1 &  N &  0 &  $<0.49$ \\
O24 &  2024-08-02T17:03:27.0 &  14401.0 &  N &  0 &  $<0.46$ \\
O25 &  2024-08-04T16:32:13.0 &  7200.5 &  N &  0 &  $<0.92$ \\
O26 &  2024-08-06T16:59:49.0 &  7200.5 &  N &  0 &  $<0.92$ \\
O27 &  2024-08-09T17:33:31.0 &  7200.5 &  N &  0 &  $<0.92$ \\
O28 &  2024-08-18T16:27:38.0 &  3600.3 &  N &  0 &  $<1.84$ \\
O29 &  2024-08-23T16:59:01.0 &  7200.5 &  N &  0 &  $<0.92$ \\
O30 &  2024-08-29T15:53:08.0 &  7200.5 &  N &  1 &  $0.50^{+1.16}_{-0.41}$ \\
O31 &  2024-08-30T16:09:52.0 &  7200.5 &  N &  1 &  $0.50^{+1.16}_{-0.41}$ \\
O32 &  2024-08-31T14:41:02.0 &  14400.0 &  N &  2 &  $0.50^{+0.66}_{-0.32}$ \\
O33 &  2024-09-03T12:38:15.0 &  14401.0 &  N &  2 &  $0.50^{+0.66}_{-0.32}$ \\
O34 &  2024-09-04T14:44:40.0 &  12600.4 &  Y &  0 &  $<0.53$ \\
O35 &  2024-09-17T15:41:02.0 &  7199.4 &  N &  1 &  $0.50^{+1.16}_{-0.41}$ \\
O36 &  2024-09-20T15:20:28.0 &  7200.5 &  N &  0 &  $<0.92$ \\
O37 &  2024-09-23T15:39:18.0 &  7200.5 &  N &  0 &  $<0.92$ \\
O38 &  2024-10-02T15:03:17.0 &  5400.9 &  N &  1 &  $0.67^{+1.55}_{-0.55}$ \\
O39 &  2024-10-08T13:49:18.0 &  7200.5 &  N &  1 &  $0.50^{+1.16}_{-0.41}$ \\
O40 &  2024-10-14T13:19:33.0 &  7200.5 &  N &  0 &  $<0.92$ \\
O41 &  2024-10-31T08:07:34.0 &  3600.3 &  N &  1 &  $1.00^{+2.32}_{-0.83}$ \\
O42 &  2024-11-05T11:25:25.0 &  7199.4 &  N &  0 &  $<0.92$ \\
O43 &  2024-11-10T08:40:04.0 &  10800.8 &  N &  0 &  $<0.61$ \\
O44 &  2024-11-17T09:10:56.0 &  7200.5 &  N &  0 &  $<0.92$ \\
O45 &  2024-12-11T11:23:15.0 &  7200.5 &  N &  0 &  $<0.92$ \\
O46 &  2024-12-21T08:14:04.0 &  7200.5 &  N &  0 &  $<0.92$ \\
O47 &  2024-12-30T08:46:08.0 &  7200.5 &  N &  1 &  $0.50^{+1.16}_{-0.41}$ \\
O48 &  2025-01-07T06:59:05.0 &  7200.5 &  N &  2 &  $1.00^{+1.33}_{-0.64}$ \\
O49 &  2025-01-08T04:13:37.0 &  4000.8 &  N &  0 &  $<1.66$ \\
O50 &  2025-01-08T09:28:01.0 &  5400.9 &  N &  1 &  $0.67^{+1.55}_{-0.55}$ \\
O51 &  2025-01-09T01:47:51.0 &  14399.9 &  Y &  1 &  $0.25^{+0.58}_{-0.21}$ \\
O52 &  2025-01-11T07:30:08.0 &  9899.9 &  Y &  0 &  $<0.67$ \\
O53 &  2025-01-12T05:26:19.0 &  14401.0 &  N &  4 &  $1.00^{+0.79}_{-0.48}$ \\
O54 &  2025-01-13T08:25:20.0 &  7200.5 &  N &  1 &  $0.50^{+1.16}_{-0.41}$ \\
O55 &  2025-01-14T05:54:08.0 &  12600.4 &  N &  1 &  $0.29^{+0.66}_{-0.24}$ \\
O56 &  2025-01-20T09:04:09.0 &  3600.3 &  N &  6 &  $6.00^{+3.60}_{-2.37}$ \\
O57 &  2025-01-23T07:09:56.0 &  5400.9 &  Y &  5 &  $3.33^{+2.26}_{-1.44}$ \\
O58 &  2025-01-25T08:11:01.0 &  6600.3 &  Y &  5 &  $2.73^{+1.85}_{-1.18}$ \\
O59 &  2025-01-26T07:46:35.0 &  9000.1 &  Y &  2 &  $0.80^{+1.06}_{-0.52}$ \\
O60 &  2025-02-05T04:12:57.0 &  10799.7 &  N &  3 &  $1.00^{+0.98}_{-0.54}$ \\
O61 &  2025-02-06T04:06:40.0 &  10800.8 &  Y &  0 &  $<0.61$ \\
O62 &  2025-02-08T15:36:04.0 &  7200.5 &  N &  0 &  $<0.92$ \\
O63 &  2025-02-09T05:18:13.0 &  10800.8 &  N &  0 &  $<0.61$ \\
O64 &  2025-02-11T01:55:15.0 &  7200.5 &  N &  0 &  $<0.92$ \\
\tablebreak
O65 &  2025-02-14T06:00:22.0 &  9000.1 &  N &  0 &  $<0.74$ \\
O66 &  2025-02-15T05:20:34.0 &  7200.5 &  N &  1 &  $0.50^{+1.16}_{-0.41}$ \\
\enddata
\tablecomments{The number of detected bursts, $N_{\rm{burst}}$, and inferred mean burst rates (or upper limits) are for 2.25~GHz observations.}
\vspace{-5pt}
\tablenotetext{a}{Topocentric at the TMRT.}
\vspace{-5pt}
\tablenotetext{b}{The uncertainties or upper limits for burst rates correspond to $1\sigma$ Poisson errors \citep{poissonerr}.}
\end{deluxetable*}

\startlongtable
\begin{deluxetable*}{lccccccc}
\tablecaption{Properties of 155 bursts from FRB~20240114A detected with the TMRT at 2.25~GHz.}
\label{table: B2}
\tablehead{
    \multicolumn{1}{l}{$\rm{ID}_{\rm{burst}}$} & \colhead{TOA\tablenotemark{a}} & \colhead{S/N$_{\rm{det}}$\tablenotemark{b}} & \colhead{S/N\tablenotemark{c}} & \colhead{Width\tablenotemark{d}} & \colhead{Fluence\tablenotemark{e, f}} & \colhead{Energy\tablenotemark{f}} & \colhead{Peak flux density\tablenotemark{f}} \\
    \colhead{} & \colhead{(MJD)} & \colhead{} & \colhead{} & \colhead{(ms)} & \colhead{$(\rm{Jy~ms})$} & \colhead{($10^{38}~\rm{erg}$)} & \colhead{(Jy)}
}
\startdata
B01 &  60490.877254813 &  9.0 &  7.8 &  $1.81\pm0.32$ &  $1.68\pm0.34$ &  $0.69\pm0.14$ &  $1.50\pm0.30$ \\
B02 &  60490.877781704 &  10.6 &  11.5 &  $4.24\pm0.46$ &  $3.27\pm0.65$ &  $1.34\pm0.27$ &  $1.47\pm0.29$ \\
B03 &  60490.877863290 &  12.9 &  15.5 &  $4.82\pm0.31$ &  $5.10\pm1.02$ &  $1.99\pm0.40$ &  $1.93\pm0.39$ \\
B04 &  60491.871781302 &  7.0 &  6.7 &  $8.50\pm1.39$ &  $2.66\pm0.53$ &  $1.09\pm0.22$ &  $1.33\pm0.27$ \\
B05 &  60500.668215640 &  7.7 &  7.1 &  $4.14\pm0.78$ &  $2.18\pm0.44$ &  $0.88\pm0.18$ &  $1.52\pm0.30$ \\
B06 &  60500.686660406 &  7.5 &  10.3 &  $9.09\pm1.60$ &  $4.43\pm0.89$ &  $1.78\pm0.36$ &  $1.71\pm0.34$ \\
B07 &  60501.614672263 &  7.4 &  5.4 &  $2.23\pm0.42$ &  $1.15\pm0.23$ &  $0.47\pm0.09$ &  $1.27\pm0.25$ \\
B08 &  60501.618415652 &  16.5 &  18.2 &  $3.18\pm0.24$ &  $4.58\pm0.92$ &  $1.88\pm0.38$ &  $1.90\pm0.38$ \\
B09 &  60501.628657695 &  11.2 &  26.1 &  $46.83\pm4.29$ &  $20.06\pm4.01$ &  $8.24\pm1.65$ &  $1.97\pm0.39$ \\
B10 &  60501.640032704 &  7.3 &  8.1 &  $5.94\pm0.91$ &  $2.73\pm0.55$ &  $1.12\pm0.22$ &  $1.39\pm0.28$ \\
B11 &  60501.674640852 &  9.3 &  7.9 &  $3.04\pm0.67$ &  $2.06\pm0.41$ &  $0.85\pm0.17$ &  $1.59\pm0.32$ \\
B12 &  60501.714895960 &  7.8 &  6.3 &  $2.14\pm0.43$ &  $1.35\pm0.27$ &  $0.56\pm0.11$ &  $1.53\pm0.31$ \\
B13 &  60501.736947322 &  7.8 &  6.7 &  $2.04\pm0.48$ &  $1.43\pm0.29$ &  $0.59\pm0.12$ &  $1.70\pm0.34$ \\
B14 &  60501.746368895 &  9.2 &  9.7 &  $2.28\pm0.25$ &  $2.08\pm0.42$ &  $0.85\pm0.17$ &  $1.32\pm0.26$ \\
B15 &  60501.750034617 &  10.8 &  8.9 &  $1.04\pm0.13$ &  $1.20\pm0.24$ &  $0.48\pm0.10$ &  $1.76\pm0.35$ \\
B16 &  60501.823491590 &  9.0 &  8.6 &  $2.52\pm0.48$ &  $2.12\pm0.42$ &  $0.83\pm0.17$ &  $1.46\pm0.29$ \\
B17 &  60501.825323426 &  29.2 &  34.7 &  $5.17\pm0.22$ &  $9.71\pm1.94$ &  $3.95\pm0.79$ &  $2.71\pm0.54$ \\
B18 &  60501.873196720 &  7.6 &  6.4 &  $3.04\pm0.56$ &  $1.68\pm0.34$ &  $0.69\pm0.14$ &  $1.54\pm0.31$ \\
B19 &  60503.591180076 &  8.2 &  6.6 &  $0.40\pm0.07$ &  $0.63\pm0.13$ &  $0.26\pm0.05$ &  $1.77\pm0.35$ \\
B20 &  60503.637351395 &  10.2 &  12.2 &  $4.74\pm0.67$ &  $3.39\pm0.68$ &  $1.35\pm0.27$ &  $1.47\pm0.29$ \\
B21 &  60503.679182276 &  7.4 &  7.1 &  $4.57\pm0.88$ &  $2.18\pm0.44$ &  $0.87\pm0.17$ &  $1.41\pm0.28$ \\
B22 &  60503.685747247 &  26.0 &  36.5 &  $8.10\pm0.36$ &  $15.49\pm3.10$ &  $6.23\pm1.25$ &  $2.65\pm0.53$ \\
B23 &  60503.691442388 &  8.2 &  10.4 &  $8.33\pm1.06$ &  $3.50\pm0.70$ &  $1.44\pm0.29$ &  $1.63\pm0.33$ \\
B24 &  60503.711994225 &  10.8 &  13.3 &  $6.10\pm0.67$ &  $4.04\pm0.81$ &  $1.66\pm0.33$ &  $1.61\pm0.32$ \\
B25 &  60503.790386689 &  9.7 &  9.2 &  $2.03\pm0.28$ &  $1.71\pm0.34$ &  $0.70\pm0.14$ &  $1.39\pm0.28$ \\
B26 &  60503.806707595 &  8.1 &  6.2 &  $5.32\pm0.93$ &  $2.19\pm0.44$ &  $0.90\pm0.18$ &  $1.45\pm0.29$ \\
B27 &  60505.667786678 &  7.7 &  6.8 &  $0.47\pm0.08$ &  $0.69\pm0.14$ &  $0.25\pm0.05$ &  $1.74\pm0.35$ \\
B28 &  60508.669378683 &  17.8 &  13.8 &  $0.37\pm0.04$ &  $1.76\pm0.35$ &  $0.72\pm0.14$ &  $3.76\pm0.75$ \\
B29 &  60508.669378752 &  16.7 &  15.2 &  $0.50\pm0.05$ &  $1.57\pm0.31$ &  $0.65\pm0.13$ &  $3.83\pm0.77$ \\
B30 &  60508.678488274 &  7.2 &  5.5 &  $2.18\pm0.48$ &  $1.19\pm0.24$ &  $0.49\pm0.10$ &  $1.58\pm0.32$ \\
B31 &  60509.604672466 &  105.7 &  215.5 &  $3.48\pm0.09$ &  $47.65\pm9.53$ &  $18.19\pm3.64$ &  $23.02\pm4.60$ \\
B32 &  60509.634631794 &  12.0 &  11.4 &  $2.70\pm0.31$ &  $2.81\pm0.56$ &  $1.10\pm0.22$ &  $1.95\pm0.39$ \\
B33 &  60509.635296208 &  8.9 &  6.2 &  $1.85\pm0.41$ &  $1.34\pm0.27$ &  $0.55\pm0.11$ &  $1.39\pm0.28$ \\
B34 &  60509.643131645 &  9.0 &  7.6 &  $4.17\pm0.90$ &  $2.15\pm0.43$ &  $0.88\pm0.18$ &  $1.36\pm0.27$ \\
B35 &  60509.651494822 &  47.0 &  48.3 &  $1.47\pm0.04$ &  $7.83\pm1.57$ &  $3.21\pm0.64$ &  $5.41\pm1.08$ \\
B36 &  60509.693539407 &  154.7 &  391.7 &  $5.29\pm0.15$ &  $277.41\pm55.48$ &  $35.31\pm7.06$ &  $124.85\pm24.97$ \\
B37 &  60509.736924515 &  11.7 &  11.0 &  $1.73\pm0.21$ &  $2.22\pm0.44$ &  $0.91\pm0.18$ &  $1.91\pm0.38$ \\
B38 &  60509.794599389 &  20.7 &  17.6 &  $0.49\pm0.03$ &  $2.44\pm0.49$ &  $1.00\pm0.20$ &  $4.36\pm0.87$ \\
B39 &  60509.818438946 &  10.1 &  10.5 &  $2.27\pm0.33$ &  $2.26\pm0.45$ &  $0.93\pm0.19$ &  $1.38\pm0.28$ \\
B40 &  60509.818716706 &  9.5 &  9.7 &  $4.05\pm0.53$ &  $2.77\pm0.55$ &  $1.14\pm0.23$ &  $1.38\pm0.28$ \\
B41 &  60509.828105766 &  7.4 &  6.6 &  $4.07\pm0.80$ &  $2.03\pm0.41$ &  $0.80\pm0.16$ &  $1.29\pm0.26$ \\
B42 &  60509.829040813 &  21.7 &  24.9 &  $5.37\pm0.47$ &  $7.65\pm1.53$ &  $3.14\pm0.63$ &  $2.09\pm0.42$ \\
B43 &  60509.853829818 &  7.6 &  6.6 &  $3.24\pm0.64$ &  $1.72\pm0.34$ &  $0.71\pm0.14$ &  $1.37\pm0.27$ \\
B44 &  60509.862573832 &  15.3 &  12.8 &  $1.80\pm0.17$ &  $2.38\pm0.48$ &  $0.98\pm0.20$ &  $2.05\pm0.41$ \\
B45 &  60509.882481974 &  13.4 &  13.2 &  $3.00\pm0.26$ &  $3.16\pm0.63$ &  $1.30\pm0.26$ &  $1.60\pm0.32$ \\
B46 &  60509.888330730 &  8.1 &  7.0 &  $7.82\pm1.33$ &  $2.82\pm0.56$ &  $1.12\pm0.22$ &  $1.37\pm0.27$ \\
B47 &  60509.890516966 &  11.6 &  11.8 &  $3.96\pm0.48$ &  $3.21\pm0.64$ &  $1.32\pm0.26$ &  $1.58\pm0.32$ \\
B48 &  60509.908105884 &  9.8 &  7.1 &  $1.25\pm0.21$ &  $1.33\pm0.27$ &  $0.55\pm0.11$ &  $1.73\pm0.35$ \\
B49 &  60511.584171533 &  10.3 &  11.3 &  $4.76\pm0.54$ &  $3.82\pm0.76$ &  $1.35\pm0.27$ &  $1.59\pm0.32$ \\
B50 &  60511.595121957 &  9.1 &  9.7 &  $3.12\pm0.48$ &  $2.64\pm0.53$ &  $1.02\pm0.20$ &  $2.08\pm0.42$ \\
B51 &  60511.626970849 &  7.4 &  6.0 &  $3.68\pm0.81$ &  $1.80\pm0.36$ &  $0.66\pm0.13$ &  $1.63\pm0.33$ \\
B52 &  60511.642859541 &  7.8 &  7.2 &  $2.82\pm0.74$ &  $2.08\pm0.42$ &  $0.83\pm0.17$ &  $1.86\pm0.37$ \\
B53 &  60511.647019470 &  24.4 &  22.3 &  $2.55\pm0.14$ &  $7.27\pm1.45$ &  $2.89\pm0.58$ &  $3.10\pm0.62$ \\
B54 &  60511.657714415 &  7.8 &  6.6 &  $1.43\pm0.28$ &  $1.15\pm0.23$ &  $0.46\pm0.09$ &  $1.56\pm0.31$ \\
B55 &  60511.684513740 &  75.0 &  102.1 &  $2.95\pm0.03$ &  $25.03\pm5.01$ &  $10.07\pm2.01$ &  $9.03\pm1.81$ \\
B56 &  60511.728715852 &  7.9 &  9.4 &  $8.57\pm1.24$ &  $4.09\pm0.82$ &  $1.51\pm0.30$ &  $1.52\pm0.30$ \\
B57 &  60511.777229725 &  7.5 &  5.9 &  $1.99\pm0.54$ &  $1.11\pm0.22$ &  $0.44\pm0.09$ &  $1.68\pm0.34$ \\
B58 &  60511.797287974 &  19.1 &  17.2 &  $3.31\pm0.23$ &  $4.42\pm0.88$ &  $1.82\pm0.36$ &  $1.88\pm0.38$ \\
B59 &  60511.816745781 &  17.0 &  20.3 &  $3.87\pm0.26$ &  $5.64\pm1.13$ &  $2.32\pm0.46$ &  $2.15\pm0.43$ \\
B60 &  60512.568038625 &  18.4 &  16.9 &  $0.87\pm0.06$ &  $2.54\pm0.51$ &  $0.95\pm0.19$ &  $3.15\pm0.63$ \\
B61 &  60512.579568329 &  9.4 &  8.5 &  $1.93\pm0.34$ &  $1.95\pm0.39$ &  $0.70\pm0.14$ &  $1.56\pm0.31$ \\
B62 &  60512.586517957 &  28.2 &  20.8 &  $2.09\pm0.10$ &  $7.80\pm1.56$ &  $3.17\pm0.63$ &  $3.43\pm0.69$ \\
B63 &  60512.633836461 &  7.8 &  8.3 &  $3.87\pm1.06$ &  $2.53\pm0.51$ &  $0.90\pm0.18$ &  $2.11\pm0.42$ \\
B64 &  60512.642024632 &  14.3 &  13.0 &  $2.25\pm0.25$ &  $2.87\pm0.57$ &  $1.11\pm0.22$ &  $2.10\pm0.42$ \\
B65 &  60512.651620303 &  14.8 &  15.0 &  $4.10\pm0.36$ &  $4.39\pm0.88$ &  $1.80\pm0.36$ &  $1.73\pm0.35$ \\
B66 &  60512.653223267 &  14.9 &  14.2 &  $3.15\pm0.31$ &  $3.68\pm0.74$ &  $1.42\pm0.28$ &  $1.91\pm0.38$ \\
B67 &  60512.654142166 &  17.3 &  18.8 &  $3.99\pm0.28$ &  $5.32\pm1.06$ &  $2.19\pm0.44$ &  $2.22\pm0.44$ \\
B68 &  60512.664815665 &  8.6 &  7.3 &  $2.30\pm0.33$ &  $1.60\pm0.32$ &  $0.62\pm0.12$ &  $1.50\pm0.30$ \\
B69 &  60512.667726336 &  86.2 &  96.1 &  $3.39\pm0.05$ &  $25.09\pm5.02$ &  $10.10\pm2.02$ &  $8.15\pm1.63$ \\
B70 &  60512.669901654 &  14.0 &  11.9 &  $2.52\pm0.26$ &  $2.65\pm0.53$ &  $1.02\pm0.20$ &  $1.81\pm0.36$ \\
B71 &  60512.671530982 &  11.9 &  12.7 &  $4.65\pm0.50$ &  $3.83\pm0.77$ &  $1.57\pm0.31$ &  $1.61\pm0.32$ \\
B72 &  60512.673234841 &  7.8 &  6.9 &  $1.72\pm0.32$ &  $1.14\pm0.23$ &  $0.41\pm0.08$ &  $1.64\pm0.33$ \\
B73 &  60512.676784417 &  12.1 &  12.8 &  $4.36\pm0.47$ &  $3.38\pm0.68$ &  $1.38\pm0.28$ &  $1.84\pm0.37$ \\
B74 &  60512.677568155 &  15.1 &  17.1 &  $4.61\pm0.41$ &  $4.85\pm0.97$ &  $1.99\pm0.40$ &  $2.07\pm0.41$ \\
B75 &  60512.689251370 &  8.0 &  5.4 &  $1.32\pm0.46$ &  $1.17\pm0.23$ &  $0.48\pm0.10$ &  $1.71\pm0.34$ \\
B76 &  60512.692032908 &  11.5 &  12.2 &  $3.94\pm0.42$ &  $6.90\pm1.38$ &  $2.58\pm0.52$ &  $1.72\pm0.34$ \\
B77 &  60512.697377568 &  9.4 &  7.3 &  $1.63\pm0.41$ &  $1.64\pm0.33$ &  $0.62\pm0.12$ &  $1.50\pm0.30$ \\
B78 &  60512.712234041 &  14.0 &  31.6 &  $53.41\pm4.64$ &  $26.00\pm5.20$ &  $10.03\pm2.01$ &  $2.30\pm0.46$ \\
B79 &  60512.712916046 &  11.3 &  10.3 &  $2.84\pm0.30$ &  $2.55\pm0.51$ &  $0.99\pm0.20$ &  $1.61\pm0.32$ \\
B80 &  60512.738444218 &  10.3 &  9.6 &  $3.42\pm0.72$ &  $2.72\pm0.54$ &  $1.12\pm0.22$ &  $1.28\pm0.26$ \\
B81 &  60512.761398345 &  8.0 &  6.9 &  $3.82\pm0.64$ &  $1.82\pm0.36$ &  $0.75\pm0.15$ &  $1.65\pm0.33$ \\
B82 &  60512.769740873 &  11.2 &  14.3 &  $4.61\pm0.48$ &  $4.83\pm0.97$ &  $1.98\pm0.40$ &  $1.67\pm0.33$ \\
B83 &  60512.770525156 &  9.7 &  8.9 &  $2.61\pm0.40$ &  $1.98\pm0.40$ &  $0.77\pm0.15$ &  $1.56\pm0.31$ \\
B84 &  60512.774487784 &  11.7 &  12.7 &  $4.09\pm0.38$ &  $4.33\pm0.87$ &  $1.74\pm0.35$ &  $1.87\pm0.37$ \\
B85 &  60512.774641899 &  12.3 &  13.3 &  $3.76\pm0.56$ &  $3.20\pm0.64$ &  $1.30\pm0.26$ &  $1.49\pm0.30$ \\
B86 &  60512.788798475 &  10.6 &  9.3 &  $3.61\pm0.51$ &  $2.53\pm0.51$ &  $1.04\pm0.21$ &  $1.40\pm0.28$ \\
B87 &  60512.837235334 &  11.2 &  10.5 &  $4.46\pm0.56$ &  $3.18\pm0.64$ &  $1.31\pm0.26$ &  $1.47\pm0.29$ \\
B88 &  60512.839860217 &  9.0 &  9.8 &  $2.52\pm0.41$ &  $2.10\pm0.42$ &  $0.86\pm0.17$ &  $1.36\pm0.27$ \\
B89 &  60512.841146996 &  8.3 &  6.3 &  $2.49\pm0.43$ &  $1.52\pm0.30$ &  $0.62\pm0.12$ &  $1.18\pm0.24$ \\
B90 &  60512.857326366 &  11.5 &  10.1 &  $2.95\pm0.35$ &  $2.55\pm0.51$ &  $1.05\pm0.21$ &  $2.07\pm0.41$ \\
B91 &  60512.864387231 &  19.0 &  17.9 &  $1.55\pm0.11$ &  $3.22\pm0.64$ &  $1.31\pm0.26$ &  $2.63\pm0.53$ \\
B92 &  60512.873368944 &  20.5 &  22.0 &  $3.97\pm0.24$ &  $6.58\pm1.32$ &  $2.68\pm0.54$ &  $2.32\pm0.46$ \\
B93 &  60514.748025821 &  9.9 &  7.2 &  $3.97\pm0.42$ &  $1.73\pm0.35$ &  $0.70\pm0.14$ &  $1.16\pm0.23$ \\
B94 &  60514.783919524 &  17.7 &  20.6 &  $2.00\pm0.12$ &  $4.16\pm0.83$ &  $1.69\pm0.34$ &  $2.70\pm0.54$ \\
B95 &  60514.796840988 &  9.8 &  7.4 &  $1.20\pm0.19$ &  $1.26\pm0.25$ &  $0.52\pm0.10$ &  $1.74\pm0.35$ \\
B96 &  60514.816124902 &  35.4 &  33.8 &  $1.85\pm0.07$ &  $13.57\pm2.71$ &  $5.57\pm1.11$ &  $5.43\pm1.09$ \\
B97 &  60514.816393025 &  15.1 &  15.2 &  $1.00\pm0.09$ &  $2.16\pm0.43$ &  $0.88\pm0.18$ &  $3.09\pm0.62$ \\
B98 &  60514.838598563 &  10.6 &  11.6 &  $3.21\pm0.37$ &  $2.92\pm0.58$ &  $1.19\pm0.24$ &  $1.71\pm0.34$ \\
B99 &  60514.849112130 &  11.7 &  13.4 &  $5.02\pm0.52$ &  $4.09\pm0.82$ &  $1.66\pm0.33$ &  $2.06\pm0.41$ \\
B100 &  60514.860204393 &  8.5 &  8.8 &  $4.54\pm0.78$ &  $2.32\pm0.46$ &  $0.94\pm0.19$ &  $1.65\pm0.33$ \\
B101 &  60514.866236488 &  15.4 &  15.5 &  $2.72\pm0.23$ &  $3.62\pm0.72$ &  $1.48\pm0.30$ &  $1.89\pm0.38$ \\
B102 &  60514.875840437 &  25.5 &  26.8 &  $2.77\pm0.14$ &  $6.26\pm1.25$ &  $2.57\pm0.51$ &  $2.88\pm0.58$ \\
B103 &  60514.878805484 &  14.5 &  15.1 &  $1.00\pm0.12$ &  $2.18\pm0.44$ &  $0.89\pm0.18$ &  $2.51\pm0.50$ \\
B104 &  60514.879354005 &  8.2 &  9.7 &  $8.30\pm1.09$ &  $4.04\pm0.81$ &  $1.64\pm0.33$ &  $1.60\pm0.32$ \\
B105 &  60516.742769399 &  7.4 &  5.8 &  $7.38\pm1.15$ &  $2.34\pm0.47$ &  $0.94\pm0.19$ &  $1.68\pm0.34$ \\
B106 &  60522.764164634 &  8.2 &  8.3 &  $2.30\pm0.40$ &  $1.80\pm0.36$ &  $0.73\pm0.15$ &  $1.26\pm0.25$ \\
B107 &  60522.783900389 &  12.6 &  14.4 &  $2.74\pm0.32$ &  $3.47\pm0.69$ &  $1.41\pm0.28$ &  $2.25\pm0.45$ \\
B108 &  60522.850074189 &  10.5 &  11.8 &  $4.98\pm0.62$ &  $3.35\pm0.67$ &  $1.36\pm0.27$ &  $1.70\pm0.34$ \\
B109 &  60522.851091723 &  8.7 &  7.8 &  $3.50\pm0.87$ &  $2.05\pm0.41$ &  $0.83\pm0.17$ &  $1.44\pm0.29$ \\
B110 &  60522.851092297 &  32.1 &  37.3 &  $5.36\pm0.18$ &  $11.61\pm2.32$ &  $4.72\pm0.94$ &  $2.86\pm0.57$ \\
B111 &  60522.896691061 &  11.8 &  15.3 &  $7.85\pm0.75$ &  $5.19\pm1.04$ &  $2.11\pm0.42$ &  $1.69\pm0.34$ \\
B112 &  60522.902640670 &  10.2 &  9.9 &  $2.17\pm0.32$ &  $2.13\pm0.43$ &  $0.87\pm0.17$ &  $1.96\pm0.39$ \\
B113 &  60551.677883704 &  7.5 &  6.1 &  $4.28\pm0.94$ &  $2.05\pm0.41$ &  $0.84\pm0.17$ &  $1.33\pm0.27$ \\
B114 &  60552.702419476 &  11.7 &  12.5 &  $2.73\pm0.27$ &  $2.69\pm0.54$ &  $1.11\pm0.22$ &  $1.49\pm0.30$ \\
B115 &  60553.709221328 &  8.7 &  7.6 &  $3.71\pm0.65$ &  $2.25\pm0.45$ &  $0.84\pm0.17$ &  $1.75\pm0.35$ \\
B116 &  60553.721187638 &  13.9 &  12.9 &  $3.94\pm0.35$ &  $4.36\pm0.87$ &  $1.79\pm0.36$ &  $2.05\pm0.41$ \\
B117 &  60556.583936821 &  21.7 &  25.4 &  $1.82\pm0.12$ &  $6.31\pm1.26$ &  $2.54\pm0.51$ &  $3.61\pm0.72$ \\
B118 &  60556.589199651 &  34.6 &  24.9 &  $1.26\pm0.05$ &  $7.27\pm1.45$ &  $2.66\pm0.53$ &  $5.79\pm1.16$ \\
B119 &  60570.699308662 &  17.8 &  16.8 &  $5.32\pm0.37$ &  $5.38\pm1.08$ &  $2.16\pm0.43$ &  $2.06\pm0.41$ \\
B120 &  60585.683098037 &  8.0 &  7.1 &  $1.88\pm0.29$ &  $1.33\pm0.27$ &  $0.54\pm0.11$ &  $1.54\pm0.31$ \\
B121 &  60591.612777770 &  8.6 &  7.9 &  $2.42\pm0.36$ &  $1.79\pm0.36$ &  $0.65\pm0.13$ &  $1.67\pm0.33$ \\
B122 &  60614.344688587 &  9.9 &  9.2 &  $7.15\pm0.89$ &  $3.51\pm0.70$ &  $1.41\pm0.28$ &  $1.54\pm0.31$ \\
B123 &  60674.366632354 &  33.3 &  35.2 &  $2.96\pm0.11$ &  $9.15\pm1.83$ &  $3.76\pm0.75$ &  $3.68\pm0.74$ \\
B124 &  60682.297361270 &  101.6 &  180.8 &  $2.89\pm0.03$ &  $50.07\pm10.01$ &  $20.56\pm4.11$ &  $18.33\pm3.67$ \\
B125 &  60682.322387620 &  14.8 &  14.8 &  $1.38\pm0.12$ &  $3.10\pm0.62$ &  $1.27\pm0.25$ &  $2.37\pm0.47$ \\
B126 &  60683.405511537 &  36.9 &  35.5 &  $0.35\pm0.01$ &  $3.14\pm0.63$ &  $1.29\pm0.26$ &  $8.55\pm1.71$ \\
B127 &  60684.213436635 &  18.6 &  20.2 &  $1.15\pm0.07$ &  $3.09\pm0.62$ &  $1.27\pm0.25$ &  $3.37\pm0.67$ \\
B128 &  60687.320886837 &  9.9 &  8.4 &  $2.27\pm0.44$ &  $1.97\pm0.39$ &  $0.70\pm0.14$ &  $1.93\pm0.39$ \\
B129 &  60687.321737199 &  13.1 &  12.1 &  $2.72\pm0.32$ &  $3.06\pm0.61$ &  $1.02\pm0.20$ &  $2.13\pm0.43$ \\
B130 &  60687.324272474 &  10.2 &  10.2 &  $2.48\pm0.34$ &  $2.20\pm0.44$ &  $0.90\pm0.18$ &  $1.41\pm0.28$ \\
B131 &  60687.325454391 &  7.5 &  7.1 &  $1.65\pm0.35$ &  $2.00\pm0.40$ &  $0.71\pm0.14$ &  $1.94\pm0.39$ \\
B132 &  60688.348701796 &  8.3 &  6.8 &  $4.65\pm0.73$ &  $2.30\pm0.46$ &  $0.76\pm0.15$ &  $1.46\pm0.29$ \\
B133 &  60689.315550539 &  8.0 &  7.8 &  $2.34\pm0.59$ &  $1.67\pm0.33$ &  $0.69\pm0.14$ &  $1.78\pm0.36$ \\
B134 &  60695.374708791 &  10.7 &  7.2 &  $2.39\pm0.51$ &  $1.56\pm0.31$ &  $0.64\pm0.13$ &  $1.36\pm0.27$ \\
B135 &  60695.389946345 &  20.5 &  22.1 &  $4.99\pm0.26$ &  $6.99\pm1.40$ &  $2.87\pm0.57$ &  $2.05\pm0.41$ \\
B136 &  60695.393007488 &  41.3 &  62.1 &  $5.76\pm0.11$ &  $22.26\pm4.45$ &  $9.14\pm1.83$ &  $5.09\pm1.02$ \\
B137 &  60695.398456532 &  28.5 &  40.9 &  $1.80\pm0.09$ &  $8.09\pm1.62$ &  $3.32\pm0.66$ &  $4.84\pm0.97$ \\
B138 &  60695.399895691 &  8.5 &  10.3 &  $19.50\pm2.85$ &  $5.61\pm1.12$ &  $2.05\pm0.41$ &  $1.84\pm0.37$ \\
B139 &  60695.400039884 &  20.3 &  21.9 &  $0.99\pm0.08$ &  $3.14\pm0.63$ &  $1.29\pm0.26$ &  $3.68\pm0.74$ \\
B140 &  60698.320643688 &  19.8 &  22.4 &  $4.61\pm0.28$ &  $7.02\pm1.40$ &  $2.68\pm0.54$ &  $2.24\pm0.45$ \\
B141 &  60698.320786123 &  20.9 &  29.2 &  $16.93\pm1.15$ &  $14.61\pm2.92$ &  $6.00\pm1.20$ &  $2.79\pm0.56$ \\
B142 &  60698.344802560 &  9.3 &  10.8 &  $2.52\pm0.29$ &  $2.46\pm0.49$ &  $1.01\pm0.20$ &  $2.15\pm0.43$ \\
B143 &  60698.354370198 &  9.9 &  8.6 &  $1.68\pm0.26$ &  $1.59\pm0.32$ &  $0.65\pm0.13$ &  $1.69\pm0.34$ \\
B144 &  60698.354370319 &  7.2 &  3.5 &  $1.29\pm0.38$ &  $0.53\pm0.11$ &  $0.22\pm0.04$ &  $1.28\pm0.26$ \\
B145 &  60700.344099273 &  11.1 &  8.6 &  $2.15\pm0.38$ &  $1.85\pm0.37$ &  $0.76\pm0.15$ &  $1.39\pm0.28$ \\
B146 &  60700.352874848 &  7.9 &  4.7 &  $1.49\pm0.41$ &  $0.91\pm0.18$ &  $0.35\pm0.07$ &  $1.49\pm0.30$ \\
B147 &  60700.356133685 &  13.3 &  14.7 &  $5.16\pm0.44$ &  $4.88\pm0.98$ &  $1.86\pm0.37$ &  $2.57\pm0.51$ \\
B148 &  60700.388086513 &  9.3 &  9.1 &  $9.82\pm1.52$ &  $3.69\pm0.74$ &  $1.51\pm0.30$ &  $1.26\pm0.25$ \\
B149 &  60700.390532284 &  22.3 &  38.2 &  $14.06\pm0.62$ &  $20.17\pm4.03$ &  $7.29\pm1.46$ &  $3.27\pm0.65$ \\
B150 &  60701.342236339 &  7.8 &  9.4 &  $5.08\pm1.11$ &  $3.02\pm0.60$ &  $1.24\pm0.25$ &  $1.21\pm0.24$ \\
B151 &  60701.378825815 &  9.6 &  8.9 &  $2.76\pm0.40$ &  $2.20\pm0.44$ &  $0.84\pm0.17$ &  $1.63\pm0.33$ \\
B152 &  60711.174810850 &  9.4 &  19.0 &  $8.36\pm0.77$ &  $7.87\pm1.57$ &  $3.23\pm0.65$ &  $2.25\pm0.45$ \\
B153 &  60711.239603997 &  7.3 &  4.5 &  $2.07\pm0.59$ &  $0.88\pm0.18$ &  $0.33\pm0.07$ &  $1.95\pm0.39$ \\
B154 &  60711.254558469 &  99.8 &  209.1 &  $3.26\pm0.10$ &  $48.90\pm9.78$ &  $20.08\pm4.02$ &  $33.65\pm6.73$ \\
B155 &  60721.295201147 &  15.0 &  18.9 &  $2.73\pm0.21$ &  $6.71\pm1.34$ &  $2.76\pm0.55$ &  $2.30\pm0.46$ \\
\enddata
\tablenotetext{a}{Corrected to the Solar System Barycenter at infinite frequency using a DM of 527.7~$\rm{pc~cm}^{-3}$ \citep{ATel16420}, a DM constant of $1/0.241~\rm{GHz^2~cm^3~pc^{-1}~ms}$, and the EVN position \citep{ATel16542}.}
\vspace{-5pt}
\tablenotetext{b}{The S/N reported by the searching pipeline.}
\vspace{-5pt}
\tablenotetext{c}{The S/N obtained after manual removal of RFI.}
\vspace{-5pt}
\tablenotetext{d}{Defined as the full width at half-maximum (FWHM) of a Gaussian function.}
\vspace{-5pt}
\tablenotetext{e}{Summing over the on-burst region (see Sec.~\ref{sec:energy}).}
\vspace{-5pt}
\tablenotetext{f}{Assuming a conservative uncertainty of 20\% dominated by the uncertainty of SEFD.}
\end{deluxetable*}

\end{CJK*}
\end{document}